\documentclass[preprint,notoc]{JHEP3}

\usepackage{axodraw}

\def\PlusBreak#1{+ \nonumber \\          
          &&  \hphantom{#1} \! \null   +}
\def\MinusBreak#1{- \nonumber \\         
          &&  \hphantom{#1} \!  \null  -}

\def\PlusBreakAdjust#1#2{+ \nonumber \\ [#1]      
          &&  \hphantom{#2} \! \null  +}    
\def\MinusBreakAdjust#1#2{- \nonumber \\ [#1]     
          &&  \hphantom{#2} \!  \null -}

\def\Tr{\mathop{\rm tr}\nolimits}

\def\BigglBl{\Biggl[}

\def\bigglB{\biggl[}
\def\biggrB{\biggr]}

\def\Xt{\tilde X}
\def\Yt{\tilde Y}

\def\ksl{\not{\hbox{\kern-2.3pt $k$}}}

\def\e{\epsilon}

\def\gluon{{\rm gluon}}

\def\Ord{{\cal O}}
\def\cm{{\cal M}}

\def\Re{{\rm Re}}
\def\Nf{{N_{\! f}}}
\def\Nfsq{{N_{\! f}^2}}
\def\la{\langle}
\def\ra{\rangle}
\def\RS{{\scriptscriptstyle\rm R\!.S\!.}}

\def\bom#1{{\mbox{\boldmath $#1$}}}
\def\MSbar{\overline{\rm MS}}
\def\DRbar{\overline{\rm DR}}
\def\HV{{\rm HV}}

\def\FDH{{\rm FDH}}
\def\lr{\leftrightarrow}
\def\li#1{{\mathop{\rm Li}\nolimits}_#1}
\def\Li{\mathop{\rm Li}\nolimits}
\def\ggtogg{{gg \to gg}}
\def\tg{\tilde{g}}

\def\alphas{\alpha_s}

\def\Susy{{\scriptscriptstyle\rm SYM}}
\def\ASusy#1#2{{A^{\Susy, #1}_{#2}}}
\def\BSusy#1#2{{B^{\Susy, #1}_{#2}}}

\def\Eqn#1{Equation~(\ref{#1})}
\def\eqn#1{eq.~(\ref{#1})}
\def\eqns#1#2{eqs.~(\ref{#1}) and~(\ref{#2})}

\def\scalar{{\rm scalar}}
\def\gluon{{\rm gluon}}
\def\fermion{{\rm fermion}}

\def\rg{r_\Gamma}

\def\Boxfour{{\rm Box}^{(4)}}
\def\Boxsix{{\rm Box}^{(6)}}
\def\Boxeight{{\rm Box}^{(8)}}

\def\Trifour{{\rm Tri}^{(4)}}
\def\Trisix{{\rm Tri}^{(6)}}

\def\Bubfour{{\rm Bub}^{(4)}}
\def\Bubsix{{\rm Bub}^{(6)}}

\def\spa#1.#2{\left\langle#1\,#2\right\rangle}
\def\spb#1.#2{\left[#1\,#2\right]}
\def\lor#1.#2{\left(#1\,#2\right)}
\def\trc{{\rm Tr}}
\def\tS{{\tt S}}
\def\tT{{\tt T}}
\def\tU{{\tt U}}
\def\I{{\cal I}}
\def\P{{\rm P}}
\def\NP{{\rm NP}}
\def\mud{\lambda}
\def\pol{\varepsilon}
\def\fig#1{figure~{\ref{#1}}}

\preprint{
  hep-ph/0201161\\
  SLAC--PUB--9103\\
  UCLA/02/TEP/1\\
  January, 2002}

\title{Two-Loop Helicity Amplitudes for Gluon-Gluon Scattering 
in QCD and Supersymmetric Yang-Mills Theory}

\author{Zvi Bern,\thanks{Research supported by the US Department of 
Energy under grant DE-FG03-91ER40662.}~ Abilio De Freitas$^*$ \\
	Department of Physics and Astronomy\\
	UCLA, Los Angeles, CA 90095-1547\\
	E-mail: \email{bern@physics.ucla.edu}, 
                \email{freitas@physics.ucla.edu}}

\author{Lance Dixon\thanks{Research supported by the US Department of 
Energy under grant DE-AC03-76SF00515.}\\
	Stanford Linear Accelerator Center, Stanford University\\
	Stanford, CA 94309\\
	E-mail: \email{lance@slac.stanford.edu}}

\abstract{
We present the two-loop helicity amplitudes for the scattering
of two gluons into two gluons in QCD, which are relevant for
next-to-next-to-leading order corrections to jet production at hadron
colliders.  We give the results in the `t Hooft-Veltman and 
four-dimensional helicity variants of dimensional regularization.
Summing our expressions over helicities and colors, and converting to 
conventional dimensional regularization, gives results in complete
agreement with those of Glover, Oleari and Tejeda-Yeomans.
We also present the amplitudes for $2 \to 2$ scattering in pure $N=1$
supersymmetric Yang-Mills theory.}


\dedicated{\centerline{Submitted to JHEP}}

\begin{document}


\section{Introduction}\label{IntroSection}

For at least the next decade, the energy frontier for accelerator-based
particle physics will be located at hadron colliders, the Tevatron at
Fermilab and the Large Hadron Collider at CERN.  At a given large momentum
transfer, the most copious events at these colliders should be hadronic
jets.  To test the Standard Model at the shortest possible distances,
therefore, the jet production cross section should be known with the
highest possible precision.  Existing calculations of jet production at
next-to-leading order (NLO) in the strong coupling constant
$\alpha_s$~\cite{Aversa,EKS,JETRAD} agree well with the data over a broad
range of transverse momentum.  Still, the NLO predictions have an
uncertainty from higher order corrections, traditionally estimated from
dependence on the renormalization and factorization scales, which is of
order 10\% or more.  For very large momentum transfer the predictions can
be improved by resumming threshold logarithms~\cite{Kidonakis}.  There are
also sizable uncertainties associated with the experimental input to the
parton distribution functions~\cite{PDFuncertainty}, even though global 
fits to the data have recently been performed~\cite{MRSTNNLO} within an
approximate next-to-next-to-leading order (NNLO) 
framework~\cite{NNLOPDFApprox}.  Nevertheless,
an exact NNLO computation of jet production rates would be very welcome.
Besides reducing the scale uncertainties for jet rates, the same numerical
program should allow a better understanding of energy flows within jets, 
as a jet may consist of up to three partons at this order.

Several types of QCD amplitudes are required for a NNLO calculation of jet
production at hadron colliders.  Both the tree amplitudes for six external
partons~\cite{TreeSixPoint,MPReview} and the one-loop amplitudes for five
external partons~\cite{OneloopFivePoint} have been known for some time
now.  Recently, in a tour de force series of calculations, Anastasiou,
Glover, Oleari, and Tejeda-Yeomans have provided the NNLO interferences 
of the two-loop amplitudes with the tree amplitudes, for all 
QCD four-parton processes, summed over all external helicities and 
colors~\cite{GOTY2to2,GOTYgggg}.

In this paper, we compute the $\ggtogg$ amplitudes directly at two loops
in the spinor helicity formalism~\cite{SpinorHelicity}, and expose their
full dependence on external colors as well.  The additional helicity and
color information provided here is not necessary for the main
phenomenological application, NNLO jet production in collisions of
unpolarized hadrons.  However, it still provides several benefits:
\begin{itemize}
\item Jet production in collisions of {\it polarized} protons, as planned
for the relativistic heavy ion collider (RHIC) at Brookhaven, may help to
determine the poorly-known polarized gluon distribution in the
proton~\cite{SofferVirey}.  Theoretical predictions of the relevant
observables require scattering amplitudes for polarized partons.
Currently, predictions are available through NLO~\cite{dFFSV}; the
helicity amplitudes presented here are a prerequisite for improving the
predictions to NNLO accuracy.
\item Many formal properties of scattering amplitudes are simpler in a 
helicity basis and/or after color decomposition.   Such properties include
supersymmetry Ward identities~\cite{SWI}, collinear 
limits~\cite{MPReview,Neq4,LoopReview}, and high-energy 
behavior~\cite{BFKL}.
\item Our results serve as a check of the results of 
ref.~\cite{GOTYgggg}, and are useful for investigating
the dependence of two-loop 
amplitudes on the variant of dimensional regularization used.
\end{itemize}

Here we also present the helicity amplitudes for $\ggtogg$ scattering in
pure $N=1$ supersymmetric $SU(N)$ gauge theory.  Such amplitudes only
differ from QCD with massless quarks in that the fermions are in the
adjoint rather than the fundamental representation; yet they obey
supersymmetry Ward identities~\cite{SWI} and are generally simpler than
their QCD counterparts.  They also provide useful auxiliary functions for
describing the QCD results.

Several versions of dimensional regularization have been used for loop
calculations in QCD, differing mainly in the number of gluon
polarization states they assign in $4-2\e$ dimensions.  The
conventional dimensional regularization (CDR) scheme~\cite{CDR}
assigns $D - 2 = 2 - 2\e$ states to all gluons, whether internal or
external, virtual or real.  This scheme is traditionally employed in
calculations of amplitude interferences, such as ref.~\cite{GOTYgggg}.
In the helicity approach, the number of external, observed gluon
states is necessarily 2 (helicity $\pm1$), but there is some freedom
in the number of virtual gluon polarizations.  The 't~Hooft-Veltman
(HV) scheme~\cite{HV} contains $2 - 2\e$ virtual gluon states, while
the four-dimensional helicity (FDH) scheme~\cite{BKgggg,TwoLoopSUSY}
assigns $2$.  The FDH scheme is related to dimensional reduction
($\DRbar$)~\cite{DR} but is more compatible with the helicity method,
because it allows 2 transverse dimensions in which to define
helicity. Of these variants, only the FDH scheme is fully compatible
with supersymmetry Ward identities for helicity amplitudes, some of
which have been verified through two loops~\cite{TwoLoopSUSY}.  Here
we work primarily in the 't Hooft-Veltman (HV) variant of dimensional
regularization~\cite{HV}, but we also discuss the conversion to the CDR
and FDH schemes.

Two-loop scattering amplitudes in massless QCD possess strong infrared
(soft and collinear) divergences.  Using dimensional regularization
with $D=4-2\e$, the amplitudes generically contain poles in $\e$ up to
$1/\e^4$.  However, these divergences have been organized by
Catani~\cite{Catani} into a relatively simple form, which is
completely predictable through at least order $1/\e^2$.  We shall use
Catani's formulae and color space notation to organize the $\ggtogg$
helicity amplitudes into singular terms (which do contain $\e^0$ terms
in their series expansion in $\e$), plus finite remainders.  We find
that the general form of the divergences given in ref.~\cite{Catani}
holds precisely in both the HV and FDH schemes; however, the numerical
value of the coefficient $K$, which appears at order $1/\e^2$, differs
in the FDH scheme from its value~\cite{Catani} in the HV (or $\MSbar$) scheme. 

The $1/\e$ poles were not predicted {\it a priori} in ref.~\cite{Catani}
for general processes at two loops.  For the $\ggtogg$ amplitude,
ref.~\cite{GOTYgggg} computed the interference of the $1/\e$ pole terms
with the tree amplitude, summed over all colors and helicities.  Here we
extract the full color and helicity dependence of the $1/\e$ pole terms.
We find a term which is independent of color and helicity, and which agrees
with that found by ref.~\cite{GOTYgggg} (when we use the HV scheme), plus
a second term with nontrivial color-dependence, which vanishes when the
color-summed interference is performed.  A term with similar color
structure has also been identified in contributions of one-loop factors
for soft radiation to NNLO processes~\cite{CataniGrazziniSoft}.  We shall
also discuss how terms in the infrared decomposition of ref.~\cite{Catani}
are modified, beginning at order $1/\e^2$, in other variants of
dimensional regularization, such as the FDH scheme.

The paper is organized as follows.  In section~\ref{IRSection}
we review the infrared and color structure of one- and 
two-loop QCD amplitudes.   In section~\ref{OneloopSection} we
describe the one-loop $\ggtogg$ amplitudes in a form that is valid
to all orders in $\e$~\cite{Neq4,BernMorgan,SelfDual,OneloopggggAllEps}, 
and show how to expand them through $\Ord(\e^2)$.  This accuracy is
required because one-loop amplitudes enter the formulae for the
singular parts of two-loop amplitudes multiplied by $1/\e^2$.
Section~\ref{OneloopSingCancel} shows that apart from this requirement, 
only finite remainder terms in the one-loop amplitudes are needed, 
because of cancellations with other NNLO contributions.  These remainder
terms are then tabulated in section~\ref{OneloopRemainderSubsection}.

In section~\ref{TwoLoopFiniteQCDSection} we describe our method for
computing the two-loop amplitudes.  Section~\ref{TensorSection} summarizes
how we evaluate loop integrals, especially those that arise only in the
helicity method.  Some consistency checks on the results are listed in
section~\ref{ChecksSubsection}.  Section~\ref{HSubsection} discusses the
additional singular term appearing at order $1/\e$ in the 
color-decomposed $\ggtogg$ amplitude, which does not contribute to
the color-summed interference with the tree amplitude.  
The finite two-loop remainder functions in the HV scheme are then 
presented in section~\ref{TwoLoopFinRemSubsection} and 
appendix~\ref{QCDRemainderAppendix}.

In section~\ref{SchemeConvertSection} we describe conversion of the HV
results to different schemes, and the comparison with
ref.~\cite{GOTYgggg}, after our results are summed over all external
colors and helicities.  In section~\ref{TwoLoopFiniteN=1Section} and
appendix~\ref{N=1RemainderAppendix} we give the two-loop amplitudes for
pure $N=1$ supersymmetric Yang-Mills theory, whose finite remainders
also serve as auxiliary functions for describing the QCD results.  In
section~\ref{ConclusionsSection} we present our conclusions.


\section{Review of infrared and color structure}\label{IRSection}

In this section we review the structure of the infrared singularities
of dimensionally regularized one- and two-loop QCD
amplitudes, using Catani's color space notation~\cite{Catani}, as a 
prelude to presenting the finite remainders of the one- and two-loop 
$\ggtogg$ amplitudes.

The process considered in this paper is
\begin{equation}
    g(-p_1,-\lambda_1) + g(-p_2,-\lambda_2)
\to g(p_3,\lambda_3) + g(p_4,\lambda_4)\,, 
\label{gggglabel} 
\end{equation}
using an ``all-outgoing'' convention for the external momentum ($p_i$) and
helicity ($\lambda_i$) labeling.  The Mandelstam variables are 
$s = (p_1+p_2)^2$, $t = (p_1+p_4)^2$, and $u = (p_1+p_3)^2$.  

We work with ultraviolet renormalized amplitudes, and employ the
$\MSbar$ running coupling for QCD, $\alphas(\mu)$.  The relation
between the bare coupling $\alphas^u$ and renormalized coupling 
$\alphas(\mu)$, through two-loop order, is~\cite{Catani}
\begin{equation}
\alphas^u \;\mu_0^{2\e} \;S_{\e} = \alphas(\mu) \;\mu^{2\e} 
\left[ 1 - { \alphas(\mu) \over 2\pi } \; { b_0 \over \e } 
         + \biggl( { \alphas(\mu) \over 2\pi} \biggr)^2 
           \left( { b_0^2 \over \e^2 } - { b_1 \over 2\e } \right) 
+ {\cal O}(\alphas^3(\mu)) \right]\,,
\label{TwoloopCoupling}
\end{equation}
where $\mu$ is the renormalization scale, 
$S_\e = \exp[\e (\ln4\pi + \psi(1))]$, and 
$\gamma = -\psi(1) = 0.5772\ldots$ is Euler's constant.
The first two coefficients appearing in the beta function for QCD,
or more generally $SU(N)$ gauge theory with $\Nf$ flavors of
massless fundamental representation quarks, are
\begin{equation}
b_0 = {11 C_A - 4 T_R \Nf \over 6} \,, \hskip 2 cm 
b_1 = {17 C_A^2 - ( 10 C_A + 6 C_F ) T_R \Nf \over 6} \,,
\label{QCDBetaCoeffs}
\end{equation}
where $C_A = N$, $C_F = (N^2-1)/(2N)$, and $T_R = 1/2$.
(Note that ref.~\cite{Catani} uses the notation 
$\beta_0 = b_0/(2\pi)$, $\beta_1 = b_1/(2\pi)^2$.)

The perturbative expansion of the $\ggtogg$ amplitude is
\begin{eqnarray}
\cm_\ggtogg(\alphas(\mu), \mu;\{p\}) &=&
4\pi\alphas(\mu) \, \bigglB \cm_\ggtogg^{(0)}(\mu;\{p\}) + 
\label{RenExpand} \\
&& \hskip 1.5 cm 
+ { \alphas(\mu) \over 2\pi } \cm_\ggtogg^{(1)}(\mu;\{p\}) +
\nonumber \\
&& \hskip 1.5 cm 
+ \biggl( { \alphas(\mu) \over 2\pi } \biggr)^2 
\cm_\ggtogg^{(2)}(\mu;\{p\}) + \Ord(\alphas^3(\mu)) \biggrB \,, 
\nonumber
\end{eqnarray}
where $\cm_\ggtogg^{(L)}(\mu;\{p\})$ is the $L^{\rm th}$
loop contribution.  
Equation~(\ref{TwoloopCoupling}) is equivalent to the following
$\MSbar$ renormalization prescriptions at one and two loops,
\begin{eqnarray}
 \cm_\ggtogg^{(1)} 
 &=&  S_\e^{-1} \, \cm_\ggtogg^{(1){\rm unren}}
    - {b_0\over\e} \, \cm_\ggtogg^{(0)} \,,
\label{OneloopCounterterm} \\
 \cm_\ggtogg^{(2)}  
 &=&  S_\e^{-2} \, \cm_\ggtogg^{(2){\rm unren}}   
  -  2 {b_0\over\e} \, S_\e^{-1} 
                    \, \cm_\ggtogg^{(1){\rm unren}}
  + \biggl( { b_0^2\over \e^2 } - {b_1 \over 2 \e} \biggr)
             \, \cm_\ggtogg^{(0)} \,.
\label{TwoloopCounterterm}
\end{eqnarray}

The infrared divergences of renormalized one- and 
two-loop $n$-point amplitudes are given by~\cite{Catani},
\begin{eqnarray}
| \cm_n^{(1)}(\mu; \{p\}) \ra_{\RS} &=& {\bom I}^{(1)}(\e, \mu; \{p\}) 
  \; | \cm_n^{(0)}(\mu; \{p\}) \ra_{\RS} 
  + |\cm_n^{(1){\rm fin}}(\mu; \{p\}) \ra_{\RS} \,,
	\label{OneloopCatani} \\
| \cm_n^{(2)}(\mu; \{p\}) \ra_{\RS} &=& {\bom I}^{(1)}(\e, \mu; \{p\}) 
  \; | \cm_n^{(1)}(\mu; \{p\}) \ra_{\RS} 
	\label{TwoloopCatani} \\ && \null
+ {\bom I}^{(2)}_{\RS}(\e, \mu; \{p\}) \;
         | \cm_n^{(0)}(\mu; \{p\}) \ra_{\RS}
+ |\cm_n^{(2){\rm fin}}(\mu; \{p\}) \ra_{\RS} \,, \hskip .5 cm 
\nonumber
\end{eqnarray}
where the ``ket'' notation $|\cm_n^{(L)}(\mu; \{p\}) \ra_{\RS}$ 
indicates that the $L$-loop amplitude is treated as a vector in 
color space.  The actual amplitude is extracted via
\begin{equation}
\cm_n(1^{a_1},\dots,n^{a_n}) \equiv \la a_1,\dots,a_n \, | \,
\cm_n(p_1,\ldots,p_n)\ra \,,
\label{MnVec}
\end{equation}
where the $a_i$ are color indices.  
The subscript $\RS$ indicates that a quantity depends on the choice of 
renormalization scheme. The divergences of $\cm_n^{(1)}$ are encoded 
in the color operator ${\bom I}^{(1)}$, while those of 
$\cm_n^{(2)}$ also involve the scheme-dependent operator 
${\bom I}^{(2)}_{\RS}$.

In QCD, the operator ${\bom I}^{(1)}$ is given by
\begin{equation}
{\bom I}^{(1)}(\e,\mu;\{p\}) = \frac{1}{2} {e^{-\e \psi(1)} \over
\Gamma(1-\e)} \sum_{i=1}^n \, \sum_{j \neq i}^n \, {\bom T}_i \cdot
{\bom T}_j \Biggl[ {1 \over \e^2} + {\gamma_i \over {\bom T}_i^2 } \,
{1 \over \e} \Biggr] \Biggl( \frac{\mu^2 e^{-i\lambda_{ij} \pi}}{2
p_i\cdot p_j} \Biggr)^{\e} \,,
\label{CataniGeneral}
\end{equation}
where $\lambda_{ij}=+1$ if $i$ and $j$ are both incoming or outgoing
partons, and $\lambda_{ij}=0$ otherwise. The color charge ${\bom T}_i =
\{T^a_i\}$ is a vector with respect to the generator label $a$, and an
$SU(N)$ matrix with respect to the color indices of the outgoing
parton $i$.  For external gluons $T^a_{cb} = i f^{cab}$, 
so ${\bom T}_i^2 = C_A = N$, and
\begin{equation}
\gamma_g = {11 C_A - 4 T_R \Nf \over 6} \,. 
\label{QCDValues}
\end{equation}
The operator ${\bom I}^{(2)}_{\RS}$ is given by~\cite{Catani}
\begin{eqnarray}
{\bom I}^{(2)}_{\RS}(\e,\mu;\{p\}) 
& =& - \frac{1}{2} {\bom I}^{(1)}(\e,\mu;\{p\})
\left( {\bom I}^{(1)}(\e,\mu;\{p\}) + {2 b_0 \over \e} \right)
  \PlusBreak{}
 {e^{+\e \psi(1)} \Gamma(1-2\e) \over \Gamma(1-\e)}
\left( {b_0 \over \e} + K_\RS \right) {\bom I}^{(1)}(2\e,\mu;\{p\})
  \PlusBreak{}
  {\bom H}^{(2)}_{\RS}(\e,\mu;\{p\}) \,,
\label{CataniGeneralI2}
\end{eqnarray}
where the coefficient $K_\RS$ in either the HV or CDR schemes is given 
by~\cite{Catani}
\begin{equation}
K_\HV = \left(\frac{67}{18} - \frac{\pi^2}{6} \right) C_A
 - \frac{10}{9} T_R \Nf \,.    \label{CataniK}
\end{equation}
Although no scheme dependence was assigned to this coefficient in
ref.~\cite{Catani}, we shall find in section~\ref{SchemeConvertSection}
that it is scheme dependent.  The function ${\bom H}^{(2)}_{\RS}$ 
contains only {\em single} poles,
\begin{equation}
{\bom H}^{(2)}_{\RS}(\e,\mu;\{p\}) \ =\ \Ord(1/\e) \,,
\label{CataniGenH}
\end{equation}
but is not predicted {\it a priori} for general processes.  
The color- and helicity-summed matrix element
$\langle \cm^{(0)} | {\bom H}^{(2)}(\e) | \cm^{(0)} \rangle$
has previously been computed in the CDR scheme for $\ggtogg$~\cite{GOTYgggg}
(and for some other multi-parton processes~\cite{GOTY2to2,TwoLoopee3Jets}).
We shall extract the full color and helicity dependence of 
${\bom H}^{(2)}_{\RS}(\e)$ 
for $\ggtogg$ in the HV scheme in section~\ref{HSubsection},
and in the FDH scheme in section~\ref{SchemeConvertSection}.

An explicit color basis for the $\ggtogg$ amplitudes is given by
\begin{equation}
 \cm^{(L)}_{\lambda_1\lambda_2\lambda_3\lambda_4} = 
  S_{\lambda_1\lambda_2\lambda_3\lambda_4} \times
  \sum_{i=1}^9 \trc^{[i]} 
        M^{(L),[i]}_{\lambda_1\lambda_2\lambda_3\lambda_4} \,,
\label{RemoveColorPhase}
\end{equation}
where
\begin{eqnarray}
&& \trc^{[1]} = \Tr(T^{a_1} T^{a_2} T^{a_3} T^{a_4})\,, \nonumber \\
&& \trc^{[2]} = \Tr(T^{a_1} T^{a_2} T^{a_4} T^{a_3})\,, \nonumber \\
&& \trc^{[3]} = \Tr(T^{a_1} T^{a_4} T^{a_2} T^{a_3})\,, \nonumber \\
&& \trc^{[4]} = \Tr(T^{a_1} T^{a_3} T^{a_2} T^{a_4})\,, \nonumber \\ 
&& \trc^{[5]} = \Tr(T^{a_1} T^{a_3} T^{a_4} T^{a_2})\,, \nonumber\\
&& \trc^{[6]} = \Tr(T^{a_1} T^{a_4} T^{a_3} T^{a_2})\,, \nonumber\\
&& \trc^{[7]} = \Tr(T^{a_1} T^{a_2}) \Tr(T^{a_3} T^{a_4})\,, \nonumber\\
&& \trc^{[8]} = \Tr(T^{a_1} T^{a_3}) \Tr(T^{a_2} T^{a_4})\,, \nonumber\\
&& \trc^{[9]} = \Tr(T^{a_1} T^{a_4}) \Tr(T^{a_2} T^{a_3})\,.
\label{TraceBasis}
\end{eqnarray}
Here $T^a$ are $SU(N)$ generators in the fundamental representation,
normalized according to the convention typically used in helicity
amplitude calculations, $\Tr(T^a T^b) = \delta^{ab}$.  (The $T^a$ used
in this color decomposition should not be confused with the $T_i^a$
appearing in ${\bom I}^{(1)}$, which are in the adjoint representation;
nor should they be confused with the generators for the quark
representation, which have the more ``standard'' normalization, 
$T_R = 1/2$, as mentioned above.)

We have also taken the opportunity in \eqn{RemoveColorPhase} to remove
some helicity-dependent overall phases, which arise because we
evaluate the amplitudes in the spinor helicity 
formalism~\cite{SpinorHelicity},
\begin{eqnarray}
S_{++++} &=& i{\spb1.2 \spb3.4 \over \spa1.2 \spa3.4} \,,
   \hskip1.25cm
S_{-+++}  =  i{\spa1.2 \spa1.4 \spb2.4 \over \spa3.4 \spa2.3 \spa2.4} \,,
  \nonumber \\
S_{--++}  &=&  i{\spa1.2 \spb3.4 \over \spb1.2 \spa3.4} \,,
   \hskip1.25cm
S_{-+-+}  =  i{\spa1.3 \spb2.4 \over \spb1.3 \spa2.4} \,. 
\label{SpinorPhases} 
\end{eqnarray}
The spinor inner products~\cite{SpinorHelicity,MPReview} are 
$\spa{i}.j = \langle i^- | j^+\rangle$ and 
$\spb{i}.j = \langle i^+| j^-\rangle$,
where $|i^{\pm}\rangle$ are massless Weyl spinors of momentum $k_i$,
labeled with the sign of the helicity.  They are anti-symmetric, with
norm $|\spa{i}.j| = |\spb{i}.j| = \sqrt{s_{ij}}$, where 
$s_{ij} = 2k_i\cdot k_j$.  It follows that the 
$S_{\lambda_1\lambda_2\lambda_3\lambda_4}$ are indeed phases.
They will cancel out from (and therefore may be freely omitted from) 
all transition probabilities involving unpolarized gluons, or
circularly polarized gluons.

In the basis~(\ref{TraceBasis}) for $\ggtogg$, 
the matrix ${\bom I}^{(1)}$ is~\cite{GOTYgggg}
\begin{eqnarray}
&&{\bom I}^{(1)}(\e) = - {e^{-\e\psi(1)} \over \Gamma(1-\e)}
 \biggl( {1\over\e^2} + {b_0\over N\e} \biggr) \times
\label{explicitI1} \\ 
&\times&\left( {\small
\begin{array}{ccccccccc}
N(\tS+\tT) & 0 & 0 & 0 & 0 & 0 & (\tT-\tU) & 0 & (\tS-\tU) \\
0 & N(\tS+\tU) & 0 & 0 & 0 & 0 & (\tU-\tT) & (\tS-\tT) & 0 \\
0 & 0 & N(\tT+\tU) & 0 & 0 & 0 & 0 & (\tT-\tS) & (\tU-\tS) \\
0 & 0 & 0 & N(\tT+\tU) & 0 & 0 & 0 & (\tT-\tS) & (\tU-\tS) \\
0 & 0 & 0 & 0 & N(\tS+\tU) & 0 & (\tU-\tT) & (\tS-\tT) & 0 \\
0 & 0 & 0 & 0 & 0 & N(\tS+\tT) & (\tT-\tU) & 0 & (\tS-\tU) \\
(\tS-\tU) & (\tS-\tT) & 0 & 0 & (\tS-\tT) & (\tS-\tU) & 2N\tS & 0 & 0 \\
0 & (\tU-\tT) & (\tU-\tS) & (\tU-\tS) & (\tU-\tT) & 0 & 0 & 2N\tU & 0 \\
(\tT-\tU) & 0 & (\tT-\tS) & (\tT-\tS) & 0 & (\tT-\tU) & 0 & 0 & 2N\tT
\end{array}
}
\right) \nonumber
\end{eqnarray}
where
\begin{equation}
\tS = \left({\mu^2\over -s}\right)^\e \,, \qquad
\tT = \left({\mu^2\over -t}\right)^\e \,, \qquad
\tU = \left({\mu^2\over -u}\right)^\e \,.
\label{STUedef}
\end{equation}

A reflection identity implies that the coefficients of two color 
structures with reversed $T^{a_i}$ ordering are identical, so that
\begin{equation}
 M^{(L),[4]}_{\lambda_1\lambda_2\lambda_3\lambda_4}
= M^{(L),[3]}_{\lambda_1\lambda_2\lambda_3\lambda_4}, \qquad
 M^{(L),[5]}_{\lambda_1\lambda_2\lambda_3\lambda_4}
= M^{(L),[2]}_{\lambda_1\lambda_2\lambda_3\lambda_4}, \qquad
 M^{(L),[6]}_{\lambda_1\lambda_2\lambda_3\lambda_4}
= M^{(L),[1]}_{\lambda_1\lambda_2\lambda_3\lambda_4}.
\label{Reflection}
\end{equation}
Also, due to Bose symmetry, parity, and time-reversal symmetry for the 
process~(\ref{gggglabel}), we only have to give results for the 
four helicity configurations
\begin{equation}
  \lambda_1\lambda_2\lambda_3\lambda_4 
\ =\ \hbox{++++}, \quad \hbox{$-$+++}, \quad \hbox{$-$$-$++}, 
   \quad \hbox{$-$+$-$+} \,.
\label{IndepHelicityConfigs}
\end{equation}
The tree amplitudes are given in the basis~(\ref{TraceBasis}) by
\begin{eqnarray}
 M^{(0),[i]}_{++++} &=& M^{(0),[i]}_{-+++} = 0, 
 \hskip 4.3 cm \hbox{for all $i$,} \nonumber \\
 M^{(0),[7]}_{\lambda_1\lambda_2\lambda_3\lambda_4} &=& 
 M^{(0),[8]}_{\lambda_1\lambda_2\lambda_3\lambda_4} =
 M^{(0),[9]}_{\lambda_1\lambda_2\lambda_3\lambda_4} = 0, 
 \qquad \qquad \hbox{for all $\lambda_i$,} \nonumber \\
 M^{(0),[1]}_{--++} &=& -{s\over t} \,, \qquad 
 M^{(0),[2]}_{--++} = -{s\over u} \,, \qquad 
 M^{(0),[3]}_{--++} = -{s^2\over tu} \,, \nonumber \\
 M^{(0),[1]}_{-+-+} &=& -{u^2\over st} \,, \qquad 
 M^{(0),[2]}_{-+-+} = -{u\over s} \,, \qquad 
 M^{(0),[3]}_{-+-+} = -{u\over t} \,. 
\label{TreeAmps}
\end{eqnarray}

A typical partonic cross section requires an amplitude interference,
summed over all external colors.  Such interferences are evaluated in
the color basis~(\ref{TraceBasis}) as
\begin{equation}
 I^{(L,L')}_{\lambda_1\lambda_2\lambda_3\lambda_4}
 \equiv
\langle \cm_{\lambda_1\lambda_2\lambda_3\lambda_4}^{(L)} 
      | \cm_{\lambda_1\lambda_2\lambda_3\lambda_4}^{(L')} \rangle
 = \sum_{i,j=1}^9 M^{(L),[i] \, *}_{\lambda_1\lambda_2\lambda_3\lambda_4}
                 {\cal C\!C}_{ij} 
                  M^{(L'),[j]}_{\lambda_1\lambda_2\lambda_3\lambda_4} \,,
\label{ColorSumGen}
\end{equation}
where the symmetric matrix 
${\cal C\!C}_{ij} \equiv \sum_{\rm colors} \trc^{[i]\,*} \trc^{[j]}$ 
is~\cite{GOTYgggg,GTYOneloopSq}
\begin{equation}
{\cal C\!C} = \frac{V}{N^2}\left(
\begin{array}{ccccccccc}
C_1 & C_2 & C_2 & C_2 & C_2 & C_3 & NV & -N & NV \\
C_2 & C_1 & C_2 & C_2 & C_3 & C_2 & NV & NV & -N \\
C_2 & C_2 & C_1 & C_3 & C_2 & C_2 & -N & NV & NV \\
C_2 & C_2 & C_3 & C_1 & C_2 & C_2 & -N & NV & NV \\
C_2 & C_3 & C_2 & C_2 & C_1 & C_2 & NV & NV & -N \\
C_3 & C_2 & C_2 & C_2 & C_2 & C_1 & NV & -N & NV \\
NV & NV & -N & -N & NV & NV & N^2 V & N^2 & N^2 \\
-N & NV & NV & NV & NV & -N & N^2 & N^2 V & N^2 \\
NV & -N & NV & NV & -N & NV & N^2 & N^2 & N^2 V
\end{array}
\right),
\label{ColorSumMatrix}
\end{equation}
with 
\begin{equation}
C_1 = N^4 - 3N^2 + 3, \qquad C_2 = 3-N^2, \qquad C_3 = 3+N^2, 
\qquad V = N^2-1.
\end{equation}
The unpolarized partonic cross section is obtained from the helicity sum
\begin{equation}
 {\bar I}^{(L,L')} \equiv
\sum_{\lambda_i = \pm 1} I^{(L,L')}_{\lambda_1\lambda_2\lambda_3\lambda_4}\,,
\end{equation}
after the usual averaging over initial spins and inclusion of flux
factors.  For example, the helicity sum for the tree-level cross
section, constructed from~\eqn{TreeAmps} in either the HV or FDH scheme, is
\begin{equation}
 {\bar I}^{(0,0)} 
= 16 \, N^2 V \biggl( 3 - {tu\over s^2} - {us\over t^2} - {st\over u^2}
              \biggr) \,.  
\label{TreeCrossSection}
\end{equation}
%


\section{One-loop amplitudes}\label{OneloopSection}

The one-loop amplitudes for $\ggtogg$ were first evaluated through
$\Ord(\e^0)$ as an interference with the tree amplitude in the 
CDR scheme~\cite{EllisSexton}.  Later they were evaluated as helicity
amplitudes in the HV and FDH schemes~\cite{BKgggg,KSTfourparton}.

Because ${\bom I}^{(1)}$ contains terms of order $1/\e^2$, the 
${\bom I}^{(1)} | \cm^{(1)} \rangle_\RS$ term in the infrared
decomposition~(\ref{TwoloopCatani}) of the two-loop $\ggtogg$ amplitudes
requires the series expansions of
the one-loop amplitudes through $\Ord(\e^2)$.
In ref.~\cite{OneloopggggAllEps}, using results from
refs.~\cite{Neq4,BernMorgan,SelfDual}, the one-loop 
$\ggtogg$ helicity amplitudes were presented in a representation 
valid to all orders in $\e$, in both the HV and FDH schemes.
These results can easily be rewritten in terms of
integral functions whose series expansions are known to the
requisite order~\cite{AllPlusTwo,BhabhaTwoLoop}.  
In section~\ref{AllOrdersSubsection} we present the all-order 
results in the color basis~(\ref{TraceBasis}),
with the normalizations implicit in \eqn{RenExpand}.

In section~\ref{OneloopSingCancel}, we show that the only place
that terms beyond $\Ord(\e^0)$ in the one-loop amplitudes are 
required in an NNLO calculation is in the infrared 
decomposition~(\ref{TwoloopCatani}) of the two-loop amplitudes.

Finally, in section~\ref{OneloopRemainderSubsection} we list the 
finite remainders of the one-loop amplitudes in the HV scheme, after the  
renormalization~(\ref{OneloopCounterterm})
and subtraction of infrared divergences~(\ref{OneloopCatani}).
The corresponding finite remainder in the one-loop/one-loop NNLO
interference has already been computed in the CDR scheme, summed over 
all colors and helicities~\cite{GTYOneloopSq}.  Our HV amplitude
remainders lead to precisely the same result.


\subsection{All orders in $\e$}\label{AllOrdersSubsection}

Here we present the renormalized one-loop $\ggtogg$ amplitudes
in the color basis~(\ref{TraceBasis}),
with the normalizations implicit in \eqn{RenExpand}, in a form
valid to all orders in $\e$.

The first coefficient in the color basis~(\ref{TraceBasis}) for $\ggtogg$
at one loop may be written in terms of ``primitive'' amplitudes for 
a gluon or quark in the loop, as~\cite{LoopColor}
\begin{equation}
  M^{(1),[1]}_{\lambda_1\lambda_2\lambda_3\lambda_4}(s,t,u)
  = N \, M^{\gluon}_{\lambda_1\lambda_2\lambda_3\lambda_4}
  + \Nf \, M^{\fermion}_{\lambda_1\lambda_2\lambda_3\lambda_4}
  - {b_0 \over \e} M^{(0),[1]}_{\lambda_1\lambda_2\lambda_3\lambda_4} \,.
\label{OneloopSingleTr}
\end{equation}
The remaining single-trace coefficients are obtained via crossing symmetry:
\begin{eqnarray}
M^{(1),[2]}_{++++}(s,t,u) 
&=& M^{(1),[1]}_{++++}(s,u,t), \qquad\qquad
M^{(1),[3]}_{++++}(s,t,u) 
= M^{(1),[1]}_{++++}(u,t,s), \nonumber \\
M^{(1),[2]}_{-+++}(s,t,u) 
&=& M^{(1),[1]}_{-+++}(s,u,t), \qquad\qquad
M^{(1),[3]}_{-+++}(s,t,u)
= M^{(1),[1]}_{-+++}(u,t,s), \nonumber \\
M^{(1),[2]}_{--++}(s,t,u) 
&=& M^{(1),[1]}_{--++}(s,u,t), \qquad\qquad
M^{(1),[3]}_{--++}(s,t,u)
= M^{(1),[1]}_{-+-+}(u,t,s), \nonumber \\
M^{(1),[2]}_{-+-+}(s,t,u) 
&=& M^{(1),[1]}_{--++}(u,s,t), \qquad\qquad
M^{(1),[3]}_{-+-+}(s,t,u)
= M^{(1),[1]}_{--++}(u,t,s), \nonumber \\
\label{OneloopCrossing}
\end{eqnarray}
where appropriate analytic continuations are required to bring 
each function into the physical region.
The double trace coefficients, to which only the gluon loops contribute,
follow from a $U(1)$ decoupling identity~\cite{LoopColor}:
\begin{equation}
  M^{(1),[7]}_{\lambda_1\lambda_2\lambda_3\lambda_4}
= M^{(1),[8]}_{\lambda_1\lambda_2\lambda_3\lambda_4}
= M^{(1),[9]}_{\lambda_1\lambda_2\lambda_3\lambda_4}
= {2\over N} 
    \Bigl( M^{(1),[1]}_{\lambda_1\lambda_2\lambda_3\lambda_4}
         + M^{(1),[2]}_{\lambda_1\lambda_2\lambda_3\lambda_4}
         + M^{(1),[3]}_{\lambda_1\lambda_2\lambda_3\lambda_4} 
    \Bigr)\Bigr|_{\Nf=0}.
\label{GDoubleTrace}
\end{equation}

It is convenient to write the gluon and fermion loop contributions,
$M^{\gluon}_{\lambda_1\lambda_2\lambda_3\lambda_4}$ and
$M^{\fermion}_{\lambda_1\lambda_2\lambda_3\lambda_4}$, in terms 
of a supersymmetric decomposition into scalar, chiral $N=1$, 
and $N=4$ supersymmetric multiplets in the loop~\cite{Fusing,LoopReview}:  
\begin{eqnarray}
 M^{\gluon}_{\lambda_1\lambda_2\lambda_3\lambda_4}
&=& (1-\e \delta_R) 
      M^{\scalar}_{\lambda_1\lambda_2\lambda_3\lambda_4}
  - 4 M^{N=1}_{\lambda_1\lambda_2\lambda_3\lambda_4}
  +   M^{N=4}_{\lambda_1\lambda_2\lambda_3\lambda_4} \,,
\label{GluonDecomp} \\
 M^{\fermion}_{\lambda_1\lambda_2\lambda_3\lambda_4}
&=& - M^{\scalar}_{\lambda_1\lambda_2\lambda_3\lambda_4}
    + M^{N=1}_{\lambda_1\lambda_2\lambda_3\lambda_4} \,,
\label{FermionDecomp}
\end{eqnarray}
where $\delta_R=1$ for the HV scheme and $\delta_R=0$ for the FDH scheme.

For ``maximally helicity violating'' configurations, the supersymmetric 
components vanish by a supersymmetry Ward identity~\cite{SWI},
\begin{equation}
 M^{N=1}_{++++} = M^{N=1}_{-+++} = M^{N=4}_{++++} = M^{N=4}_{-+++} = 0.
\label{MHVVanish}
\end{equation}
The remaining independent components 
are~\cite{Neq4,BernMorgan,OneloopggggAllEps}
%
%
%
\begin{eqnarray}
M^{\scalar}_{++++} & = &
- \e {} (1-\e) \,\Boxeight(s,t)
\,, \nonumber\\
M^{\scalar}_{-+++} & = &
  { t {} (u-s) \over s u } \e \, \Trisix{(s)}
+ { s {} (u-t) \over t u } \e \,  \Trisix{(t)} 
	\PlusBreak{} 
 {t-u \over s^2 } \e \, \Bubsix{(s)}
+ {s-u \over t^2 } \e \, \Bubsix{(t)}
- {s t \over 2 u } \e \, \Boxsix(s,t)
	\MinusBreak{} 
  \e {} (1-\e) \Boxeight(s,t)
\,, \nonumber \\
M^{\scalar}_{--++} & = &
- { s - \e t \over t^2}  \Bubsix{(t)} - \e {} (1-\e) \, \Boxeight(s,t)
\,, \nonumber \\
M^{\scalar}_{-+-+} & = &
- { s-t \over u } \e \, \Trisix{(t)}
- { t-s \over u } \e \, \Trisix{(s)}
+ { s \over u } \Bubfour{(t)}
+ { t \over u } \Bubfour{(s)}
	\PlusBreak{ }
  { u - \e t \over t^2 }  \Bubsix{(t)}
+ { u - \e s \over s^2 }  \Bubsix{(s)}
- \Trisix{(t)}
- \Trisix{(s)}
	\MinusBreak{ }
 { s t \over u } \Boxsix(s,t)
- \e {} (1-\e) \, \Boxeight(s,t) 
\,, 
\label{OneloopGlueAmplitudes} \\
M^{N=1}_{--++} & = &
- {1\over 2} s\, \e \, \Boxsix(s,t) - {s \over 2 t}\Bubfour{(t)} \,,
\nonumber \\
M^{N=1}_{-+-+} & = &
{1\over 2} {u \over s}  \Bubfour{(s)}
+  {1\over 2} {u \over t} \Bubfour{(t)}
-  {1\over 2} u {} (1-\e) \, \Boxsix(s,t)
\,, \label{OneloopNeq1GlueAmplitudes} \\
M^{N=4}_{--++} & = &
     {1\over 2}\, s^2 \, \Boxfour(s,t) 
\,, 
\nonumber \\ 
M^{N=4}_{-+-+} & = &
     {1\over 2} \, u^2 \, \Boxfour(s,t) 
\,.
\label{OneloopNeq4GlueAmplitudes}
\end{eqnarray}
Here ${\rm Bub}^{(n)}(s)$, ${\rm Tri}^{(n)}(s)$ and 
${\rm Box}^{(n)}(s,t)$ are the one-loop bubble, triangle and box scalar
integrals, evaluated in $D = n - 2\e$ dimensions.  
The bubble and box integrals are 
\begin{eqnarray}
\Bubfour(s)  &=& {\rg \over \e {} (1-2 \e)} (-s)^{-\e} \,, 
\nonumber\\
\Bubsix(s) &=& - {\rg \over 2\e {} (1-2\e) (3-2\e)}
(-s)^{1-\e} \,, 
\nonumber\\
\Trifour(s) &=& -{\rg \over \e^2} \, (-s)^{-1-\e}\,, 
\nonumber\\
\Trisix(s) &=& -{\rg (-s)^{-\e} \over 2 \e {} (1- 2\e)
                     (1 - \e)} \,, 
\label{IntDefs}
\end{eqnarray}
where
\begin{eqnarray}
\rg & = & e^{-\e \psi(1)} \,
{\Gamma(1+\e) \Gamma^2(1-\e) \over \Gamma(1-2\e)} 
\nonumber \\
& = & 1 - {1\over2} \zeta_2 \, \e^2 - {7\over3} \zeta_3 \, \e^3 -
{47\over16} \zeta_4 \, \e^4 + \Ord(\e^5) \,,
\end{eqnarray}
with
\begin{equation}
\zeta_s \equiv \sum_{n=1}^\infty n^{-s} \,, \qquad \quad
\zeta_2 = {\pi^2\over6} \,, \qquad \zeta_3 = 1.202057\ldots, \qquad
\zeta_4 = {\pi^4\over90} \,,
\label{ZetaValues}
\end{equation}
and we have kept the full dependence on $\e$ in the integrals.  In
the $s$-channel ($s>0$), $\e$-expansions of the functions~(\ref{IntDefs}) 
are given by using the analytic continuation 
$\ln(-s) \to \ln s - i\pi$.

The box integrals in various dimensions appearing in
\eqns{OneloopGlueAmplitudes}{OneloopNeq4GlueAmplitudes} 
are related via a dimension-shifting formula~\cite{DimShift} valid 
to all orders in $\e$,
\begin{eqnarray}
\Boxsix(s,t) & =&  {1\over 2 \, (-1+2\e) u} \, 
\Bigl(s t \, \Boxfour(s,t) - 2 t\, \Trifour(t)
- 2 s \, \Trifour(s) \Bigr) \,, 
\nonumber \\
\Boxeight(s,t) & = &  {1\over 2 \, (-3+2\e) u} \,
\Bigl(s t \, \Boxsix(s,t) - 2 t\, \Trisix(t)
  - 2 s \, \Trisix(s) \Bigr) \,. \qquad
\label{DimShiftFormula}
\end{eqnarray}
\looseness=1Because the $D=6-2\e$ scalar box integral is completely finite as
$\e\to0$, it is convenient to express the other box integrals in terms
of it. This isolates all divergences to triangle and bubble integrals.
To expand the six-dimensional box to higher orders in $\e$, one could use
an expression for $\Boxfour(s,t)$ valid to all orders in $\e$,
in terms of hypergeometric functions~\cite{EarlyBox,DimShift}, and 
the dimension-shifting formula~(\ref{DimShiftFormula}).  Or one
can expand the Feynman parameter integrand for $\Boxsix(s,t)$
in $\e$ directly.  

In the $u$-channel ($s<0$, $t<0$), where the functions are manifestly
real, the expansion of the six-dimensional box through $\Ord(\e^2)$
is~\cite{AllPlusTwo,BhabhaTwoLoop}
\begin{eqnarray}
\Boxsix(s,t) &= &
{\rg u^{-1-\e} \over 2 (1-2\e) } \Biggl[ {1\over2} 
\Bigl( (V-W)^2 +\pi^2 \Bigr)
	\PlusBreakAdjust{1pt plus 1pt}{{\rg u^{-1-\e} \over 2 (1-2\e)} \Biggl[}
  2 \e {} \Biggl( \li3(-v) - V \li2(-v) - 
{V^3\over3} - {\pi^2\over2} V\Biggr)
  	\MinusBreakAdjust{1pt plus 1pt}{{\rg u^{-1-\e}\over 2 (1-2\e)} \Biggl[}
  2 \e^2 \Biggl( \li4(-v) + W \li3(-v) - {1\over2} V^2 \li2(-v)-\qquad
\nonumber \\[1pt plus 1pt]&& 
	\hphantom{{ \rg u^{-1-\e} \over 2 (1-2\e) } \BigglBl- 2 \e^2 \biggl(}\!
- {1\over8}\, V^4 - {1\over6} \, V^3 W + {1\over4} \, V^2 W^2
- {\pi^2\over4} \, V^2 - 
\nonumber \\[1pt plus 1pt]&& 
	\hphantom{{ \rg u^{-1-\e} \over 2 (1-2\e) } \BigglBl- 2 \e^2 \biggl(}\!
-{\pi^2\over3} \, V W - 2 \zeta_4 \Biggr)
+ \hbox{$(s \lr t)$}\Biggr] + \Ord(\e^3)\,, 
\label{D6BoxEuclid}
\end{eqnarray}
where
\begin{equation}
v = {s\over u} \, , \qquad 
w = {t\over u} \, , \qquad 
V = \ln\biggl(-{s\over u}\biggr) \, , \qquad 
W = \ln\biggl(-{t\over u}\biggr) \, .
\label{uChannelvwVWdef}
\end{equation}
In the $s$-channel ($s>0$, $t<0$) an
analytic continuation of the box integral yields,
\begin{eqnarray}
\Boxsix(s,t) &= &
{ \rg |s|^{-\e} \over u (1-2\e) }\times
\nonumber \\ && 
\times\Biggl\{ {1\over 2} X^2 \!+
\e \biggr( - \li3(-x) +  X \li2(-x) -{1\over 3} X^3 \!+  \zeta_3
+ {1\over 2} \, Y X^2 \!- {1\over 2}\, \pi^2 X \biggl) -
\nonumber \\ && 
	\hphantom{\times\Biggl\{}\!
- \e^2 \biggl( \li4\Bigl(-\frac xy\Bigr) - \li4(-y) + \li3(-y) X+
\nonumber \\ && 
	\hphantom{\times\Biggl\{- \e^2 \biggl(}\!
+ {1\over 2} \li2(-x) (X^2 + \pi^2) + {1\over 24} (Y^2 + \pi^2)^2 -
{1\over 6} Y^3 X + {1\over 4} Y^2 X^2+
\nonumber \\ && 
	\hphantom{\times\Biggl\{- \e^2 \biggl(}\!
+ {1\over 3} X^3 Y - {1\over 8} (X^2 + \pi^2)^2 + {\pi^2\over 3} X Y +
{7\over 360} \pi^4 \biggr)+
\nonumber \\ && 
	\hphantom{\times\Biggl\{}\!
+ i \pi \Biggl[ X + \e \Bigl( \li2(-x) + Y X - {1\over 2}\, X^2
- {\pi^2 \over 6} \Bigr)+
\nonumber \\ && 
	\hphantom{\times\Biggl\{+ i \pi \BigglBl}\!
+ \e^2 \Bigl(- \li3(-x) - \li3(-y) - {1\over 2} Y X^2 + {1\over 6} X^3
+ \zeta_3 \Bigr) \Biggr] \Biggr\}+ 
\nonumber \\ &&  
\null + \Ord(\e^3)\,, 
\label{D6Box}
\end{eqnarray}
where
\begin{equation}
x = {t\over s} \, , \qquad y = {u\over s} \, , \qquad 
X = \ln\biggl(-{t\over s}\biggr) \, , \qquad 
Y = \ln\biggl(-{u\over s}\biggr) \, .
\label{VariableNames}
\end{equation}
Also define, for future use,
\begin{equation}
\Xt = X + i \pi\,, \qquad\qquad
\Yt = Y + i \pi\,. 
\label{XtYtDef}
\end{equation}
In general, both expansions of box integrals~(\ref{D6BoxEuclid}) 
and (\ref{D6Box}) will appear in~\eqn{TwoloopCatani} for the divergences
of the two-loop amplitudes.


\subsection{NNLO cancellations involving $\cm_\ggtogg^{(1)}$}
\label{OneloopSingCancel}

The NNLO $\ggtogg$ cross section has a term proportional to the 
square of the one-loop amplitude,
${\bar I}^{(1,1)} \equiv 
\langle \cm_\ggtogg^{(1)} | \cm_\ggtogg^{(1)} \rangle$.
One might expect to need the $\Ord(\e^2)$ terms in $\cm_\ggtogg^{(1)}$
here, because $\cm_\ggtogg^{(1)}$ also contains $1/\e^2$ terms.
Here we show this is not the case, for a generic NNLO calculation of
an infrared-safe observable, because of cancellations against 
contributions involving radiation of additional partons.

If one uses the one-loop infrared 
decomposition~(\ref{OneloopCatani}), one can rewrite 
${\bar I}^{(1,1)}$ as~\cite{GTYOneloopSq}
\begin{eqnarray}
{\bar I}^{(1,1)} &=& 
\langle \cm_\ggtogg^{(0)} | {\bom I}^{(1) \dagger}(\e) {\bom I}^{(1)}(\e)
| \cm_\ggtogg^{(0)} \rangle
+ 2 \, \Re \langle \cm_\ggtogg^{(1){\rm fin}} | {\bom I}^{(1)}(\e)
| \cm_\ggtogg^{(0)} \rangle
 \PlusBreak{ }
 \langle \cm_\ggtogg^{(1){\rm fin}} | \cm_\ggtogg^{(1){\rm fin}} \rangle
\nonumber \\
 &=& 
- \langle \cm_\ggtogg^{(0)} | {\bom I}^{(1) \dagger}(\e) {\bom I}^{(1)}(\e)
| \cm_\ggtogg^{(0)} \rangle
+ 2 \, \Re \langle \cm_\ggtogg^{(1)} | {\bom I}^{(1)}(\e) 
| \cm_\ggtogg^{(0)} \rangle
 \PlusBreak{ }
 \langle \cm_\ggtogg^{(1){\rm fin}} | \cm_\ggtogg^{(1){\rm fin}} \rangle
\,. 
\label{OneloopSqExpr}
\end{eqnarray}
Similarly, the contribution of the two-loop/tree interference to the NNLO
$\ggtogg$ cross section is
\begin{eqnarray}
2 \, \Re {\bar I}^{(2,0)} &=& 
2 \, \Re \langle \cm_\ggtogg^{(0)} | {\bom I}^{(2)}(\e)
| \cm_\ggtogg^{(0)} \rangle
+ 2 \, \Re \langle \cm_\ggtogg^{(1)} | {\bom I}^{(1) \dagger}(\e)
| \cm_\ggtogg^{(0)} \rangle
 \PlusBreak{ }
 2 \, \Re \langle \cm_\ggtogg^{(2){\rm fin}} | \cm_\ggtogg^{(0)} \rangle
\,. 
\label{TwoloopTreeExpr}
\end{eqnarray}
(For clarity, we have dropped $\RS$ subscripts from equations in this 
subsection.)

Thus the two singular terms containing the one-loop amplitude in the NNLO
$\ggtogg$ cross section combine to form
\begin{equation}
2 \, \Re \langle \cm_\ggtogg^{(1)} | 
\bigl( {\bom I}^{(1)}(\e) + {\bom I}^{(1) \dagger}(\e) \bigr)
| \cm_\ggtogg^{(0)} \rangle \,.
\label{OneloopSingSum}
\end{equation}
These terms will partially cancel, in an NNLO cross section for 
an infrared-safe quantity, against phase-space integration of 
certain terms arising from the one-loop/tree interference for the 
processes with one additional parton radiated 
(in this case, $gg\to ggg$, $gq\to ggq$, $gg\to gq\bar{q}$, {\it etc.}).
The ``radiation'' terms may be written generically as
\begin{equation}
2 \, \Re \langle \cm_{5, {\rm rad}}^{(1)} | 
              \cm_{5, {\rm rad}}^{(0)} \rangle \,.
\label{OneloopRadSum}
\end{equation}

To see this NNLO cancellation, it is useful to recall the corresponding
cancellation at NLO, where the singular part of the virtual correction
\begin{equation}
\Re \langle \cm_\ggtogg^{(0)} | 
\bigl( {\bom I}^{(1)}(\e) + {\bom I}^{(1) \dagger}(\e) \bigr)
| \cm_\ggtogg^{(0)} \rangle
\label{NLOvirtual}
\end{equation}
is cancelled by phase-space integration of the real radiation terms,
\begin{equation}
\Re \langle \cm_{5, {\rm rad}}^{(0)} | \cm_{5, {\rm rad}}^{(0)} \rangle \,.
\label{NLOreal}
\end{equation}
The singular phase-space behavior, soft or collinear, of the one-loop 
five-point amplitude factorizes 
as~\cite{Neq4,BernChalmers,OneloopggggAllEps,KUBDKS}
\begin{equation}
 \cm_{5, {\rm rad}}^{(1)}
\longrightarrow 
\cm_{4}^{(1)} {\cal S}^{(0)} + \cm_{4}^{(0)} {\cal S}^{(1)} \,,
\label{LoopSoftColFact}
\end{equation}
where ${\cal S}^{(0)}$ (${\cal S}^{(1)}$) represents a universal
tree-level (one-loop) soft or collinear factor, which contains all the
dependence on the unresolved phase-space variables that have to be
integrated over.  The tree-level factorization is of course
\begin{equation}
 \cm_{5, {\rm rad}}^{(0)}
\longrightarrow 
 \cm_{4}^{(0)} {\cal S}^{(0)} \,.
\label{TreeSoftColFact}
\end{equation}

Using~\eqns{LoopSoftColFact}{TreeSoftColFact}, one sees that the 
$\cm_{4}^{(1)} {\cal S}^{(0)}$ terms in the singular 
behavior of~\eqn{OneloopRadSum} have 
exactly the same form as the real NLO terms~(\ref{NLOreal}), 
but with $\langle \cm_{4}^{(0)} |$ replaced by $\langle \cm_{4}^{(1)} |$, 
and an overall factor of 2 from the interference.  
Thus the result of integrating the $\cm_{4}^{(1)} {\cal S}^{(0)}$ 
terms in~\eqn{OneloopRadSum} over phase space 
must be cancelled by the virtual NLO terms~(\ref{NLOvirtual}),
but with the corresponding replacements, {\it i.e.} by \eqn{OneloopSingSum}.
The NLO cancellation is good to $\Ord(\e^0)$ (after factorizing
initial-state collinear singularities in the usual manner);
see {\it e.g.} ref.~\cite{CataniSeymourBig}.
The NNLO cancellation is at the same order, in the sense that $\Ord(\e^1)$
and higher terms in $\cm_{4}^{(1)}$ no longer contribute.

In summary, the only place the terms beyond $\Ord(\e^0)$ in 
$\cm_{4}^{(1)}$ are really required at NNLO is in the infrared 
decomposition of the two-loop amplitude.  Once the two-loop finite
remainders $\cm_{4}^{(2){\rm fin}}$ are given, the higher-order terms in 
$\cm_{4}^{(1)}$ are no longer needed.


\subsection{Finite remainders}\label{OneloopRemainderSubsection}

Next we tabulate the finite remainders of the one-loop $\ggtogg$
amplitudes at $\Ord(\e^0)$, defined by $\cm_\ggtogg^{(1){\rm fin}}$
in~\eqn{OneloopCatani} and color decomposed into 
$M^{(1),[i]{\rm fin}}_{\lambda_1\lambda_2\lambda_3\lambda_4}$ in
\eqn{RemoveColorPhase}. We write, in the HV scheme,
\begin{eqnarray}
 M^{(1),[i]{\rm fin}}_{\lambda_1\lambda_2\lambda_3\lambda_4} &=& 
 - b_0 (\ln(s/\mu^2) - i\pi) 
    M^{(0),[i]}_{\lambda_1\lambda_2\lambda_3\lambda_4}
 + N \, a^{[i]}_{\lambda_1\lambda_2\lambda_3\lambda_4}
 + \Nf \, c^{[i]}_{\lambda_1\lambda_2\lambda_3\lambda_4} \,, 
  \quad i = 1,2,3,
\nonumber \\
 M^{(1),[i]{\rm fin}}_{\lambda_1\lambda_2\lambda_3\lambda_4} &=& 
  g^{[i]}_{\lambda_1\lambda_2\lambda_3\lambda_4}
 + {\Nf \over N} \, h^{[i]}_{\lambda_1\lambda_2\lambda_3\lambda_4} \,, 
  \quad i = 7,8,9,
\label{OneloopRemainderDef}
\end{eqnarray}
and the $M^{(1),[i]{\rm fin}}$ for $i=4,5,6$ follow from \eqn{Reflection}.
The one-loop $U(1)$ decoupling identity (\ref{GDoubleTrace}) implies that
\begin{equation}
  g^{[i]}_{\lambda_1\lambda_2\lambda_3\lambda_4}
  = 2 \Bigl( a^{[1]}_{\lambda_1\lambda_2\lambda_3\lambda_4}
           + a^{[2]}_{\lambda_1\lambda_2\lambda_3\lambda_4}
           + a^{[3]}_{\lambda_1\lambda_2\lambda_3\lambda_4} \Bigr),
 \qquad i = 7,8,9.
\label{ggElim}
\end{equation}
For the ++++, $-$+++, and $-$$-$++ helicity configurations, Bose symmetry
under exchange of legs 3 and 4 ($t \lr u$) implies that
\begin{eqnarray}
a^{[2]}_{\lambda_1\lambda_2\lambda_3\lambda_4}(s,t,u) &=&
a^{[1]}_{\lambda_1\lambda_2\lambda_3\lambda_4}(s,u,t) \,,
\nonumber \\
c^{[2]}_{\lambda_1\lambda_2\lambda_3\lambda_4}(s,t,u) &=&
c^{[1]}_{\lambda_1\lambda_2\lambda_3\lambda_4}(s,u,t) \,.
\label{acSym}
\end{eqnarray}

For the ++++ helicity amplitude, the independent remainder functions 
$a$, $c$, $g$ and $h$ are
\begin{eqnarray}
a^{[1]}_{++++} &=& 
- {1\over 6} 
\,, \label{a1pppp}\\[1pt plus 4pt]
%
%
a^{[3]}_{++++} &=&  
- {1\over 6} 
\,, \label{a3pppp}\\[1pt plus 4pt]
%
%
c^{[1]}_{++++} &=&  
{1\over 6}
\,, \label{c1pppp}\\[1pt plus 4pt]
%
%
c^{[3]}_{++++} &=&  
 {1\over 6}
\,, \label{c3pppp}\\[1pt plus 4pt]
%
%
h^{[7]}_{++++} &=& h^{[8]}_{++++} = h^{[9]}_{++++} = 
 0
\,. \label{hpppp}
\end{eqnarray}

For $-$+++, they are
\begin{eqnarray}
a^{[1]}_{-+++} &=& 
- {y^2\over 6 x} 
\,, \label{a1mppp}\\[1pt plus 4pt]
%
%
a^{[3]}_{-+++} &=&  
- {1\over 6 x y} 
\,, \label{a3mppp}\\[1pt plus 4pt]
%
%
c^{[1]}_{-+++} &=&  
{y^2\over 6 x}
\,, \label{c1mppp}\\[1pt plus 4pt]
%
%
c^{[3]}_{-+++} &=&  
 {1\over 6 x y}
\,, \label{c3mppp}\\[1pt plus 4pt]
%
%
h^{[7]}_{-+++} &=& h^{[8]}_{-+++} = h^{[9]}_{-+++} = 
 0
\,, \label{hmppp}
\end{eqnarray}
where $x$ and $y$ are defined in \eqn{VariableNames}.

For $-$$-$++, they are
\begin{eqnarray}
a^{[1]}_{--++} &=& 
- {1 \over 2 x} (\Xt^2 + \pi^2) + {67 \over 18 x}
\,, \label{a1mmpp}\\[1pt plus 4pt]
%
%
a^{[3]}_{--++} &=&  
 - {(1 - x y)^2 \over 4 x y} \Bigl((X-Y)^2 + \pi^2 \Bigr)
 - \biggl( {11\over 6 y} - {x-y \over 2} \biggr) \Xt
 + {67 \over 36 x y} + {1 \over 4} 
  \PlusBreak{ }
 \Bigl\{t \leftrightarrow u \Bigr\} 
\,, \label{a3mmpp}\\[1pt plus 4pt]
%
%
c^{[1]}_{--++} &=&  
- {5 \over 9 x}
\,, \label{c1mmpp}\\[1pt plus 4pt]
%
%
c^{[3]}_{--++} &=&  
- { x^2 \over 4 } \Bigl((X-Y)^2 + \pi^2 \Bigr)
+ {1 \over 6} \biggl( 2 { x^2 \over y } + x + 5 y \biggr) \Xt
- { 29 \over 36 } - {5 \over 9} \, {x \over y}  
  \PlusBreak{ } 
 \Bigl\{t \leftrightarrow u \Bigr\} 
\,, \label{c3mmpp}\\[1pt plus 4pt]
%
%
h^{[7]}_{--++} &=& h^{[8]}_{--++} = h^{[9]}_{--++} = 
 {2 \over 3 y} \Xt
+ \Bigl\{t \leftrightarrow u \Bigr\} 
\,. \label{hmmpp}
\end{eqnarray}

For $-$+$-$+, the required functions are
\begin{eqnarray}
a^{[1]}_{-+-+} &=& 
- { (1-x y)^2 \over 2 x y^2 } (\Xt^2 + \pi^2)
- \biggl( { 4 \over 3 } \, y - { x \over 2 } + {1 \over y } \biggr) \Xt
+ { 67 \over 18 } \, { y^2 \over x } + { 1 \over 2 }
\,, \label{a1mpmp}\\[1pt plus 4pt]
%
%
a^{[2]}_{-+-+} &=& 
- { y \over 2 } (\Yt^2 + \pi^2) + { 11 \over 6 }\, y \Yt 
+ {67 \over 18 }\, y
\,, \label{a2mpmp}\\[1pt plus 4pt]
%
%
a^{[3]}_{-+-+} &=&  
- { y \over 2 x } \Bigl((X-Y)^2 + \pi^2 \Bigr) 
   + { 11 \over 6 }\, {y \over x } \Yt
+ {67 \over 18 } \,{ y \over x }
\,, \label{a3mpmp}\\[1pt plus 4pt]
%
%
c^{[1]}_{-+-+} &=&  
\biggl( { x \over 2 y^2 } - { 1 \over 4 } \biggr) (\Xt^2 + \pi^2)
+ { x {} (x - y) + 5 \over 6 y } \Xt
- { (3 - 2 y) (3 x - 2 y) \over 18 x }
\,, \label{c1mpmp}\\[1pt plus 4pt]
%
%
c^{[2]}_{-+-+} &=&  
- { y \over 3 } \, \Yt - { 5 \over 9 } y
\,, \label{c2mpmp}\\[1pt plus 4pt]
%
%
c^{[3]}_{-+-+} &=&  
- { y \over 3 x }\, \Yt - {5 \over 9} \, {y \over x} 
\,, \label{c3mpmp}\\[1pt plus 4pt]
%
%
h^{[7]}_{-+-+} &=& h^{[8]}_{-+-+} = h^{[9]}_{-+-+} = 
 {2 \over 3} y \Xt
+ {2 \over 3} \, { y^2 \over x } \Yt
\,. \label{hmpmp}
\end{eqnarray}

The contribution of the one-loop finite remainders to the
NNLO $\ggtogg$ cross section is
\begin{equation}
 {\bar I}^{(1,1){\rm fin}} \equiv
 \sum_{\lambda_i = \pm 1} 
\langle \cm_{\lambda_1\lambda_2\lambda_3\lambda_4}^{(1){\rm fin}} 
      | \cm_{\lambda_1\lambda_2\lambda_3\lambda_4}^{(1){\rm fin}} \rangle
\,.
\end{equation}
Using the color sum matrix ${\cal C\!C}_{ij}$ in~\eqn{ColorSumMatrix},
the color and helicity sum in ${\bar I}^{(1,1){\rm fin}}$ may be
evaluated in terms of the above explicit
expressions~(\ref{OneloopRemainderDef})--(\ref{hmpmp}) for
$M^{(1),[i]{\rm fin}}_{\lambda_1\lambda_2\lambda_3\lambda_4}$.  It
reproduces precisely the finite remainder function $Finite(s,t,u)$
given in eq.~(3.22) of ref.~\cite{GTYOneloopSq} for the corresponding
quantity evaluated in the CDR scheme; the HV/CDR scheme difference for
${\bar I}^{(1,1)}$ has been completely absorbed into the first
two of the three terms in~\eqn{OneloopSqExpr}.


\section{Two-loop amplitudes and finite remainders}
\label{TwoLoopFiniteQCDSection}

A generic sample of two-loop Feynman diagrams for $\ggtogg$ is
shown in figure~\ref{DiagramSampleFigure}.   However, we did not 
evaluate the diagrams directly.  Instead we computed the unitarity cuts 
in various channels, working to all orders in the dimensional regularization
parameter $\e = (4-D)/2$~\cite{CutBasedvanN,Neq4,Fusing,LoopReview}.  
Essentially we followed the approach first employed at two loops for 
the pure gluon four-point amplitude with all 
helicities identical~\cite{AllPlusTwo} and for $N=4$ supersymmetric 
amplitudes~\cite{BRY}.  These amplitudes were
simple enough that a compact expression for the integrand could be
given.  The fermion loop contributions with all helicities identical are
about as simple~\cite{TwoLoopSUSY}.  For other helicity
configurations, the integrands become rather complicated. We therefore
used the general integral reduction algorithms developed for the
all-massless four-point
topologies~\cite{PBReduction,IntegralsAGO,NPBReduction,Lorentz}, in order 
to reduce the loop integrals to a minimal basis of master integrals.  
To efficiently incorporate polarization vectors of gluons with definite 
helicity requires some minor extensions of these techniques, which 
we now discuss.

\FIGURE[t]{\begin{picture}(380,192)(0,0)
\Text(10,110)[r]{1} \Text(10,179)[r]{2} 
\Text(102,179)[l]{3} \Text(102,110)[l]{4} 
\Gluon(54,127)(29,127){2.5}{3} \Gluon(79,127)(54,127){2.5}{3} 
\Gluon(29,162)(54,162){2.5}{3} \Gluon(54,162)(79,162){2.5}{3}
\Gluon(29,127)(29,162){2.5}{4}
\Gluon(54,162)(54,127){2.5}{4} \Gluon(79,162)(79,127){2.5}{4}
\Gluon(29,127)(12,110){2.5}{3} \Gluon(12,179)(29,162){2.5}{3} 
\Gluon(96,110)(79,127){2.5}{3} \Gluon(79,162)(96,179){2.5}{3} 
\Text(152,110)[r]{1} \Text(152,179)[r]{2} 
\Text(244,179)[l]{3} \Text(244,110)[l]{4} 
\Line(196,127)(221,127) \Gluon(196,127)(171,127){2.5}{3} 
\Gluon(171,162)(196,162){2.5}{3} \ArrowLine(196,162)(221,162)
\Gluon(171,127)(171,162){2.5}{4} \Line(196,127)(221,162)
\Line(196,162)(207,146) \Line(210,143)(221,127)
\Gluon(171,127)(154,110){2.5}{3} \Gluon(154,179)(171,162){2.5}{3}
\Gluon(238,110)(221,127){2.5}{3} \Gluon(221,162)(238,179){2.5}{3} 
\Text(294,110)[r]{1} \Text(294,179)[r]{2} 
\Text(371,179)[l]{3} \Text(371,110)[l]{4} 
\Gluon(348,127)(313,127){2.5}{4} \Gluon(313,162)(348,162){2.5}{4}
\Gluon(313,127)(313,162){2.5}{4} \Gluon(348,162)(348,127){2.5}{4}
\Gluon(348,142)(326,164){2.5}{3}
\Gluon(313,127)(296,110){2.5}{3} \Gluon(296,179)(313,162){2.5}{3}
\Gluon(365,110)(348,127){2.5}{3} \Gluon(348,162)(365,179){2.5}{3} 
\Text(10,5)[r]{1} \Text(10,74)[r]{2}
\Text(102,74)[l]{3} \Text(102,5)[l]{4}
\ArrowLine(29,22)(29,57) \Line(79,22)(79,57)
\Line(29,22)(79,22) \Line(29,57)(44,57) \Line(64,57)(79,57) 
\GlueArc(54,57)(10,0,180){2.5}{4} \CArc(54,57)(10,180,0)
\Gluon(29,22)(12,5){2.5}{3} \Gluon(12,74)(29,57){2.5}{3}
\Gluon(96,5)(79,22){2.5}{3} \Gluon(79,57)(96,74){2.5}{3}
\Text(142,22.5)[r]{1} \Text(142,56.5)[r]{2}
\Text(254,74)[l]{3} \Text(254,5)[l]{4}
\Gluon(161,39.5)(186,39.5){2.5}{3}
\Gluon(186,39.5)(211,57){2.5}{3} 
\Gluon(211,22)(186,39.5){2.5}{3}
\Gluon(211,22)(211,57){2.5}{4}
\Gluon(231,22)(211,22){2.5}{2} \Gluon(211,57)(231,57){2.5}{2}
\Gluon(231,57)(231,22){2.5}{4}
\Gluon(161,39.5)(145,22.5){2.5}{3} \Gluon(145,56.5)(161,39.5){2.5}{3}
\Gluon(248,5)(231,22){2.5}{3} \Gluon(231,57)(248,74){2.5}{3}
\Text(294,5)[r]{1} \Text(294,74)[r]{2} 
\Text(371,74)[l]{3} \Text(371,5)[l]{4} 
\Line(313,22)(348,22) \Line(335,57)(348,57)
\Line(313,22)(313,35) \ArrowLine(348,22)(348,57)
\Line(313,35)(335,57) \Gluon(313,57)(335,57){2.5}{2}
\Gluon(313,35)(313,57){2.5}{2}
\Gluon(313,22)(296,5){2.5}{3} \Gluon(296,74)(313,57){2.5}{3}                  
\Gluon(365,5)(348,22){2.5}{3} \Gluon(348,57)(365,74){2.5}{3}
\end{picture}%
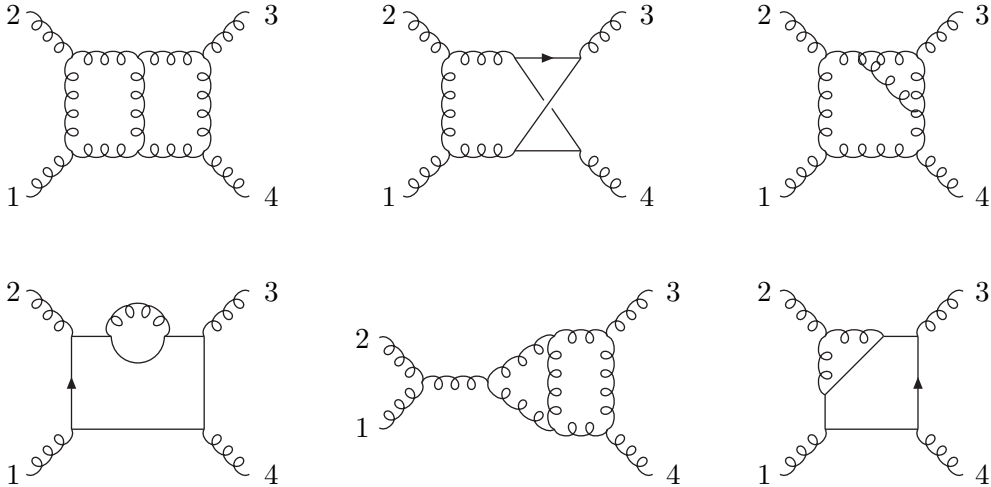
\caption{Some of the two-loop diagrams for $gg \to
gg$.\label{DiagramSampleFigure}}}

\subsection{Tensor loop integrals}\label{TensorSection}

Here we discuss techniques for evaluating the loop integrals
required for the two-loop amplitudes for $gg\to gg$ and related processes,
with an emphasis on the additional types of integrands encountered in 
the helicity amplitude method.

In calculating a typical two-loop scattering amplitude in QCD,
a large number of two-loop integrals are encountered.  The most
complicated topologies are the planar and non-planar double box integrals,
displayed in figure~\ref{ParentsFigure}, which are given by
\begin{eqnarray}
&&\hskip-0.7cm
\I_4^\P [{\cal P}] (s,t) \equiv \nonumber \\ 
 & \equiv & \int {d^{D}p\over (2\pi)^{D}} \, {d^{D}q\over (2\pi)^{D}}\, 
 { {\cal P} \over 
     p^2\, q^2\, (p+q)^2 (p - k_1)^2 \,(p - k_1 - k_2)^2 \,
        (q - k_4)^2 \, (q - k_3 - k_4)^2 }  \,, \nonumber \\
\label{PlanarInt} \\
&&\hskip-0.7cm
\I_4^\NP [{\cal P}] (s,t) \equiv \nonumber \\ 
 & \equiv & \int {d^{D} p \over (2\pi)^{D}} \, {d^{D} q \over (2\pi)^{D}}\, 
 { {\cal P} \over 
     p^2\, q^2\, (p+q)^2 \, (p-k_1)^2 \,(q-k_2)^2\,
   (p+q+k_3)^2 \, (p+q+k_3+k_4)^2 } \,. \nonumber \\ \label{NonPlanarInt} 
\end{eqnarray}
Here $p$ and $q$ are the loop momenta, and $k_i$, $i=1,2,3,4$, are the
external (outgoing) momenta.   ${\cal P}$ is a polynomial (or tensor)
in the loop momenta $p$ and $q$, 
which accompanies the scalar propagator factors shown in the figure.  
It is generated by the numerator algebra of the Feynman diagram, 
or unitarity cut, that is being evaluated.

%
\FIGURE{
\begin{picture}(380,75)(-65,0)
\Text(13,10)[r]{1} \Text(13,70)[r]{2} 
\Text(117,70)[l]{3} \Text(117,10)[l]{4} 
\Line(15,10)(25,20) \Line(15,70)(25,60) 
\Line(115,70)(105,60) \Line(115,10)(105,20)
\Line(25,20)(25,60) \ArrowLine(65,60)(65,20) \Line(105,20)(105,60)
\Line(25,60)(105,60) \ArrowLine(65,20)(25,20) \ArrowLine(65,20)(105,20)
\Text(69,40)[l]{$p+q$} \Text(45,10)[c]{$p$} \Text(85,10)[c]{$q$}
\Text(143,40)[r]{1} \Text(207,40)[l]{2} 
\Text(247,70)[l]{3} \Text(247,10)[l]{4} 
\Line(145,40)(155,40) \Line(195,40)(205,40) 
\Line(245,70)(235,60) \Line(245,10)(235,20)
\Line(235,20)(235,60)
\ArrowLine(235,60)(175,60) \Line(175,20)(235,20)
\ArrowLine(175,60)(195,40) \ArrowLine(175,60)(155,40) 
\Line(175,20)(195,40) \Line(175,20)(155,40) 
\Text(205,70)[c]{$p+q$} 
\Text(155,50)[c]{$p$} \Text(195,50)[c]{$q$}
\end{picture}
\caption{The planar and non-planar double box integrals.}
\label{ParentsFigure}}


In the {\bf interference method}, as recently applied to two-loop QED and QCD
scattering 
amplitudes~\cite{BhabhaTwoLoop,GOTY2to2,GOTYgggg,TwoLoopee3Jets}, 
one sums over all external polarization states in $D$ dimensions.  
In this case, ${\cal P}$ can only depend on the loop momenta, $p$ and $q$, 
and external momenta, $k_i$.  By Lorentz invariance, this dependence 
is only through scalar products,
\begin{equation}
{\cal P}_{\rm interf.} = {\cal P}(p^2,p\cdot q,q^2,p\cdot k_i, q\cdot k_i).
\label{InterfP}
\end{equation}

In contrast, in the {\bf helicity amplitude method}~\cite{SpinorHelicity} 
used in the present paper --- and previously applied to two-loop amplitudes 
in $N=4$ super-Yang-Mills theory~\cite{BRY}, $\ggtogg$ for identical
helicities in pure Yang-Mills theory~\cite{AllPlusTwo}, and the QCD processes 
$gg\to\gamma\gamma$~\cite{gggamgamPaper} and $\gamma\gamma\to\gamma\gamma$
~\cite{PhotonPaper} ---
${\cal P}$ also depends on the polarization vectors $\pol_i$ for the 
external gluons.  We take $D>4$ in the calculation, {\it i.e.}, $\e<0$,
in order to have two transverse dimensions in which to define helicities.
Because the polarization vectors are intrinsically four-dimensional, 
their Lorentz products with the loop momenta distinguish
between the four-dimensional and $(-2\e)$-dimensional components of $p$
and $q$. 
Write 
\begin{equation}
p^\mu = p_{[4]}^\mu + \vec{\mud}_p, \qquad
q^\mu = q_{[4]}^\mu + \vec{\mud}_q, 
\label{pqDecompose}
\end{equation}
where $p_{[4]}$, $q_{[4]}$ are the four-dimensional components
and $\vec{\mud}_p$, $\vec{\mud}_q$ are the $(-2\e)$-dimen\-sional components.
We use the Minkowski metric with signature $(1,-1,-1,-1,\ldots)$, and
write
\begin{equation}
p^2 = p_{[4]}^2 - \mud_p^2, \qquad
q^2 = q_{[4]}^2 - \mud_q^2, \qquad
(p+q)^2 = (p_{[4]} + q_{[4]})^2 - \mud_{p+q}^2,
\label{pqDotProducts}
\end{equation}
where 
$\mud_p^2 \equiv \vec{\mud}_p \cdot \vec{\mud}_p \geq 0$, 
$\; \; \mud_{p+q}^2 \equiv (\vec{\mud}_p + \vec{\mud}_q)^2 
= \mud_p^2 + \mud_q^2 + 2 \vec{\mud}_p \cdot \vec{\mud}_q$.  
Then the generic polynomial encountered in the helicity amplitude method
has the form
\begin{equation}
{\cal P}_{\rm hel.}
 = {\cal P}(p^2,p\cdot q,q^2,p\cdot k_i, q\cdot k_i;
     \mud_p^2, \mud_q^2, \mud_{p+q}^2, \pol_i \cdot p, \pol_i \cdot q).
\label{HelP}
\end{equation}

We rely on reduction algorithms developed to handle general tensor
integrals for the all-massless planar~\cite{PBReduction} and
non-planar~\cite{NPBReduction} double boxes, and related topologies such
as the pentabox~\cite{IntegralsAGO}.   These algorithms were derived
using integration by parts~\cite{IBP} and (for the non-planar double box)
Lorentz invariance~\cite{Lorentz} identities, which act in the space of
integrals with ${\cal P} = 1$, but with the scalar propagators raised to 
arbitrary integer powers $\nu_i$.  For example, for the planar 
double box topology one considers
\begin{equation}
I_4^\P (\nu_1,\nu_2,\nu_3,\ldots,\nu_7)
\equiv (4\pi)^D \int {d^{D}p\over (2\pi)^{D}} \, {d^{D}q\over (2\pi)^{D}}\, 
 \prod_{i=1}^7 { 1 \over \bigl( p_i^2 \bigr)^{\nu_i} }
\label{PlanarMultiInt} \,,
\end{equation}
where
\begin{eqnarray}
&&p_1 = q, \qquad p_2 = q-k_3-k_4, \qquad p_3 = p, \qquad p_4 = p-k_1-k_2, 
\nonumber \\
&& p_5 = p-k_1, \qquad p_6 = p+q, \qquad p_7 = q-k_4, 
\label{pidef}
\end{eqnarray}
and $\nu_i \in \{ 0, 1, 2, \ldots \}$.
The reduction algorithms reduce any such integral to a linear combination
of simpler ``boundary'' integrals, where at least one of the $\nu_i$
vanishes, plus one or two {\it master integrals} with the same topology.
All told, there are 10 different master integrals for the massless 2~$\to$~2
processes~\cite{PBReduction,NPBReduction,IntegralsAGO}.

Given an integral with ${\cal P}$ of the form (\ref{InterfP}) or 
(\ref{HelP}), it is simple to convert it to integrals of the form 
(\ref{PlanarMultiInt}) using Schwinger parametrization~\cite{Tarasov}.
For the scalar integrals, using
\begin{equation}
 {1\over p_i^2} = \int_0^\infty dt_i \exp(-t_i p_i^2)
\label{Schwinger1}
\end{equation}
and performing the $p$ and $q$ integrals, leads to 
\begin{eqnarray}
I_4^X(1,1,\ldots,1) & = & 
 \prod_{i=1}^7 \int_0^\infty dt_i \Bigl[ \Delta(T) \Bigr]^{-D/2}
   \exp\Biggl[- {Q_X(s,t,t_i) \over \Delta(T)} \Biggr] \,,
\label{SchwingerScalar} \\
 \Delta(T) & \equiv & T_p T_q + T_p T_{pq} + T_q T_{pq} \,,
\label{DeltaDef}
\end{eqnarray}
where $X$ labels the topology of the integral ($\P$, $\NP$, {\it etc.}).
In~\eqn{DeltaDef}, $T_p$, $T_q$, $T_{pq}$ are the sums of Schwinger
parameters along the lines carrying loop momenta $p$, $q$, $p+q$,
respectively.  Equivalently, they are Schwinger parameters for the
two-loop vacuum graph obtained by omitting all the external lines, as
shown in \fig{VacuumFigure}.  For the planar double box integral, with
propagators numbered by \eqn{pidef}, the $T_a$ are given by
\begin{equation} 
T_p = t_3 + t_4 + t_5, \qquad T_q = t_1 + t_2 + t_7, \qquad T_{pq} = t_6.
\label{TiPDef}
\end{equation}

The quantity $Q_X$ is more cumbersome.  For the planar double box
integral, its expression is
\begin{equation} 
Q_\P(s,t,t_i) = 
-s \, \Bigl[ t_1 t_2 (t_3+t_4+t_5) + t_3 t_4 (t_1+t_2+t_7)
        + t_6 (t_1+t_3) (t_2+t_4) \Bigr]
-t \ t_5 t_6 t_7 \,.
\label{QPDef}
\end{equation}
However, the precise form of $Q_X$ will not matter in the following.  

%
\FIGURE{
\begin{picture}(360,80)(-130,10)
\Text(3,50)[l]{$T_p$} \Text(48,50)[r]{$T_{pq}$}
\Text(98,50)[r]{$T_q$}
\CArc(50,50)(30,0,360) \Line(50,20)(50,80)
\end{picture}
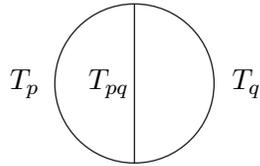
\caption{The two-loop vacuum graph obtained by omitting external momenta,
and its three Schwinger parameters, which are relevant for general 
two-loop integrals containing $v\cdot p$, $v\cdot q$, 
$\vec{\mud}_p$, and $\vec{\mud}_q$.}
\label{VacuumFigure}}


If a polynomial ${\cal P}$ is present, Schwinger parametrization converts
it to a polynomial in the Schwinger parameters, along with inverse powers
of $\Delta$.   Then the more general parametrization,
\begin{equation}
 {1\over \bigl( p_i^2 \bigr)^{\nu_i} } 
= {1\over \Gamma(\nu_i)} \int_0^\infty dt_i \, t_i^{\nu_i-1} \exp(-t_i p_i^2),
\label{Schwinger2}
\end{equation}
can be used to rewrite Schwinger parameter monomials as integrals 
of the form~(\ref{PlanarMultiInt}), typically in shifted dimensions, 
$D\ \to\ D+2n$, $n = 0,1,2,\ldots$ (to account for the inverse powers
of $\Delta$).  (Shifted-dimension integrals pose no problem; equations for 
them can be found by rewriting the factor $\Delta^{-D/2}$ in 
\eqn{SchwingerScalar} as
\begin{equation}
 \Delta^{-D/2} = (T_p T_q + T_p T_{pq} + T_q T_{pq}) 
                  \times  \Delta^{-(D+2)/2} \,,
\end{equation}
and reducing the latter, shifted-dimension representation.)
In principle, this approach gives a prescription to handle
any polynomial in the loop momentum, for either the interference or 
helicity method.

However, as the degree $\sum_{i} \nu_i$ increases, the number of 
integrals of the form~(\ref{PlanarMultiInt}) grows rapidly, and the 
reduction algorithm can become rather time-consuming.  We have found 
it useful to instead use simple algebraic relations, {\it e.g.} for the planar 
double box integral,
\begin{eqnarray}
2q\cdot k_4 &=& p_1^2 - p_7^2, 
\quad 2q\cdot k_3 = p_7^2 - p_2^2 + 2 k_3\cdot k_4,  \nonumber \\
2p\cdot k_1 &=& p_3^2 - p_5^2, 
\quad 2p\cdot k_2 = p_5^2 - p_4^2 + 2 k_1\cdot k_2, 
\nonumber \\ 
2p\cdot q &=& p_6^2 - p_1^2 - p_3^2,\label{PDBAlgebra}
\end{eqnarray}
to quickly reduce integrals with polynomials of the form 
${\cal P}_{\rm interf.}$ to a relatively small set of ``irreducible'' 
integrals for each topology, plus boundary integrals generated when the
$p_i^2$ factors cancel propagators.
Of course the ``irreducible'' integrals are only irreducible with respect 
to~(\ref{PDBAlgebra}), and not with respect to the integration-by-parts 
and Lorentz identities.  We compute the ``irreducible'' integrals once 
and store them.

For the planar double box, \eqn{PDBAlgebra} 
and momentum conservation, $k_1+k_2+k_3+k_4=0$,
show that the ``irreducible'' monomials needed to generate all 
${\cal P}_{\rm interf.}$ are 
\begin{equation}
{\cal P}^{\P,\, {\rm irred}}_{\rm interf.}(m,n) 
  = (2q\cdot k_1)^m \, (2p\cdot k_4)^n,
\qquad m+n \leq6.
\label{PlanarIrred}
\end{equation}
The restriction on the sum of $m$ and $n$ comes from gauge theory 
--- at most six powers of the loop momentum can appear in the Feynman 
diagram numerator algebra.  The planar double box also has a symmetry, 
reflection about the central vertical line, so that 
${\cal P}_{\rm interf.}(m,n) = {\cal P}_{\rm interf.}(n,m)$.  This
leaves only 16 integrals~(\ref{PlanarIrred}) to store.  Not all of the
highest degree integrals actually appear in the amplitude.
For the non-planar box, we store the integrals of the monomials,
\begin{equation}
{\cal P}^{\NP,\,{\rm irred}}_{\rm interf.}(m,n) 
  = (2(p+q)\cdot k_1)^m \, (2p\cdot k_4)^n,
\qquad m+n \leq6.
\label{NonPlanarIrred}
\end{equation}
We also store those ``irreducible'' boundary integrals
with six (instead of seven) propagators; these monomials
are generated by three independent factors.
For example, for the planar double box boundary integral obtained by 
setting $\nu_3 \to 0$, one requires
\begin{equation}
(2q\cdot k_1)^m \, (2p\cdot k_4)^n \, (2p\cdot k_1)^r \,,
\qquad m+n+r \leq5.
\label{PlanarBdyIrred}
\end{equation}

For the helicity amplitude approach, the more general loop momentum
polynomial ${\cal P}_{\rm hel.}$ given in \eqn{HelP} requires a 
bit more work before the above method can be used.
Consider the product $\pol_i \cdot p$.  Because
$\pol_i^\mu$ is a four-dimensional vector, this can also be written
as $\pol_i \cdot p_{[4]}$.  We can expand $p_{[4]}^\mu$ in terms of a 
basis of four different four-dimensional vectors.  Because of momentum
conservation, there are only three independent external momenta, but
we can use the Levi-Civita tensor to construct a fourth one,
\begin{equation}
v^\mu \equiv 
\pol^{\mu}_{~\nu_1\nu_2\nu_3} k_1^{\nu_1} k_2^{\nu_2} k_3^{\nu_3} \,.
\label{vdef}
\end{equation}
Then 
\begin{equation}
p_{[4]}^\mu \equiv 
  c_1^p \, k_1^\mu + c_2^p \, k_2^\mu + c_3^p \, k_3^\mu 
+ c_v^p \, v^\mu \,, 
\label{pExpansion}
\end{equation}
where
\begin{eqnarray}
c_1^p &=& {1\over2su} \Bigl[ 
-t \, (2p\cdot k_1) + u \, (2p\cdot k_2) + s \, (2p\cdot k_3) \Bigr] \,, 
\nonumber \\
c_2^p &=& {1\over2st} \Bigl[ 
 t \, (2p\cdot k_1) - u \, (2p\cdot k_2) + s \, (2p\cdot k_3) \Bigr] \,, 
\nonumber \\
c_3^p &=& {1\over2tu} \Bigl[ 
 t \, (2p\cdot k_1) + u \, (2p\cdot k_2) - s \, (2p\cdot k_3) \Bigr] \,, 
\nonumber \\
c_v^p &=& -{4\over stu} \pol_{\mu\nu_1\nu_2\nu_3} 
   p^\mu k_1^{\nu_1} k_2^{\nu_2} k_3^{\nu_3}
     \ =\  -{4\over stu} v\cdot p\,.
\label{cpEqns}
\end{eqnarray}

Thus we can write
\begin{eqnarray}
\pol_i \cdot p = c_1^p \, \pol_i \cdot k_1 + c_2^p \, \pol_i \cdot k_2 
               + c_3^p \, \pol_i \cdot k_3 + c_v^p \, \pol_i \cdot v \,. 
\label{epExpansion}
\end{eqnarray}
This equation, and the analogous one for $\pol_i \cdot q$,
reduce the problem of handling helicity amplitude polynomials~(\ref{HelP}) 
to those of the form
\begin{equation}
\tilde{\cal P}_{\rm hel.}
 = {\cal P}(p^2,p\cdot q,q^2,p\cdot k_i, q\cdot k_i;
               \mud_p^2, \mud_q^2, \mud_{p+q}^2, v\cdot p, v\cdot q),
\label{NewHelP}
\end{equation}
where $v\cdot p$ and $v\cdot q$ come from the $c_v$ coefficients.

The effect of inserting factors of $v \cdot p$ and $v \cdot q$ into the
integral is very similar to inserting factors involving the
$(-2\e)$-dimensional components of the loop momenta, $\vec{\mud}_p$ and
$\vec{\mud}_q$.  In either case, shifts of $p$ and $q$ by amounts
proportional to the external momenta $k_i$, as required to perform the 
Gaussian integrals over $p$ and $q$, have no effect, because
$v\cdot k_i = \mud_j \cdot k_i = 0$.  Thus both types of factors result 
only in polynomials in the ``vacuum graph'' Schwinger parameters $T_a$.  

Insertions of $\mud_p^2$, $\mud_q^2$ and $\mud_{p+q}^2$ may be handled 
by differentiating the $(-2\e)$-dimensional part of the (Wick rotated)
integral,
\begin{equation}
\int d^{-2\e} \mud_p \ d^{-2\e} \mud_q 
\exp\Bigl[ - \mud_p^2 \, T_p - \mud_q^2 \, T_q 
           - \mud_{p+q}^2 \, T_{pq} \Bigr] \ \propto\ \Delta^\e \,,
\label{extraDIntegral}
\end{equation}
with respect to $T_p$, $T_q$ and $T_{pq}$.  They lead to parameter
insertions such as
\begin{eqnarray} 
 \mud_p^2 &\to&  -\e {T_q+T_{pq} \over \Delta} \,, \nonumber \\ 
 \mud_{p+q}^2 
         &\to&  -\e {T_p+T_q \over \Delta}  \,, \nonumber \\ 
 ( \mud_p^2 )^2 
     &\to&  -\e (1-\e) {(T_q+T_{pq})^2 \over \Delta^2} \,, \nonumber \\ 
 \mud_p^2 \, \mud_q^2 
    &\to&  {\e^2\over\Delta} -\e (1-\e) {T_{pq}^2 \over \Delta^2} \,, 
                 \nonumber \\ 
 ( \mud_p^2 )^2 \, \mud_q^2 
    &\to&  \e (1-\e) \biggl[ \e {T_q+T_{pq}\over\Delta^2} 
                           - (2-\e) {T_{pq}^2 (T_q+T_{pq}) \over \Delta^3} 
                     \biggr] \,, \nonumber \\ 
 \mud_p^2 \, \mud_q^2 \, \mud_{p+q}^2 
    &\to&  \e (1-\e) \biggl[ \e {T_p+T_q+T_{pq}\over\Delta^2} 
                           + (2-\e) { T_p T_q T_{pq} \over \Delta^3} 
                     \biggr] \,.
\label{muParam}
\end{eqnarray}
Similarly, the polynomials in $v\cdot p$ and $v\cdot q$ are easily
parametrized:
\begin{eqnarray} 
 (v\cdot p)^2 &\to&  {stu\over8} {T_q+T_{pq} \over \Delta} \,, \nonumber \\ 
 (v\cdot (p+q))^2
         &\to&  {stu\over8}  {T_p+T_q \over \Delta}  \,, \nonumber \\ 
 (v\cdot p)^4
     &\to& 3 \biggl( {stu\over8} \biggr)^2 {(T_q+T_{pq})^2 \over \Delta^2} 
           \,, \nonumber \\ 
 (v\cdot p)^2 (v\cdot q)^2
    &\to&  \biggl( {stu\over8} \biggr)^2  \biggl[ {1\over\Delta} 
                  + 3 {T_{pq}^2 \over \Delta^2} \biggr] \,, 
                 \nonumber \\ 
 (v\cdot p)^4 (v\cdot q)^2
    &\to&  \biggl( {stu\over8} \biggr)^3 \biggl[ 3 {T_q+T_{pq}\over\Delta^2} 
                           + 15 { T_{pq}^2 (T_q+T_{pq}) \over \Delta^3 } 
                     \biggr] \,, \nonumber \\ 
 (v\cdot p)^2 (v\cdot q)^2 (v\cdot (p+q))^2
    &\to&  \biggl( {stu\over8} \biggr)^3 \biggl[ 
                                             3 {T_p+T_q+T_{pq}\over\Delta^2} 
                           - 15 { T_p T_q T_{pq} \over \Delta^3 } 
                     \biggr] \,.
\label{pvqvParam}
\end{eqnarray}
These equations apply to any two-loop integral, independent of
the external momenta.  They also apply in the presence of any additional
numerator factor of the form $f(p\cdot k_i, q\cdot k_i)$, since cross 
contractions are forbidden by the orthogonality of $v$ to the $k_i$.

Finally, polynomials in $v\cdot p$, $v\cdot q$ can be related to those in
$\vec{\mud}_i$, using the expansion~(\ref{pExpansion}).  For example,
\begin{equation}
p^2 + \mud_p^2 = p_{[4]} \cdot p_{[4]}
= s \, c_1^p c_2^p + t \, c_2^p c_3^p + u \, c_1^p c_3^p 
- {1\over4} stu \, (c_v^p)^2,
\label{psqExpansion}
\end{equation}
or
\begin{equation}
\mud_p^2 = - {4\over stu} (v\cdot p)^2 + \hat{\cal P}_{p} \,,  
\label{vpsqtomupsq}
\end{equation}
where
\begin{equation}
\hat{\cal P}_{p} \equiv - p^2
 + s \, c_1^p c_2^p + t \, c_2^p c_3^p + u \, c_1^p c_3^p \,.
\label{mud_nov}
\end{equation}
Similarly,
\begin{eqnarray}
\mud_q^2 &=& - {4\over stu} (v\cdot q)^2 + \hat{\cal P}_{q} \,, 
\nonumber \\ 
\mud_{p+q}^2 &=& - {4\over stu} (v\cdot (p+q))^2
                          + \hat{\cal P}_{pq} \,,  
\label{vpsqtomupsq_more}
\end{eqnarray}
where
\begin{eqnarray}
\hat{\cal P}_{q} &\equiv& - q^2
 + s c_1^q c_2^q + t c_2^q c_3^q + u c_1^q c_3^q \,, \nonumber \\ 
\hat{\cal P}_{pq} &\equiv& - (p+q)^2
 + s (c_1^p+c_1^q)(c_2^p+c_2^q) 
 + t (c_2^p+c_2^q)(c_3^p+c_3^q)
 + u (c_1^p+c_1^q)(c_3^p+c_3^q) \,. \nonumber \\
\label{mud_nov_more}
\end{eqnarray}
Note that $\hat{\cal P}_{p}$, $\hat{\cal P}_{q}$ and 
$\hat{\cal P}_{pq}$ only contain the types of Lorentz products
which already appear in ${\cal P}_{\rm interf.}$.

Because \eqns{muParam}{pvqvParam} are so similar, and taking into account
the relations between the integrands, we can solve for the 
additional ``new'' integrals required for the helicity method, in terms 
of the ``old'' integrals needed for the interference method.
For example, for a general function $f(p\cdot k_i, q\cdot k_i)$,
we have
\begin{eqnarray}
\int \mud_p^2 \, f &=& - \e {8\over stu} \int (v\cdot p)^2 \, f 
   = - {2\e \over 1- 2 \e} \int  \hat{\cal P}_{p} \, f \,, \nonumber \\
\int \mud_{p+q}^2 \, f &=& 
    - \e {8\over stu} \int (v\cdot (p+q))^2 \, f 
   = - {2\e \over 1- 2 \e} \int  \hat{\cal P}_{pq} \, f \,, \nonumber \\
\int ( \mud_p^2 )^2 \, f &=& 
  - { \e (1-\e) \over 3 } \biggl( {8\over stu} \biggr)^2
        \int (v\cdot p)^4 \, f 
   = - {4\e (1-\e) \over (1-2\e) (3-2\e) } 
          \int \hat{\cal P}_{p}^2 \, f \,, \nonumber \\
\int \mud_p^2 \mud_q^2 \, f &=& 
    - { 4 \e (1-\e) \over (1-2\e) (3-2\e) } 
          \int \hat{\cal P}_{p} \hat{\cal P}_{q} \, f 
    + { \e \over 3-2\e } \int {f\over\Delta}
\,, \nonumber \\
\int (v\cdot p)^2 (v\cdot q)^2 \, f &=& 
\biggl( {stu\over4} \biggr)^2 \biggl[ 
      { 3 \over (1-2\e) (3-2\e) } 
          \int \hat{\cal P}_{p} \hat{\cal P}_{q} \, f 
    + { \e \over 3-2\e } \int {f\over\Delta}  \biggr] \,, \nonumber \\
\int ( \mud_p^2 )^2 \mud_q^2 \, f &=& 
    - { 8 \e (1-\e) (2-\e) \over (1-2\e) (3-2\e) (5-2\e) } 
          \int \hat{\cal P}_{p}^2 \hat{\cal P}_{q} \, f
  \PlusBreak{ }
      { 4 \e (1-\e) \over (3-2\e) (5-2\e) } 
          \int { \hat{\cal P}_{p} \, f \over\Delta } \,. 
\label{SolveFormupvqv}
\end{eqnarray}
A factor of $1/\Delta$ indicates that a shift of the dimension of 
the integral is required: $D\to D+2$, $\e \to \e-1$ 
($\e$'s in prefactors should {\it not} be shifted, however).

In practice, we used \eqns{vpsqtomupsq}{vpsqtomupsq_more} to eliminate
$v\cdot p$ and $v\cdot q$ in favor of $\mud_p^2$, $\mud_q^2$ 
and $\mud_{p+q}^2$ in the loop momentum polynomial.  We used equations
like~(\ref{SolveFormupvqv}) to compute the 
``irreducible'' monomials including the $\vec{\mud}_i$, which we then
stored.  {\it E.g.}, for the planar double box integral, we stored values for
\begin{eqnarray}
&&{\cal P}^{\P,\, {\rm irred}}_{\rm hel.}(l_1,l_2,l_3,m,n) 
  = ( \mud_p^2 )^{l_1} \, ( \mud_q^2 )^{l_2} \, ( \mud_{p+q}^2 )^{l_3} 
      \, (2q\cdot k_1)^m \, (2p\cdot k_4)^n, \nonumber \\
&& \hskip 6 cm  2 l_1 + 2 l_2 + 2 l_3 + m + n \leq6.
\label{PlanarIrredHel}
\end{eqnarray}

Having reduced all the tensor loop integrals in the amplitudes to a linear
combination of master integrals, the next step is to expand the master 
integrals in a Laurent series in $\e$, beginning at order $1/\e^4$,
using results from 
refs.~\cite{PBScalar,NPBScalar,PBReduction,NPBReduction,IntegralsAGO}.
Many of these master integral expansions are given in terms of Nielsen 
functions~\cite{NielsenRef}, usually denoted by $S_{n,p}(x)$.  
However, it is straightforward~\cite{NielsenIds} to 
express the results solely in terms of polylogarithms~\cite{Lewin},
\begin{eqnarray}
\Li_n(x) &=& \sum_{i=1}^\infty { x^i \over i^n }
= \int_0^x {dt \over t} \Li_{n-1}(t)\,,  
\\
\Li_2(x) &=& -\int_0^x {dt \over t} \ln(1-t) \,, 
\label{PolyLogDef}
\end{eqnarray}
with $n=2,3,4$.  The analytic properties of the non-planar double 
box integrals appearing in the amplitudes are somewhat
intricate~\cite{AllPlusTwo,NPBScalar}; there is no Euclidean
region in any of the three kinematic channels, $s$, $t$ or $u$.
So we do not attempt to give a crossing-symmetric representation, but
instead quote all our results in the physical $s$-channel 
$(s > 0; \; t, \, u < 0)$ for the $\ggtogg$ 
kinematics~(\ref{gggglabel}).


\subsection{Checks on results}\label{ChecksSubsection}

We performed a number of consistency checks on the amplitudes to
ensure their reliability:
\begin{enumerate}
\item As a check of gauge invariance, we verified that the amplitudes
vanish when a gluon polarization vector is replaced with a
longitudinal one.

\item The agreement of the explicitly computed infrared divergences
with the expected form~(\ref{TwoloopCatani}) provides a stringent check 
on the amplitudes.  Most of the master integrals contain divergent as well
as finite terms, so the finite remainders are checked indirectly in this way.

\item Using supersymmetry Ward identities~\cite{SWI}, we evaluated the
identical-helicity case, including fermion loops~\cite{TwoLoopSUSY}, by 
relating it to the already known identical-helicity pure-glue $\ggtogg$ 
amplitude~\cite{AllPlusTwo}.  The integration in ref.~\cite{AllPlusTwo} was 
done by a completely different technique, thus checking the programs and 
integration methods used to obtain the general helicity cases.

\item As described in more detail at the end of
section~\ref{SchemeConvertSection}, we compared our results for $\ggtogg$
to those of ref.~\cite{GOTYgggg}.  The interference of the two-loop
$\ggtogg$ helicity amplitudes with the tree amplitudes, after summing over
all external helicities and colors, and accounting for the different
schemes used (HV {\it vs.}~CDR), agrees precisely with the calculation of
ref.~\cite{GOTYgggg}.
\end{enumerate}


\subsection{${\bom H}^{(2)}$ operator}\label{HSubsection}

As mentioned in section~\ref{IRSection}, the function 
${\bom H}^{(2)}_{\RS}(\e)$, which contains only $1/\e$ poles,
has not been predicted {\it a priori} for general processes.  
However, there is accumulating evidence from explicit 
calculations~\cite{GOTY2to2,GOTYgggg,TwoLoopee3Jets} in the CDR scheme
that the color- and helicity-summed matrix element
$\langle \cm^{(0)} | {\bom H}^{(2)}(\e) 
| \cm^{(0)} \rangle$ is a sum of terms for each 
external colored leg in the process, namely 
\begin{equation}
\langle \cm^{(0)} | {\bom H}^{(2)}(\e) | \cm^{(0)} \rangle
 = 
{ e^{-\e\psi(1)} \over 4\e \, \Gamma(1-\e) }
  ( n_g H_g^{(2)} + n_q H_q^{(2)} ) 
   \langle \cm^{(0)} | \cm^{(0)} \rangle \,,
\label{GOTYH}
\end{equation}
where $n_g$ is the number of external gluons, and $n_q$ is the number of
external quarks plus anti-quarks, with
\begin{eqnarray}
H_g^{(2)} &=& 
\biggl( {\zeta_3 \over 2} + {11 \over 144} \pi^2 + {5\over 12} \biggr) N^2
+ \biggl( - {\pi^2\over 72} - {89 \over 108} \biggr) N \Nf
- {\Nf \over 4N}
+ {5 \over 27} \, \Nfsq  \,, 
\label{Hgluon} \\
H_q^{(2)} &=& 
\biggl( {7 \over 4} \zeta_3 - {11 \over 96} \pi^2 + {409\over 864} \biggr) N^2
+ \biggl( - {1 \over 4} \zeta_3 - {\pi^2 \over 96} - {41\over 108} \biggr)
+ \biggl( - {3 \over 2} \zeta_3 + {\pi^2 \over 8} - {3\over 32} \biggr) 
   {1 \over N^2}
\PlusBreak{}
 \biggl( {\pi^2\over 48} - {25 \over 216} \biggr) {N^2 -1 \over N} \, \Nf \,.
\label{Hquark}
\end{eqnarray}
Note that $H_g^{(2)}$ and $H_q^{(2)}$ are constants, independent of the 
kinematic variables.

We find that the full color and helicity dependence of 
${\bom H}^{(2)}_{\RS}(\e)$ for $\ggtogg$ is the sum of two terms,
\begin{equation}
{\bom H}^{(2)}(\e) = 
{ e^{-\e\psi(1)} \over 4\e \, \Gamma(1-\e) }
 \biggl( { \mu^2 \over -s } \biggr)^{2\e} 
  \Bigl( 4 H_g^{(2)} \, {\bom 1} + \hat{\bom H}^{(2)} \Bigr) \,,
\label{OurH}
\end{equation}
where
\begin{equation}
\hat{\bom H}^{(2)} = - 4 \, \ln\biggl( {-s\over-t} \biggr)
                            \ln\biggl( {-t\over-u} \biggr)
                            \ln\biggl( {-u\over-s} \biggr)
      \times \Bigl[ {\bom T}_1 \cdot {\bom T}_2 \,,
                    {\bom T}_2 \cdot {\bom T}_3 \Bigr] \,,
\label{OurHExtra}
\end{equation}
with $\ln((-s)/(-t)) \to \ln s - \ln(-t) - i\pi$ in the $s$-channel, {\it etc.}
(The overall factor of $(\mu^2/(-s))^{2\e}$ is a choice of convention, because 
$(\mu^2/(-s))^{2\e} - 1$ is of order $\e$.  Including it cleans up 
the finite remainder $\cm_\ggtogg^{(2){\rm fin}}$ a bit.)
The first term in the sum is proportional to the identity matrix in 
both helicity and color spaces.   In the HV scheme, $H_g^{(2)}$ is given 
by precisely the same value~(\ref{Hgluon}) found in the CDR 
scheme~\cite{GOTYgggg}.  The value in the FDH scheme is different; 
see~\eqn{GeneralH}.

The second term in \eqn{OurH} is also independent of the helicity 
configuration, but it is a nontrivial commutator matrix in color space.  
(The possibility of nontrivial color structure in ${\bom H}^{(2)}(\e)$ 
was pointed out in ref.~\cite{Catani}.)  Indeed, it vanishes when 
sandwiched between tree amplitudes, after performing the color sum,
\begin{equation}
\langle \cm^{(0)} | \hat{\bom H}^{(2)} | \cm^{(0)} \rangle
 \propto 
\langle \cm^{(0)} | \Bigl( {\bom T}_1 \cdot {\bom T}_2 \,
                           {\bom T}_2 \cdot {\bom T}_3
                         - {\bom T}_2 \cdot {\bom T}_3 \,
                           {\bom T}_1 \cdot {\bom T}_2 \Bigr)
                   | \cm^{(0)} \rangle
 = 0,
\label{HExtraTrace}
\end{equation}
using hermiticity of the ${\bom T}_i$.  \Eqn{HExtraTrace} ensures
that the result~(\ref{OurH}) is perfectly compatible with the previous
color-summed results~(\ref{GOTYH}).

Actually, in the course of the $q\bar{q} \to q\bar{q}$ calculation,
the authors of ref.~\cite{GOTY2to2} evaluated the matrix element
\begin{equation}
\langle \cm_{q\bar{q}'\to q\bar{q}'}^{(0)} | \hat{\bom H}^{(2)} 
| \cm_{q\bar{q}\to q'\bar{q}'}^{(0)} \rangle \,,
\label{HExtraOffDiagonal}
\end{equation}
relevant for the interference term in identical-quark scattering.
Because the color structures for the $s$-channel and $t$-channel tree
amplitudes, $\cm_{q\bar{q}\to q'\bar{q}'}^{(0)}$ and 
$\cm_{q\bar{q}'\to q\bar{q}'}^{(0)}$, are different, \eqn{HExtraTrace}
does not apply, and the non-vanishing CDR result is completely consistent
with our HV result~\eqn{OurHExtra}, 
including all normalization factors~\cite{BabisPrivate}.   

We find that the commutator term is the same in FDH scheme as well.
Note that it can be rewritten as a ``triple product'' in color space,
because
\begin{equation}
\Bigl[ {\bom T}_1 \cdot {\bom T}_2 \,,
       {\bom T}_2 \cdot {\bom T}_3 \Bigr]
 = i f_{abc} T_1^a T_2^b T_3^c \,.
\label{CGForm}
\end{equation}
This form of the color operator has previously appeared in analysis of
the contributions of one-loop factors for soft radiation 
({\it i.e.}, ${\cal S}^{(1)}$ in \eqn{LoopSoftColFact}) 
at NNLO~\cite{CataniGrazziniSoft}.
This fact, and the lack of scheme dependence for
$\hat{\bom H}^{(2)}$, leads one to suspect that it arises from soft, not
collinear, virtual contributions.  The $1/\e$ divergence would presumably
cancel against the contributions discussed in
ref.~\cite{CataniGrazziniSoft}, in a color-resolved approach to a NNLO
computation.  In a fully color-summed approach, however, such contributions
should cancel individually, thanks to~\eqn{HExtraTrace}.

The factorization of $\hat{\bom H}^{(2)}$ in~\eqn{OurHExtra} into 
a product of kinematic and color factors is clearly an accident of
having exactly four external colored partons.  Color conservation,
${\bom T}_1 + {\bom T}_2 + {\bom T}_3 + {\bom T}_4 = 0$ in the color-space
notation, implies that there are only three independent 
${\bom T}_i \cdot {\bom T}_j$ factors, say 
${\bom T}_1 \cdot {\bom T}_2$, ${\bom T}_2 \cdot {\bom T}_3$, and
${\bom T}_1 \cdot {\bom T}_3$.  But their sum is a $c$-number,
\begin{equation}
   {\bom T}_1 \cdot {\bom T}_2
 + {\bom T}_2 \cdot {\bom T}_3
 + {\bom T}_1 \cdot {\bom T}_3
 = {1\over2} \Bigl[ {\bom T}_4^2 
                  - {\bom T}_1^2 - {\bom T}_2^2 - {\bom T}_3^2 \Bigl] \,,
\label{TTSum}
\end{equation}
hence there is only one independent commutator.  (A similar argument
holds in the $f_{abc} T_i^a T_j^b T_k^c$ 
representation~\cite{CataniGrazziniSoft}.)
For three or less external colored partons, all such structures vanish;
whereas for five or more partons there are multiple independent ones.
 

\subsection{Finite remainders}\label{TwoLoopFinRemSubsection}

The two-loop finite remainders are defined in~\eqn{TwoloopCatani} and
are color decomposed into $M^{(2),[i]{\rm
fin}}_{\lambda_1\lambda_2\lambda_3\lambda_4}$ in
\eqn{RemoveColorPhase}.  Their dependence on the renormalization scale
$\mu$, $N$ and $\Nf$ may be extracted as
\begin{eqnarray}
 M^{(2),[i]{\rm fin}}_{\lambda_1\lambda_2\lambda_3\lambda_4} 
&=& 
 - \Bigl[ b_0^2 \, (\ln(s/\mu^2) - i\pi)^2 
        + b_1 \, (\ln(s/\mu^2) - i\pi) \Bigr]
    M^{(0),[i]}_{\lambda_1\lambda_2\lambda_3\lambda_4}
 \MinusBreak{ }
  2 b_0 \, (\ln(s/\mu^2) - i\pi) \, 
    M^{(1),[i]{\rm fin}}_{\lambda_1\lambda_2\lambda_3\lambda_4}
 \PlusBreak{ }
   N^2 \, A^{[i]}_{\lambda_1\lambda_2\lambda_3\lambda_4}
 +        B^{[i]}_{\lambda_1\lambda_2\lambda_3\lambda_4}
 + N \Nf \, C^{[i]}_{\lambda_1\lambda_2\lambda_3\lambda_4} 
 + {\Nf \over N} \, D^{[i]}_{\lambda_1\lambda_2\lambda_3\lambda_4} 
 \PlusBreak{ }
  \Nfsq \, E^{[i]}_{\lambda_1\lambda_2\lambda_3\lambda_4} 
 + {\Nfsq \over N^2} \, F^{[i]}_{\lambda_1\lambda_2\lambda_3\lambda_4} 
\,, 
  \hskip3cm   i = 1,2,3,
\label{TwoloopSingleTrace} \\
 M^{(2),[i]{\rm fin}}_{\lambda_1\lambda_2\lambda_3\lambda_4} 
&=& 
 - 2 b_0 \, (\ln(s/\mu^2) - i\pi) \, 
    M^{(1),[i]{\rm fin}}_{\lambda_1\lambda_2\lambda_3\lambda_4}
 \PlusBreak{ }
   N \, G^{[i]}_{\lambda_1\lambda_2\lambda_3\lambda_4}
 + \Nf \, H^{[i]}_{\lambda_1\lambda_2\lambda_3\lambda_4} 
 + {\Nfsq \over N} \, I^{[i]}_{\lambda_1\lambda_2\lambda_3\lambda_4} 
\,, 
  \hskip0.6cm i = 7,8,9.
\label{TwoloopDoubleTrace} 
\end{eqnarray}
The $\mu$-dependence is a consequence of renormalization group invariance.
The tree and one-loop functions,
$M^{(0),[i]}_{\lambda_1\lambda_2\lambda_3\lambda_4}$
and $M^{(1),[i]{\rm fin}}_{\lambda_1\lambda_2\lambda_3\lambda_4}$,
are given in~\eqn{TreeAmps} and 
eqs.~(\ref{OneloopRemainderDef})--(\ref{hmpmp}),
respectively, while $b_0$ and $b_1$ are given
in~\eqn{QCDBetaCoeffs}.

The coefficient functions $A,B,C,D,E,F,G,H,I$, which depend only on
the Mandelstam variables, obey several relations.  
Group theory ({\it e.g.} $U(1)$ decoupling identities)
implies that the $G$ functions are not independent of the others,
\begin{eqnarray}
G^{[7]}_{\lambda_1\lambda_2\lambda_3\lambda_4} & = &
 2 \Bigl( A^{[1]}_{\lambda_1\lambda_2\lambda_3\lambda_4}
   + A^{[2]}_{\lambda_1\lambda_2\lambda_3\lambda_4}
   + A^{[3]}_{\lambda_1\lambda_2\lambda_3\lambda_4} \Bigr) 
   - B^{[3]}_{\lambda_1\lambda_2\lambda_3\lambda_4}
\,, \label{G7elim}\\[1pt plus 4pt]
%
%
G^{[8]}_{\lambda_1\lambda_2\lambda_3\lambda_4} & = &
 2 \Bigl( A^{[1]}_{\lambda_1\lambda_2\lambda_3\lambda_4}
   + A^{[2]}_{\lambda_1\lambda_2\lambda_3\lambda_4}
   + A^{[3]}_{\lambda_1\lambda_2\lambda_3\lambda_4} \Bigr) 
   - B^{[1]}_{\lambda_1\lambda_2\lambda_3\lambda_4}
\,, \label{G8elim}\\[1pt plus 4pt]
%
%
G^{[9]}_{\lambda_1\lambda_2\lambda_3\lambda_4} & = &
 2 \Bigl( A^{[1]}_{\lambda_1\lambda_2\lambda_3\lambda_4}
   + A^{[2]}_{\lambda_1\lambda_2\lambda_3\lambda_4}
   + A^{[3]}_{\lambda_1\lambda_2\lambda_3\lambda_4} \Bigr) 
   - B^{[2]}_{\lambda_1\lambda_2\lambda_3\lambda_4} 
\,, \label{G9elim}
\end{eqnarray}
and that the sum of the $B$ coefficients vanishes,
\begin{equation}
B^{[3]}_{\lambda_1\lambda_2\lambda_3\lambda_4} = 
    - B^{[1]}_{\lambda_1\lambda_2\lambda_3\lambda_4}
    - B^{[2]}_{\lambda_1\lambda_2\lambda_3\lambda_4} \,.
\label{B3elim}
\end{equation}

As at one loop, for the ++++, $-$+++, and $-$$-$++ helicity
configurations, Bose symmetry under exchange of legs 3 and 4 ($t \lr u$) 
implies further relations,
\begin{eqnarray}
X^{[2]}_{\lambda_1\lambda_2\lambda_3\lambda_4}(s,t,u)
 &=& X^{[1]}_{\lambda_1\lambda_2\lambda_3\lambda_4}(s,u,t) \,,
 \hskip1.2cm X \in \{ A,B,C,D,E,F \}, 
\label{X2sym} \\
Y^{[9]}_{\lambda_1\lambda_2\lambda_3\lambda_4}(s,t,u)
 &=& Y^{[8]}_{\lambda_1\lambda_2\lambda_3\lambda_4}(s,u,t) \,,
 \hskip1.3cm Y \in \{ G,H,I \}. 
\label{Y9sym}
\end{eqnarray}

In appendix~\ref{QCDRemainderAppendix},
we give the explicit forms for the independent finite remainder 
functions appearing in~\eqns{TwoloopSingleTrace}{TwoloopDoubleTrace}.
For the two complicated helicity configurations, $-$$-$++ and $-$+$-$+,
these functions also involve auxiliary functions, 
$A^{\Susy,[i]}$ and $B^{\Susy,[i]}$,
which will be presented in appendix~\ref{N=1RemainderAppendix}.
The latter functions also serve as the finite remainders for $\ggtogg$ 
in $N=1$ super-Yang-Mills theory, as discussed in 
section~\ref{TwoLoopFiniteN=1Section}.


\section{Scheme conversion at two loops}\label{SchemeConvertSection}

The preceding helicity amplitudes were presented in the HV
variant of dimensional regularization and renormalization.  
As mentioned in the introduction, the HV scheme contains 
$D - 2 = 2 - 2\e$ virtual (unobservable) gluon states, and 2 external 
(observable) gluon states.  However, it is possible to alter the number 
of virtual states.  In the FDH scheme~\cite{BKgggg,TwoLoopSUSY},
one adjusts the number of virtual gluon states to be 2, matching the 
number of external states, and also matching the number of fermionic 
degrees of freedom in a supersymmetric theory.  This scheme is quite
similar to dimensional reduction ($\DRbar$)~\cite{DR}.  

Dimensional reduction is usually thought of as having $D<4$, {\it i.e.},
$\e>0$, and contains $D - 2 = 2 - 2\e$ gluon states, plus $2\e$ scalar
states, for a total of 2 bosonic states.  On the other hand, the helicity
of a particle is its angular momentum eigenvalue for a rotation in the
two-dimensional plane normal to its momentum vector.  If $D$ is less than
four, this plane does not exist, making the definition of helicity
obscure.  The FDH scheme can be regarded as an analytic continuation of
$\DRbar$ to $D>4$, to make it compatible with helicity amplitudes.  
No scalars are required, however.  For both the HV and FDH schemes, 
helicity amplitudes with fermions and gluons are computed in the same 
fashion, with $D$-dimensional loop momenta and four-dimensional gluon
polarization vectors (see section~\ref{TensorSection}).
In performing the algebra leading to the loop-momentum polynomial ${\cal P}$, 
when the trace of the Minkowski metric is encountered, one sets 
\begin{eqnarray}
\eta^{\mu}{}_{\mu} &\equiv& D_s \equiv 4 - 2 \e \, \delta_R  \,, 
\label{EtaTrace} \\ \nonumber \\
\delta_R &=& \cases{ 1, \quad \hbox{HV scheme}, \cr 
                     0, \quad \hbox{FDH scheme}. }
\label{deltaR}
\end{eqnarray}
This procedure is gauge invariant because the terms proportional to 
$D_s$ are related to loops containing scalar fields in the adjoint 
representation~\cite{TwoLoopSUSY}.  We allow $\delta_R$ to be arbitrary
below, although only the HV and FDH cases in~\eqn{deltaR} seem 
well motivated.  

The CDR scheme has $D_s - 2 = 2 - 2\e$ virtual gluon states, just as
in the HV scheme; but in addition there are $D - 2 = 2 - 2\e$ external
gluon states.  To convert from the HV to the CDR scheme within the helicity
method, one could in principle compute additional amplitudes where
some external states have $\e$-helicities (explicit polarization
vectors that point into the extra
$(-2\e)$-dimensions)~\cite{KosowerHelicity}.  Since the CDR result is
already available via the interference method~\cite{GOTYgggg}, we have
not done that computation.  Instead we shall check the conversion
between schemes expected from experience at one loop.

A given scheme has implications for regularization of both ultraviolet
and infrared singularities.  These implications have been discussed
extensively at one loop~\cite{BKgggg,KSTfourparton,CST}.  Let us first
consider the ultraviolet situation.  Renormalization by modified
minimal subtraction, as
in~\eqns{OneloopCounterterm}{TwoloopCounterterm}, leads to different
renormalized coupling constants, related by finite shifts.  In the
class of schemes we are considering, the ultraviolet behavior only
depends on the number of virtual gluon states.  Thus the CDR and HV
schemes imply the same coupling constant, the standard $\MSbar$
coupling, $\alpha_s(\mu)$.  The FDH and $\DRbar$ schemes also are the
same in the ultraviolet (the ability to accommodate helicity, and the
sign of $\e$, are irrelevant here), and so they both define the
$\DRbar$ coupling, $\alpha_s^{\DRbar}(\mu)$.

To shift from either pair of schemes to the other, {\it in 
the ultraviolet}, amounts to using the following relations between 
coupling constants~\cite{MartinVaughn,KSTfourparton}, 
recently improved to two-loop accuracy~\cite{TwoLoopSUSY},
\begin{eqnarray}
\alpha_s^{\DRbar}(\mu) &=& \alpha_s(\mu) \biggl[ 1 
  + {C_A \over 6} {\alpha_s(\mu) \over 2\pi} 
  + {11 C_A^2 - 9 C_F T_R \Nf \over 18 }       
          \biggl( {\alpha_s(\mu) \over 2\pi} \biggr)^2 
  \PlusBreak{ \alpha_s(\mu) \biggl[ }
    \Ord([\alpha_s(\mu)]^3) \biggr] \,,
\label{DRMSconv} \\
\alpha_s(\mu) &=& \alpha_s^{\DRbar}(\mu) \biggl[ 1 
  - {C_A \over 6} {\alpha_s^{\DRbar}(\mu) \over 2\pi} 
  - {10 C_A^2 - 9 C_F T_R \Nf \over 18 }       
        \biggl( {\alpha_s^{\DRbar}(\mu) \over 2\pi} \biggr)^2 
  \PlusBreak{ \alpha_s^{\DRbar}(\mu) \biggl[ }
  \Ord([\alpha_s^{\DRbar}(\mu)]^3) \biggr] \,.
\label{MSDRconv} 
\end{eqnarray}
(Recall that the three-loop running coupling enters into any NNLO 
computation.  The three-loop beta-function coefficient $b_2$ in 
$\DRbar$ differs from the value in $\MSbar$~\cite{MSbarb2}, but it can be 
obtained simply from the coupling shift~(\ref{DRMSconv})~\cite{TwoLoopSUSY}.)
For completeness, we give the two-loop relation~\cite{TwoLoopSUSY} 
between the $\MSbar$ coupling and that defined by an arbitrary value of 
$\delta_R$,
\begin{eqnarray}
\alpha_s^{\delta_R}(\mu) &=& \alpha_s(\mu) \biggl[ 1 
  + {C_A \over 6} (1-\delta_R) {\alpha_s(\mu) \over 2\pi} 
  \PlusBreak{ \alpha_s(\mu) \biggl[ }
    \biggl( {C_A^2 \over 36} (1-\delta_R)^2
        + {7 C_A^2 - 6 C_F T_R \Nf \over 12 } (1-\delta_R) \biggr)   
          \biggl( {\alpha_s(\mu) \over 2\pi} \biggr)^2 
  \PlusBreak{ \alpha_s(\mu) \biggl[ }
    \Ord([\alpha_s(\mu)]^3) \biggr] \,,
\label{deltaRMSconv} \\
\alpha_s(\mu) &=& \alpha_s^{\delta_R}(\mu) \biggl[ 1 
  - {C_A \over 6} (1-\delta_R) {\alpha_s^{\delta_R}(\mu) \over 2\pi} 
  \PlusBreak{ \alpha_s^{\delta_R}(\mu) \biggl[ }
    \biggl( {C_A^2 \over 36} (1-\delta_R)^2
        - {7 C_A^2 - 6 C_F T_R \Nf \over 12 } (1-\delta_R) \biggr)   
        \biggl( {\alpha_s^{\delta_R}(\mu) \over 2\pi} \biggr)^2 
  \PlusBreak{ \alpha_s^{\delta_R}(\mu) \biggl[ }
  \Ord([\alpha_s^{\delta_R}(\mu)]^3) \biggr] \,,
\label{MSdeltaRconv} 
\end{eqnarray}
which reduces to~\eqns{DRMSconv}{MSDRconv} for $\delta_R=0$.
Inserting the appropriate coupling relation into the perturbative 
expansion of the amplitude~(\ref{RenExpand}) leads to simple, finite 
ultraviolet conversion relations between renormalized amplitudes 
$\cm_n^{(L)}$.

Because the ultraviolet shifts are so simple to implement, in the rest 
of this paper (with one exception to be discussed below) we take the 
ultraviolet scheme to be the same as the infrared scheme.  That is, 
when we report results for $\cm_\ggtogg^{(L)}$ in the FDH scheme, 
they correspond to coefficients of a perturbative expansion defined as 
in~\eqn{RenExpand}, but where $\alpha_s(\mu)$ is the FDH/$\DRbar$ 
coupling, $\alpha_s^{\DRbar}(\mu)$.  In the more general $\delta_R$ scheme,
the expansion parameter would be $\alpha_s^{\delta_R}(\mu)$.

In this convention, the one-loop relation between $\ggtogg$ helicity
amplitudes is~\cite{BKgggg,KSTfourparton}
\begin{equation}
| \cm_\ggtogg^{(1)} \rangle_{\delta_R} 
= | \cm_\ggtogg^{(1)} \rangle_\HV
  + {C_A\over6} \, (1-\delta_R) \, | \cm_\ggtogg^{(0)} \rangle \,,
\label{OneLoopFiniteShift}
\end{equation}
which only involves a finite shift.  The corresponding relation at two
loops also requires shifts of the divergent terms in the infrared 
decomposition~(\ref{TwoloopCatani}).  We find that
\begin{eqnarray}
K_{\delta_R} &=& K_\HV 
   - C_A \biggl( {1\over6} + {4\over9} \, \e \biggr) \, (1-\delta_R) \,,
\label{GeneralK} \\
(H_g^{(2)})_{\delta_R} &=& H_g^{(2)} 
    - {C_A \over 6} \, b_0 \, (1-\delta_R) \,,
\label{GeneralH} \\
| \cm_\ggtogg^{(2){\rm fin}} \rangle_{\delta_R} 
&=& | \cm_\ggtogg^{(2){\rm fin}} \rangle_\HV
   \PlusBreak{ }
    \Biggl[ C_A^2 \biggl( - {5 \over 144} \pi^2 + {5 \over 12} \biggr)
    + T_R \, \Nf \biggl( {8 \over 27} C_A - {1 \over 2} C_F \biggr) 
    \Biggr] \, (1-\delta_R) \, | \cm_\ggtogg^{(0)} \rangle \,, 
\nonumber \\
\label{GeneralFinite}
\end{eqnarray}
where $K_\HV$ is given in~\eqn{CataniK} and $H_g^{(2)}$ (the value in 
the HV or CDR schemes) is given in~\eqn{Hgluon}.
Because ${\bom I}^{(1)}(2\e,\mu;\{p\})$ contains at most $1/\e^2$ poles,
the term proportional to $\e \times (1-\delta_R)$ in $K_{\delta_R}$ 
clearly could be shifted into $(H_g^{(2)})_{\delta_R}$ if desired.  However,
the assignment we have chosen makes \eqn{GeneralFinite} simpler.

Also, the interpretation of $K_\RS$ as
the integral of a splitting function\footnote{We thank Henry Wong for
clarifying this point.}~\cite{CataniSeymourBig,CST} 
leads to both the $\e^0$ and $\e$ terms proportional to $(1-\delta_R)$ 
in \eqn{GeneralK}:  The azimuthally-averaged $g \to gg$ splitting kernel
is given in the general $\delta_R$ scheme by~\cite{CST}
\begin{equation}
\langle \hat{P}^{\delta_R}_{gg}(z;\e) \rangle 
  = 2 C_A \biggl[
  { z \over 1-z } + { 1-z \over z } 
   + \biggl( 1 + {\e \over 1 - \e} (1-\delta_R) \biggr) z (1-z) \biggr]
  \,.
\label{Pgg}
\end{equation}
With the identification~\cite{CST}
\begin{equation}
- {1\over2} \sum_b \int_0^1 dz\ (z(1-z))^{-\e}
 \ \langle \hat{P}^{\delta_R}_{gg}(z;\e) \rangle = 
  { 2 \, {\bom T}_a^2 \over \e } + \gamma_a 
 + \biggl( K_a - {\pi^2\over6} {\bom T}_a^2 \biggr) \e \,,
\label{Ksplitdef}
\end{equation}
one sees that the $\delta_R$-dependent term of $K \equiv K_g$ is
\begin{equation}
- { C_A \over 1 - \e } (1-\delta_R) \int_0^1 dz\ (z(1-z))^{1-\e} 
 = - C_A \biggl( {1\over6} + {4\over9} \, \e \biggr) \, (1-\delta_R) \,,
\label{KdeltaCalc}
\end{equation}
in agreement with~\eqn{GeneralK}.

One can also present results for the two-loop $\ggtogg$ amplitudes
using the $\delta_R$ scheme as the infrared regulator, but switching to 
the $\MSbar$ coupling constant with the aid of~\eqn{deltaRMSconv}.
For the infrared decomposition~(\ref{TwoloopCatani}) to hold,
assuming that ${\bom I}^{(1)}(\e)$ is scheme-independent, we find that
the quantity $K_\RS$ must be set to 
\begin{equation}
\tilde{K}_{\delta_R} = K_\HV + {C_A\over2} (1-\delta_R) + \Ord(\e).
\label{GeneralKTilde}
\end{equation}
Thus such a definition of $K$ would be scheme dependent too.
Also, in contrast to the simplicity of~\eqn{GeneralH}, 
the scheme-dependent part of ${\bom H}^{(2)}$ will contain logarithms
and will no longer be proportional to the identity matrix in color space.
Hence we refrain from presenting such a decomposition explicitly.

Finally we discuss conversion from the HV scheme results
reported in section~\ref{TwoLoopFiniteQCDSection} to the CDR scheme used
in ref.~\cite{GOTYgggg}.  In the CDR scheme, one usually computes 
the interference of amplitudes, summed over all external colors and 
($2-2\e$) polarizations.  The generic one-loop/tree interference encountered
at NLO is
\begin{equation}
2 \, \Re \, \bar{I}^{(1,0)}_\RS 
\equiv 2 \, \Re \, \sum_{\rm color, hel.} 
\Bigl[ \langle \cm_n^{(1)} | \cm_n^{(0)} \rangle \Bigr]_\RS
 \,.
\label{OneLoopInterference}
\end{equation}
Inserting the infrared decomposition~(\ref{OneloopCatani}) for 
$\cm_n^{(1)}$ into~\eqn{OneLoopInterference} gives
\begin{equation}
\bar{I}^{(1,0)}_\RS = 2 \, \Re \, \sum_{\rm color, hel.} 
\Bigl[ \langle \cm_n^{(0)} | {\bom I}^{(1)} | \cm_n^{(0)} \rangle\Bigr]_\RS
+ \bar{I}^{(1,0){\rm fin}}_\RS 
 \,,
\label{OneloopInterfExp}
\end{equation}
where
\begin{equation}
\bar{I}^{(1,0){\rm fin}}_\RS = 
2 \, \Re \, \sum_{\rm color, hel.} 
\Bigl[ \langle \cm_n^{(1){\rm fin}} | \cm_n^{(0)} \rangle 
\Bigr]_\RS \,.
\label{OneloopInterfFinite}
\end{equation}
It is well-established from explicit calculations and general 
arguments~\cite{BKgggg,KSTfourparton,CST} that the finite 
remainder~(\ref{OneloopInterfFinite}) has the same value in the HV and 
CDR schemes, in the limit $\e \to 0$.  Essentially, the treatment of
unobserved partons is the same in both schemes, so the infrared 
divergences should take the same form, when expressed in terms of the 
lower-order-in-$\alpha_s$ amplitudes.  

It is natural to expect the same pattern to hold at two loops.
The two-loop/tree interference is
\begin{eqnarray}
2 \, \Re \, \bar{I}^{(2,0)}_\RS 
&\equiv& 2 \, \Re \, \sum_{\rm color, hel.} 
\Bigl[ \langle \cm_n^{(2)} | \cm_n^{(0)} \rangle \Bigr]_\RS
\label{Interference} \\
&=&  2 \, \Re \sum_{\rm color, hel.}
\Bigl[ \langle \cm_n^{(0)} | {\bom I}^{(2)} | \cm_n^{(0)} \rangle 
     + \langle \cm_n^{(1)} | {\bom I}^{(1) \dagger} 
                                 | \cm_n^{(0)} \rangle 
\Bigr]_\RS
+ \bar{I}^{(2,0){\rm fin}}_\RS ,
\label{InterfExp}
\end{eqnarray}
where
\begin{equation}
\bar{I}^{(2,0){\rm fin}}_\RS = 
2 \, \Re \, \sum_{\rm color, hel.} 
\Bigl[ \langle \cm_n^{(2){\rm fin}} | \cm_n^{(0)} \rangle 
\Bigr]_\RS \,.
\label{InterfFinite}
\end{equation}
Note that ${\bom I}^{(1)}$ and ${\bom I}^{(2)}$ are the 
same operators in the HV scheme as in the CDR scheme.

We have interfered the color-decomposed finite remainders of the two-loop
$\ggtogg$ helicity amplitudes in the HV scheme, as given in 
section~\ref{TwoLoopFiniteQCDSection},
with the tree amplitudes given in~\eqn{TreeAmps}, summing over all
external helicities and colors with the help of~\eqn{ColorSumMatrix}.
This sum gives precisely the same result as the corresponding
quantity~(\ref{InterfFinite}) in the CDR scheme, as evaluated in 
ref.~\cite{GOTYgggg}, after accounting for the slightly different 
definition of ${\bom H}^{(2)}$ that we used in~\eqn{OurH}.  We conclude 
that~\eqn{InterfFinite} should be the same in the HV or CDR schemes for
general two-loop QCD scattering amplitudes.


\section{Two-loop amplitudes in pure $N=1$ super-Yang-Mills theory}
\label{TwoLoopFiniteN=1Section}

The quarks of QCD are (massless) fermions transforming in the fundamental
representation of $SU(N)$.   If one replaces the quarks by a gluino, 
a massless Majorana fermion transforming in the adjoint representation, 
one obtains a supersymmetric theory, pure $N=1$ super-Yang-Mills theory.
The amplitudes for this theory, when it is regularized in a
supersymmetry preserving fashion, obey supersymmetry Ward
identities~\cite{SWI}, and from experience at one loop they
are expected to be simpler than the corresponding QCD amplitudes.  
On the other hand, $SU(N)$ group theory generates linear relations 
between amplitudes of the two theories, so one can use the two-loop 
super-Yang-Mills amplitudes to simplify the presentation of the 
two-loop QCD amplitudes, as we do in 
appendix~\ref{QCDRemainderAppendix}.  

In this section we discuss the supersymmetry Ward identities and infrared
decomposition for the two-loop amplitudes in pure $N=1$ super-Yang-Mills
theory.  Then we describe the finite remainder functions for these
amplitudes, deferring the most complicated formulas to 
appendix~\ref{N=1RemainderAppendix}.

Here we work in the FDH scheme discussed
in section~\ref{SchemeConvertSection}, in order that the Ward identities
are valid.  One set of identities implies that ``maximal helicity violating''
amplitudes vanish for any supersymmetric theory and any number of loops,
\begin{eqnarray}
\cm_n^{\rm SUSY}(g_1^\pm,g_2^+,g_3^+,\ldots, g_n^+) &=& 0,
\label{SUSYVanish} \\
\cm_n^{\rm SUSY}(\tg_1^-,\tg_2^+,g_3^+,\ldots, g_n^+) &=& 0,
\label{SUSYVanishGluinos}
\end{eqnarray}
where $g$ ($\tg$) denotes a gluon (gluino), and the superscripts denote
helicities in the all-outgoing convention.  In addition
to~\eqns{SUSYVanish}{SUSYVanishGluinos}, all other super-Yang-Mills
$n$-point amplitudes containing either zero or one negative-helicity
particle vanish trivially, by using gluino helicity conservation.  We have
checked that~\eqn{SUSYVanish} is indeed obeyed for the four-point 
amplitude $\ggtogg$ at two loops in the FDH scheme~\cite{TwoLoopSUSY}.

Other identities relate the non-vanishing supersymmetric 
helicity amplitudes for external gluons alone, to amplitudes where some 
of the gluons are replaced by gluinos.  For the four-point amplitudes,
all the non-vanishing amplitudes in pure $N=1$ super-Yang-Mills theory 
can be related to the $\ggtogg$ amplitudes~\cite{KSTfourparton}:
\begin{eqnarray}
\cm_4^\Susy(\tg_1^-,\tg_2^+,g_3^-,g_4^+) &=& 
{ \spa2.3 \over \spa1.3 } \, \cm_4^\Susy(g_1^-,g_2^+,g_3^-,g_4^+),
\label{SUSYNonVanish1} \\
\cm_4^\Susy(\tg_1^-,\tg_2^+,\tg_3^-,\tg_4^+) &=& 
{ \spa2.4 \over \spa1.3 } \, \cm_4^\Susy(g_1^-,g_2^+,g_3^-,g_4^+).
\label{SUSYNonVanish2}
\end{eqnarray}
These relations are crossing symmetric, when a crossing symmetric
definition~\cite{GunionKunszt} of the spinor products is used.  Thus,
to obtain all the $\tg\tg\to gg$, $\tg g\to \tg g$, $gg\to\tg\tg$, and
$\tg\tg\to\tg\tg$ amplitudes
from~\eqns{SUSYNonVanish1}{SUSYNonVanish2}, it suffices to give the
two independent non-vanishing helicity amplitudes for $\ggtogg$, namely
the $-$$-$++ and $-$+$-$+ configurations.

First we present the infrared decomposition of the pure $N=1$
super-Yang-Mills amplitudes at two loops.  The equations in
section~\ref{IRSection} hold with a few modifications for the
super-Yang-Mills case.  We use the perturbative
expansion~(\ref{RenExpand}) but in terms of the FDH (or $\DRbar$)
coupling~(\ref{DRMSconv}).  The group-theoretic replacements required
to convert the quarks to gluinos are 
\begin{equation}
C_F \to C_A, 
\hskip 2 cm  
T_R \Nf \to C_A/2 \, .
\label{CasimirSub}
\end{equation}
Some of the previous equations, such as~\eqn{Hgluon} for $H_g^{(2)}$,
are given for gauge group $SU(N)$ in terms of $N$ and $\Nf$, with
$T_R=1/2$, rather than in terms of general Casimir operators.  In such
equations, to recover the Casimir representation, one should first
substitute $\Nf \to 2 T_R \Nf$ and $1/N \to (C_A-2C_F)$, followed by
$N \to C_A$.  Then one can apply the substitutions~(\ref{CasimirSub}).

The first two coefficients of the beta function for pure $N=1$
super-Yang-Mills theory are
\begin{equation}
b_0^\Susy = {3 \over 2} C_A , \hskip 2 cm 
b_1^\Susy = {3 \over 2} C_A^2 ,
\label{SUSYQCDBetaCoeffs}
\end{equation}
and $\gamma_g$ is similarly modified,
\begin{equation}
\gamma_g^\Susy = {3 \over 2} C_A . 
\label{SUSYQCDValues}
\end{equation}
The coefficients $K$ and $H_g^{(2)}$ are given in the FDH scheme, 
via~\eqns{GeneralK}{GeneralH}, by
\begin{eqnarray}
K_\FDH^\Susy = \left( 3 - {\pi^2 \over 6} - {4 \over 9} \e \right) C_A ,   
\label{SUSYK} \\
(H_g^{(2)})_\FDH^\Susy =
 \biggl( {\zeta_3 \over 2} + {\pi^2 \over 16} - {2 \over 9} \biggr) C_A^2 . 
\label{SUSYH}
\end{eqnarray}

With these replacements, the one- and two-loop infrared 
decompositions~(\ref{OneloopCatani}) and~(\ref{TwoloopCatani}) hold 
in the super-Yang-Mills case.  We also color decompose the 
amplitudes and strip off the helicity phases exactly as 
in~\eqn{RemoveColorPhase} for the QCD case.
The dependence of the one- and two-loop finite remainders on $\mu$ and $N$ 
are then extracted as
\begin{eqnarray}
 M^{(1),\Susy,[i]{\rm fin}}_{\lambda_1\lambda_2\lambda_3\lambda_4} &=& 
 - b_0^\Susy (\ln(s/\mu^2) - i\pi) 
    M^{(0),[i]}_{\lambda_1\lambda_2\lambda_3\lambda_4}
 + N \, a^{\Susy,[i]}_{\lambda_1\lambda_2\lambda_3\lambda_4} \,, 
  \quad i = 1,2,3,
\nonumber \\
 M^{(1),\Susy,[i]{\rm fin}}_{\lambda_1\lambda_2\lambda_3\lambda_4} &=& 
  g^{\Susy,[i]}_{\lambda_1\lambda_2\lambda_3\lambda_4} \,, 
  \hskip6.6cm i = 7,8,9,
\label{SUSYOneloopRemainderDef}
\end{eqnarray}
\begin{eqnarray}
 M^{(2),\Susy,[i]{\rm fin}}_{\lambda_1\lambda_2\lambda_3\lambda_4} 
&=& 
 - \Bigl[ \bigl( b_0^\Susy \bigr)^2 \, (\ln(s/\mu^2) - i\pi)^2 
        + b_1^\Susy \, (\ln(s/\mu^2) - i\pi) \Bigr]
    M^{(0),[i]}_{\lambda_1\lambda_2\lambda_3\lambda_4}
 \MinusBreak{ }
  2 b_0^\Susy \, (\ln(s/\mu^2) - i\pi) \, 
    M^{(1),\Susy,[i]{\rm fin}}_{\lambda_1\lambda_2\lambda_3\lambda_4}
 + N^2 \, A^{\Susy,[i]}_{\lambda_1\lambda_2\lambda_3\lambda_4}
 +        B^{\Susy,[i]}_{\lambda_1\lambda_2\lambda_3\lambda_4} \,, 
\nonumber \\
&&  \hskip8cm   i = 1,2,3,
\label{SUSYTwoloopSingleTrace} \\
 M^{(2),\Susy,[i]{\rm fin}}_{\lambda_1\lambda_2\lambda_3\lambda_4} 
&=& 
 - 2 b_0^\Susy \, (\ln(s/\mu^2) - i\pi) \, 
    M^{(1),\Susy,[i]{\rm fin}}_{\lambda_1\lambda_2\lambda_3\lambda_4}
 + N \, G^{\Susy,[i]}_{\lambda_1\lambda_2\lambda_3\lambda_4} \,, 
\nonumber \\
&&  \hskip8cm i = 7,8,9.
\label{SUSYTwoloopDoubleTrace} 
\end{eqnarray}

The one-loop supersymmetric remainder functions are given in
terms of the QCD ones,
\begin{eqnarray}
 a^{\Susy,[i]}_{\lambda_1\lambda_2\lambda_3\lambda_4} &=&
  a^{[i]}_{\lambda_1\lambda_2\lambda_3\lambda_4}
+ c^{[i]}_{\lambda_1\lambda_2\lambda_3\lambda_4} 
+ {1 \over 6} M^{(0),[i]}_{\lambda_1\lambda_2\lambda_3\lambda_4} 
\,,  \qquad i = 1,2,3,
\label{aSUSY} \\
  g^{\Susy,[i]}_{\lambda_1\lambda_2\lambda_3\lambda_4} &=& 
    2 \Bigl( a^{\Susy,[1]}_{\lambda_1\lambda_2\lambda_3\lambda_4}
           + a^{\Susy,[2]}_{\lambda_1\lambda_2\lambda_3\lambda_4}
           + a^{\Susy,[3]}_{\lambda_1\lambda_2\lambda_3\lambda_4} \Bigr)
\,,
 \qquad i = 7,8,9.
\label{gSUSY}
\end{eqnarray}
The $M^{(0),[i]}$ correction term in~\eqn{aSUSY} is a consequence of
using two different schemes --- FDH for super-Yang-Mills theory {\it
vs.} HV for QCD.  The two-loop analogues of~\eqn{aSUSY} are
eqs.~(\ref{B1mmpp}), (\ref{D1mmpp}), (\ref{D3mmpp}), (\ref{B1mpmp}),
(\ref{B2mpmp}), (\ref{D1mpmp}), (\ref{D2mpmp}), and (\ref{D3mpmp}) in
appendix~\ref{QCDRemainderAppendix}.  These equations also have
correction terms due to the different schemes used, as well as
feed-down from the subtracted singular terms, which depend on the
fermion representation.  The correction terms are more complicated
than at one loop, involving $M^{(1),[i]{\rm fin}}$ as well as
$M^{(0),[i]}$, but still they contain no special functions, only
logarithms.

The two-loop supersymmetric remainder functions $A^\Susy$, $B^\Susy$, 
$G^\Susy$, obey the same types of identities as the corresponding 
QCD functions.  The group theory relations are,
\begin{eqnarray}
G^{\Susy,[7]}_{\lambda_1\lambda_2\lambda_3\lambda_4} & = &
 2 \Bigl( A^{\Susy,[1]}_{\lambda_1\lambda_2\lambda_3\lambda_4}
   + A^{\Susy,[2]}_{\lambda_1\lambda_2\lambda_3\lambda_4}
   + A^{\Susy,[3]}_{\lambda_1\lambda_2\lambda_3\lambda_4} \Bigr) 
   - B^{\Susy,[3]}_{\lambda_1\lambda_2\lambda_3\lambda_4}
\,, \label{GSUSY7elim}\\[1pt plus 4pt]
%
%
G^{\Susy,[8]}_{\lambda_1\lambda_2\lambda_3\lambda_4} & = &
 2 \Bigl( A^{\Susy,[1]}_{\lambda_1\lambda_2\lambda_3\lambda_4}
   + A^{\Susy,[2]}_{\lambda_1\lambda_2\lambda_3\lambda_4}
   + A^{\Susy,[3]}_{\lambda_1\lambda_2\lambda_3\lambda_4} \Bigr) 
   - B^{\Susy,[1]}_{\lambda_1\lambda_2\lambda_3\lambda_4}
\,, \label{GSUSY8elim}\\[1pt plus 4pt]
%
%
G^{\Susy,[9]}_{\lambda_1\lambda_2\lambda_3\lambda_4} & = &
 2 \Bigl( A^{\Susy,[1]}_{\lambda_1\lambda_2\lambda_3\lambda_4}
   + A^{\Susy,[2]}_{\lambda_1\lambda_2\lambda_3\lambda_4}
   + A^{\Susy,[3]}_{\lambda_1\lambda_2\lambda_3\lambda_4} \Bigr) 
   - B^{\Susy,[2]}_{\lambda_1\lambda_2\lambda_3\lambda_4} 
\,, \label{GSUSY9elim}
\end{eqnarray}
and
\begin{equation}
B^{\Susy,[3]}_{\lambda_1\lambda_2\lambda_3\lambda_4} = 
    - B^{\Susy,[1]}_{\lambda_1\lambda_2\lambda_3\lambda_4}
    - B^{\Susy,[2]}_{\lambda_1\lambda_2\lambda_3\lambda_4} \,.
\label{BSUSY3elim}
\end{equation}
The Bose symmetry relations, which hold only for the $-$$-$++ helicity
configuration, are
\begin{equation}
X^{\Susy,[2]}_{--++}(s,t,u)
 = X^{\Susy,[1]}_{--++}(s,u,t) \,,
 \hskip1.2cm X \in \{ A,B \}. 
\label{XSUSY2sym}
\end{equation}

In appendix~\ref{N=1RemainderAppendix},
we give the explicit forms for the independent $N=1$ supersymmetric 
finite remainder functions appearing 
in~\eqns{SUSYTwoloopSingleTrace}{SUSYTwoloopDoubleTrace}.
This completes the description of the two-loop four-point amplitudes
for $N=1$ super-Yang-Mills theory, and simultaneously of the auxiliary
functions required for the QCD amplitudes presented in 
section~\ref{TwoLoopFiniteQCDSection}.


\section{Conclusions}\label{ConclusionsSection}

In this paper we have presented the two-loop amplitudes for
gluon-gluon scattering in QCD, and for all of the 2~$\to$~2 scattering
processes in pure $N=1$ super-Yang-Mills theory, including the full
dependence on external colors and helicities.  We found that there is
an additional $1/\e$ pole term, $\hat{\bom H}^{(2)}$
in~\eqn{OurHExtra}, which has nontrivial color dependence, and which
vanishes after interfering it with the tree amplitude and summing over
colors.  We investigated the dependence of the amplitudes on the
flavor of dimensional regularization employed.  The QCD results, when
summed over all external colors and helicities and converted to the CDR
scheme, are in complete agreement with the previous results of Glover,
Oleari, and Tejeda-Yeomans~\cite{GOTYgggg}.  We also expressed the
one-loop-squared contribution to the NNLO $\ggtogg$ cross section in
terms of one-loop finite remainders.  Again the appropriate
interference, converted to CDR scheme, is in complete agreement with
previous results~\cite{GTYOneloopSq}.

Much numerical work still remains in order to implement the two-loop
amplitudes of this paper, or those of refs.~\cite{GOTY2to2,GOTYgggg}, 
in a numerical program for NNLO jet production at hadron colliders.  
When that is accomplished, however, the intrinsic precision on the 
QCD predictions should reach the few percent level, providing
a stringent test of the Standard Model at short distances.

\acknowledgments We thank Babis Anastasiou, Adrian Ghinculov,
Massimiliano Grazzini and Henry Wong for helpful comments.
Z.B. thanks SLAC, and L.D. thanks UCLA, for hospitality while this
paper was being completed.


\appendix

\section{Finite remainder functions for QCD}
\label{QCDRemainderAppendix}

In this appendix, we present the explicit forms for the independent
finite remainder functions for $\ggtogg$ in QCD, which appear
in~\eqns{TwoloopSingleTrace}{TwoloopDoubleTrace}.  For the ++++
helicity configuration, the functions are
\begin{eqnarray}
A^{[1]}_{++++} &=& 
{1\over 36}  \biggl(11  \Xt - {y^2 \over x} - 8 \biggr) 
\,, \label{A1pppp}\\[1pt plus 4pt]
%
%
A^{[3]}_{++++} &=&  
 {1\over 72} \biggl(22 \Xt - {1\over x y} - 8 \biggr) 
           + \Bigl\{t \leftrightarrow u \Bigr\} 
\,, \label{A3pppp}\\[1pt plus 4pt]
%
%
B^{[1]}_{++++} &=& 
  {11\over 6} (\Xt - 2 \Yt )
\,, \label{B1pppp}\\[1pt plus 4pt]
%
%
C^{[1]}_{++++} &=&  
{1\over 36}  \biggl( -13 \Xt + 2 {y^2\over x} + 16 \biggr) 
\,, \label{C1pppp}\\[1pt plus 4pt]
%
%
C^{[3]}_{++++} &=&  
 {1\over 36} \biggl( - 13 \Xt + 2 \, {y\over x} + 10 \biggr) 
+ \Bigl\{t \leftrightarrow u \Bigr\} 
\,, \label{C3pppp}\\[1pt plus 4pt]
%
%
D^{[1]}_{++++} &=&
 - {1\over 3}  (\Xt- 2 \Yt ) + {1\over 4} 
\,, \label{D1pppp} \\[1pt plus 4pt]
%
%
D^{[3]}_{++++} &=&
 - {1\over 3}\, \Xt + {1\over 8} + \Bigl\{t \leftrightarrow u \Bigr\} 
\,, \label{D3pppp} \\[1pt plus 4pt]
%
%
E^{[1]}_{++++} &=&
{1\over 36} \biggl( 2  \Xt - {y^2\over x} + 1 \biggr) 
\,, \label{E1pppp} \\[1pt plus 4pt]
%
%
E^{[3]}_{++++} &=&
{1\over 72} \biggl( 4  \Xt - {1\over x y} + 1 \biggr) 
   + \Bigl\{t \leftrightarrow u \Bigr\} 
\,, \label{E3pppp} \\[1pt plus 4pt]
%
%
F^{[1]}_{++++} &=&
0 
\,, \label{F1pppp} \\[1pt plus 4pt]
%
%
F^{[3]}_{++++} &=&
0 
\,, \label{F3pppp} \\[1pt plus 4pt]
%
%
H^{[7]}_{++++} &=&
 {13 \over 18}\, \Xt + {1\over 4}  +
             \Bigl\{t \leftrightarrow u \Bigr\} 
\,, \label{H7pppp} \\[1pt plus 4pt]
%
%
H^{[8]}_{++++} &=&
{1\over 18}   ( 13 \Xt - 38 \Yt ) + {1\over 2} 
\,, \label{H8pppp} \\[1pt plus 4pt]
%
I^{[7]}_{++++} &=&
- {1\over 9}\,  \Xt + \Bigl\{ t \leftrightarrow u \Bigr\} 
\,, \label{I7pppp} \\[1pt plus 4pt]
%
%
I^{[8]}_{++++} &=&
 - {1\over 9}  (\Xt - 2 \Yt )
\,, \label{I8pppp}
\end{eqnarray}
where $x$, $y$, $X$, $Y$, $\Xt$ and $\Yt$ are defined 
in~\eqns{VariableNames}{XtYtDef}.

For $-$+++, the functions are
\begin{eqnarray}
A^{[1]}_{-+++} &=& 
 - {1\over 12}  \biggl( 15 \, {x\over y^2} - { 1+x^3 \over x y} \biggr) 
                (\Xt^2 + \pi^2)
           + \biggl( {11 \over 36}\, {y^2 \over x} 
           - {5\over 4}\, {1-x \over y} \biggr) \Xt 
	\MinusBreak{ }
             {5\over 4} \, {1 - x y \over x} 
\,, \label{A1mppp}\\[1pt plus 4pt]
%
%
A^{[3]}_{-+++} &=&  
  - {1 \over 24} \biggl(15  x  y - {y^2 \over x} - {x^2 \over y} \biggr) 
           \Bigl((X-Y)^2 + \pi^2\Bigr)
	\PlusBreak{}
   {11 \over 36} \, {\Xt \over x y}  + {5 \over 4} (x-y) X 
  + {5 \over 4} \, { 1 - x y \over x}  + \Bigl\{t \leftrightarrow u \Bigr\}
 \,, \label{A3mppp}\\[1pt plus 4pt]
%
%
B^{[1]}_{-+++} &=& 
  -  {y^2 + 2 x \over 2 y^2} (\Xt^2 + \pi^2)
         -  {1 + 2 x y \over 2 x^2} (\Yt^2 + \pi^2)
	\MinusBreak{}
           {1 \over 2}  (x^2 + 2 y) \Bigl((X-Y)^2 + \pi^2 \Bigr)
         + {11 \over 6} (\Xt - 2 \Yt ) 
\,, \label{B1mppp}\\[1pt plus 4pt]
%
%
C^{[1]}_{-+++} &=&  
\biggl( {11 \over 8}\, {x \over y^2} - {5 \over 16} 
      + {y^2 \over 12 x } \biggr)   (\Xt^2 + \pi^2) 
          + {11 \over 8} \, {1 - x y \over x}
	\PlusBreak{}
           \biggl({11\over 8} \, {1-x \over y} 
          - {13 \over 36} \, {y^2 \over x} \biggr) \Xt 
\,, \label{C1mppp}\\[1pt plus 4pt]
%
%
C^{[3]}_{-+++} &=&  
    {x \over 48 y} (4 x^2 + x y + 30 y^2 ) 
    \Bigl((X-Y)^2 +\pi^2\Bigr)
	\MinusBreak{}
    \biggl({11\over 8} (x-y) + {13\over 36} \, {1\over x y} \biggr) \Xt
          + {11 \over 16}\, {1 - x y \over x y} 
            + \Bigl\{t \leftrightarrow u \Bigr\} 
\,, \label{C3mppp}\\[1pt plus 4pt]
%
%
D^{[1]}_{-+++} &=&
  -  {1+x^2 \over 16 y^2}  (\Xt^2 + \pi^2)
             + {6-5 y\over 24 y}\, \Xt 
  +             {2\over 3}\, \Yt
             +  {1- x y \over 8 x} 
\,, \label{D1mppp} \\[1pt plus 4pt]
%
%
D^{[3]}_{-+++} &=&
 - {1\over 16} \, y^2 \Bigl((X-Y)^2 + \pi^2 \Bigr)
                     + {1 \over 24} (6 y - 5) \Xt 
	\PlusBreak{}
                        {1 - x y \over 16 x y} 
      + \Bigl\{t \leftrightarrow u \Bigr\} 
\,, \label{D3mppp} \\[1pt plus 4pt]
%
%
E^{[1]}_{-+++} &=&
{y^2 \over 18 x} \, \Xt 
\,, \label{E1mppp} \\[1pt plus 4pt]
%
%
E^{[3]}_{-+++} &=&
  {1 \over 18 x y} \, \Xt 
 + \Bigl\{t \leftrightarrow u \Bigr\} 
\,, \label{E3mppp} \\[1pt plus 4pt]
%
%
F^{[1]}_{-+++} &=&
0 
\,, \label{F1mppp} \\[1pt plus 4pt]
%
%
F^{[3]}_{-+++} &=&
0
 \,, \label{F3mppp} \\[1pt plus 4pt]
%
%
H^{[7]}_{-+++} &=&
         {1 \over 12} \biggl( 2 {y^2\over x} 
                    + {3 x {} (1 - 2 x)  \over y^2 } \biggr) 
                        (\Xt^2 + \pi^2)
            + \biggl({11 \over 9} \, {x^2 \over y} 
                    - {1\over 2} \biggr) \Xt
	\PlusBreak{}
            {1 \over 24} \biggl(- 4 {y^2 \over x} + 3 x y \biggr)  
                                              \Bigl( (X-Y)^2 + \pi^2\Bigr)
        + \Bigl\{t \leftrightarrow u \Bigr\} 
\,, \label{H7mppp} \\[1pt plus 4pt]
%
%
H^{[8]}_{-+++} &=&
  {1 \over 12} \biggl( 2 {y^2 \over x} - 3 {2 y^2 - x \over y^2} \biggr)
                                          (\Xt^2 + \pi^2)
           + {1 \over 12} \biggl(2 {x^2 \over y} - 3 {2-y \over x^2} \biggr) 
                                                    (\Yt^2 + \pi^2) 
	\MinusBreak{}
            {1 \over 12} \biggl(2 {x^2 \over y} - 15 x y - 8 y^2 - 
                                          2\, {y^3 \over x} \biggr)
                    \Bigl((X-Y)^2 + \pi^2\Bigr)
	 \PlusBreak{}
            {1 \over 18} \biggl(22 \, {x^2 \over y} - 9 \biggr)  \Xt
           + {1 \over 9} \biggl(11 \, {y^2 \over x} - 30 \biggr) \Yt 
\,, \label{H8mppp} \\[1pt plus 4pt]
%
I^{[7]}_{-+++} &=&
   - {x^2\over 9 y}\, \Xt
 + \Bigl\{t \leftrightarrow u \Bigr\} 
\,, \label{I7mppp} \\[1pt plus 4pt]
%
%
I^{[8]}_{-+++} &=&
- {1\over 9} \biggl( {x^2\over y}\, \Xt 
+ \biggl({y^2 \over x} - 3 \biggr) \Yt \biggr) 
\,. \label{I8mppp} 
%
\end{eqnarray}

For $-$$-$++, the functions are
\begin{eqnarray}
A^{[1]}_{--++} & = & 
       (x^2 + 6 y^2-3 x y) \Biggl[ \li4\biggl(-{x \over y}\biggr)
                                     - \li4(-x) - \li4(-y)
  	\PlusBreak{(x^2 + 6 y^2-3 x y) \BigglBl}
               \Xt {} ( \li3(-x) + \li3(-y) ) 
  	\MinusBreak{(x^2 + 6 y^2-3 x y) \BigglBl}
               {\pi^2 \over 6} \li2(-x)
             + {1 \over 12} \, \Xt^4 + {1 \over 6} (\Xt^2 + \pi^2) \Xt Y 
  	\PlusBreak{(x^2 + 6 y^2-3 x y) \BigglBl}
              {1 \over 24} \, Y^2  ( 6   X^2 - 4  X Y + Y^2 + 2 \pi^2) 
            - {17 \over 720} \, \pi^4 \Biggr]  
      	\PlusBreak{}
            y {} \,  {y^2-3 x y+6 x^2 \over x}  \biggl( 
               {1 \over 8} \, \Xt^4 
                   + {\pi^2 \over 6}\, \Xt^2 + {\pi^4 \over 80} \biggr) 
        + {\zeta_3  \over 2 x} \, \Xt 
     	\PlusBreak{}
         {1 \over 6} \,  {11 + 60 x y \over x}   \Biggl[ 
           \li3(-x) - \Xt \li2(-x) 
         + {1 \over 3} \, \Xt^3 - {1 \over 2} (\Xt^2 + \pi^2) Y 
     	\PlusBreak{\null + {1 \over 6} \,  { 11 + 60  x  y \over x} \BigglBl}
         {17 \over 24}  \, \pi^2  \Xt - {19 \over 6}\, \zeta_3 \Biggr] 
      - {5 \over 12} \,  y {} (4 \Xt^3 + 9 \pi^2   \Xt - 52   \zeta_3 ) 
       	\MinusBreak{}
          {5 \over 4} \biggl( {x^2 \over y^2} 
          - {4 \over y} + 1 \biggr)  (\Xt^2 + \pi^2)
          - {\pi^2 \over 36 x} \Bigl(53 + 15 y {} (3 x-y) \Bigr)
       	\MinusBreak{}
         \biggl[ {5 \over 2} \biggl({x \over y} - 2  {y \over x} \biggr) 
        + {67 \over 54 x} \biggr]   \Xt 
      - {x \over 36} - {25 \over 18} + {11093 \over 648 x} 
\,, \label{A1mmpp} \\[1pt plus 4pt]
%
%
A^{[3]}_{--++} & = & 
      -  x^2  {3 - 2 x y \over 24 x y} (X-Y)^4 
      - {\pi^2 \over 12} \, {(1 - x y)^2 \over x y} (X-Y)^2
      +     {\zeta_3 \over 2 x y} \Xt - {\pi^4 \over 160 x y} 
       	\MinusBreak{}
          {11 \over 6 x y} \biggl( \li3(-x) - X \li2(-x) - {1 \over 3} X^3 
           - {7 \over 8} \,\pi^2 X + {13 \zeta_3  \over 12}
               - i  {\pi^3 \over 24} \biggr)     
       	\PlusBreak{}
        \biggl( {11 \over 12}\, {x^3 \over y} - {2 \over 9} (x-y) \biggr) 
              \Bigl( (X-Y)^2 X - {\pi^2 \over 3} Y \Bigr)  
       	\MinusBreak{}
     {11 \over 6} \, i \pi {} {(1-x y)^2 \over y} \Bigl( (X-Y)^2 + \pi^2 \Bigr)
      - {\pi^2 \over 27} ( 16 x^2 - 17 x y + 58 ) X 
       	\PlusBreak{}
      {11 \over 36} \biggl( {11 \over y} 
         - 3 (x-y) \biggr) ( \Xt^2 + \pi^2 )
      + {1 \over 24} (41 - 30 x y ) \Bigl((X-Y)^2 + \pi^2\Bigr)  
       	\MinusBreak{}
       \pi^2 \, {1 + 41 (1-x) \over 36  y}
      + \biggl( {101 \over 27 y} + 5  x  
              + {19 \over 12} + {337 \over 54 x} \biggr) \Xt
      + {5 \over 4} - {11093 \over 648 x}  
       	\PlusBreak{}
    \Bigl\{ t \leftrightarrow u \Bigr\} 
\,, \label{A3mmpp} \\[1pt plus 4pt]
%
%
B^{[1]}_{--++} & = & 
B^{\Susy,[1]}_{--++} - H^{[7]}_{--++} + 2   H^{[8]}_{--++} - H^{[9]}_{--++}
                     + I^{[7]}_{--++} - 2 I^{[8]}_{--++} + I^{[9]}_{--++} 
       	\MinusBreak{}
  {1 \over 3} (x-y)  \biggl( 2 {x^2 \over y} - 3 \biggr) \Xt^3
         - {4 \over 3} \biggl( 2 {x^2 \over y} - 3 x y 
                                   - {y^3 \over x} \biggr) \Xt^2 \Yt   
       	\PlusBreak{}
       {1 \over 3} \biggl(8 \, {x^2 \over y} + x - 20 x y + 3 y 
                             + 10 \, {y^2 \over x} \biggr) 
                 \Xt \Yt^2
        - {2 \over 3} (x-y) \biggl( 2 {y^2 \over x} - 3 \biggr) \Yt^3 
       	\PlusBreak{}
         {\pi^2 \over 3} (5 - 4 x y) (\Xt-2 \Yt)
        - {2 \over 9} \biggl( 13 {x^2 \over y} + x - 19 \biggl) \Xt^2
       	\PlusBreak{}
          {4 \over 9} \biggl( 13 {y^2 \over x} + y - 19 \biggr) \Yt^2  
         + {2 \over 9} \biggl( {26 \over y} 
                    - 18 (x-y) - {13 \over x} \biggr) \Xt \Yt
       	\PlusBreak{}
          {4 \over 3}  (\Xt - 2 \Yt) 
\,, \label{B1mmpp} \\[1pt plus 4pt]
%
%
C^{[1]}_{--++} & = & 
   (2 x^2 - 6 x y + 3 y^2) \Biggl[
            \li4(-x) + \li4(-y) - \li4\biggl(-{x \over y}\biggr)
          - \Xt \li3(-x) 
       	\MinusBreak{(2 x^2-6 x y+3 y^2) \BigglBl}
            \Xt \li3(-y) 
          + {\pi^2 \over 6}\, \li2(-x)
          + {1 \over 24}\,  \Xt^4 
       	\MinusBreak{(2 x^2-6 x y+3 y^2) \BigglBl}
            {1 \over 4}\, X^2 (\Yt^2 + \pi^2)
          - {1 \over 6} X Y {} (X^2-Y^2) - {1 \over 24}\, Y^4  
       	\PlusBreak{(2 x^2-6 x y+3 y^2) \BigglBl}
            {\pi^2 \over 12} (2 \Xt^2 + 4 X Y - Y^2 )
          + {13 \over 360}  \, \pi^4 \Biggr]
       	\MinusBreak{}
            {1 \over 6} \biggl( {2 \over x} - 3 (9 x - 13 y) \biggr)
                  \Biggl[ \li3(-x) - \Xt \li2(-x)
                  + {1 \over 3} \Xt^3 - {1 \over 2} (\Xt^2 + \pi^2) Y 
       	\PlusBreak{\null - {1 \over 6}\Bigl( {2 \over x} - 
                              3 (9 x - 13 y) \Bigr) \BigglBl }
                   {17 \over 24}\, \pi^2 \Xt - {\zeta_3 \over 6} \Biggr]  
       	\MinusBreak{}
          {1 \over 12} (9 x - 13 y) \biggl( \Xt^3 + {9 \over 4} \pi^2 \Xt 
                                  + 5 \zeta_3 \biggr)
       	\PlusBreak{}
           {1 \over 8} \biggl( {11 \over y^2} 
                     + 4 \, {x \over y} + 26 \biggr) (\Xt^2 + \pi^2)  
        - {\pi^2 \over 216} 
           \biggl( 38 {y^2 \over x} - {85 \over x} - 358 y - 101 \biggr) 
       	\PlusBreak{}
            {1 \over 108} \biggl( 297 \, {x + 2 y \over y} 
                + {137 \over x} \biggr) \Xt 
       - {y \over 18} + {115 \over 72} - {4849 \over 1296 x} 
\,, \label{C1mmpp} \\
%
%
C^{[3]}_{--++} & = & 
          - {1 \over 48}\, x^2   \Bigl( (X-Y)^2 + \pi^2 \Bigr) 
                     \Bigl(3 (X-Y)^2 - \pi^2 \Bigr)  
	\PlusBreak{}
           {1 \over 3 x y}  \Biggl[ \li3(-x) - X \li2(-x) 
                    - {1 \over 3} X^3 - i {\pi \over 2} X {} (X-Y)
            - {3 \over 8} \, \pi^2  X 
	\MinusBreak{\null + {1 \over 3 x y} \BigglBl}
                  {5 \over 12} \, \zeta_3 
                 - i \, {\pi^3 \over 8} \Biggr] 
           - {1 \over 12} \biggl( 2 {y^3 \over x} 
                + {x^2 \over y} (11 y - 4 x) \biggr) \Xt^2 \Yt
  	\MinusBreak{}
            {1 \over 36} \biggl( 6 \, {x^3 \over y} - 11 (2 x^2+y^2) \biggr) 
                   \Xt {} (\Xt^2 + \pi^2)   
   	\PlusBreak{}
           {\pi^2 \over 18} \biggl( 2 {y^2 \over x} - 1 + 4 x \biggr) \Xt
          - {1 \over 36}   \biggl( {44 \over y} - 39 (x-y) \biggr) 
                ( \Xt^2 + \pi^2 )   
	\MinusBreak{}
            {1 \over 24}  ( 37 x^2 + 4 x y) \Bigl((X-Y)^2 + \pi^2 \Bigr)
          - \pi^2  \, {13 + 204 (1-x y) \over 432 x y} 
	\MinusBreak{}
      {1 \over 108} \biggl( - {56 \over x y} 
            + {9 \over x} (55 x^2 - 2 x y + 9 y^2 ) \biggr) \Xt  
          - {11 \over 8} + {4849 \over 1296 x} 
   	\PlusBreak{}
          \Bigl\{ t \leftrightarrow u \Bigr\} 
\,, \label{C3mmpp} \\[1pt plus 4pt]
%
%
D^{[1]}_{--++} & = & 
 - A^{\Susy,[1]}_{--++}  + A^{[1]}_{--++} 
      + C^{[1]}_{--++}  + E^{[1]}_{--++} 
           + {1 \over 3} \biggl( {x^2 \over y} + y^2 \biggr) \Xt^3 
	\PlusBreak{}
            {2 \over 3} \biggl( {x^3 \over y} - {1 \over y} 
                        + {y^3 \over x} \biggr) \Xt^2 \Yt 
           - {1 \over 3} \biggl(5 {y \over x} - 1 + 5 x y 
                       + 4 {x \over y}  \biggr) \Xt \Yt^2   
	\PlusBreak{}
            {\pi^2 \over 3} (2-x y) (\Xt - 2 \Yt)
           - {2 \over 3} \biggl( {y^2 \over x} + x^2 \biggr)   \Yt^3 
	\MinusBreak{}
             {1 \over 18} \biggl( {11 \over x} + 3 (x-y) \biggr) 
                             (\Xt^2 + 4 \pi^2)   
   	\PlusBreak{}
           {1 \over 18} \biggl( {33 \over x} 
                       + 3 (x-y) + {22 \over y} \biggr) (\Yt^2 + \pi^2) 
	\MinusBreak{}
          {1 \over 18} \biggl( - {11 \over x} - 9 (x-y) 
                      + {22 \over y} \biggr) \Bigl( (X-Y)^2 + \pi^2 \Bigr)
           + {5 \over 144 x} \, \pi^2 
   	\PlusBreak{}
        {1 \over 3} ( \Xt - 2 \Yt ) - {17 \over 54 x} 
\,, \label{D1mmpp} \\[1pt plus 4pt]
%
%
D^{[3]}_{--++} & = & 
     - A^{\Susy,[3]}_{--++} 
     + A^{[3]}_{--++} + C^{[3]}_{--++} + E^{[3]}_{--++}
       	\PlusBreak{}
           \Biggl[ {1 \over 3} \biggl( {x^2 \over y} + y^2 \biggr) \Xt^3   
          + {1 \over 3} \biggl({1 \over x y} + x 
                   - y + {y^3 \over x} \biggr) \Xt^2 \Yt 
           + {\pi^2 \over 3} (2-x y) \Xt
       	\MinusBreak{\null + \BigglBl}
           {1 \over 9} \biggl({11 \over y} - 3 (x-y) \biggr)
                               ( \Xt^2 + \pi^2 )  
        - {11 \over 9 y}\, \Xt \Yt + {19 \over 16 y}\, \pi^2
       	\PlusBreak{\null + \BigglBl}
           {1 \over 3} \, \Xt + {17 \over 54 y} 
  +   \Bigl\{ t \leftrightarrow u \Bigr\}  \Biggr]
\,, \label{D3mmpp} \\[1pt plus 4pt]
%
%
E^{[1]}_{--++} &=& 
 - {\pi^2 \over 108 x} - {1\over 72} (x-y) 
\,, \label{E1mmpp}\\[1pt plus 4pt]
%
%
E^{[3]}_{--++} &=& 
     - {1\over 12} (x^2+y^2) \Xt {} \Bigl( (X-Y)^2 + \pi^2 \Bigr)
     + {1\over 18} \biggl(2 \, {x^2\over y} + x + 5 y \biggr) \Xt^2 
 \MinusBreak{}
      { \pi^2 \over 216 x y}
     - {1\over 6} \, \Xt +
             \Bigl\{ t \leftrightarrow u \Bigr\} 
\,, \label{E3mmpp}\\[1pt plus 4pt]
%
%
F^{[1]}_{--++} &=& 
  {1\over 18} \Biggl[ {1 \over x} (\Xt^2 + 4 \pi^2)
                     + {2-y \over x y} (\Yt^2 + \pi^2)
 	\PlusBreak{ {1\over 18}  \BigglBl }
                   {2 x - y \over x y} \Bigl((X-Y)^2 + \pi^2 \Bigr) \Biggr] 
\,, \label{F1mmpp}\\[1pt plus 4pt]
%
%
F^{[3]}_{--++} &=& 
    {1 \over 9} \biggl( {1\over y} (\Xt^2 + \pi^2) + {X Y \over x}
                            - i \pi {} {\Xt \over x y} \biggr) +
             \Bigl\{ t \leftrightarrow u \Bigr\} 
\,, \label{F3mmpp}\\[1pt plus 4pt]
%
%
H^{[7]}_{--++} & = & 
            {4 \over 3}   (2 - x y)   \biggl( \li3(-x) - \Xt  \li2(-x)
             + {1 \over 6}   \Xt^3 - {3 \over 4}   \Xt^2   \Yt
             + i   {\pi \over 2} (\Xt^2 + \pi^2) \biggr) 
   	\PlusBreak{}
             {1 \over 36 y}\, \Xt {} \Bigl( 8 \Xt^2 + 29   \pi^2 \Bigr)
           + {1 \over 6} \biggl( 2 \, {x^3 \over y} - 6 (x-y) 
                         - 4 \, {y^3 \over x} \biggr) \Xt^2 \Yt 
   	\MinusBreak{}
         {1 \over 18} \Biggl[ \biggl( 18 \,{x \over y^2} 
                     - 3 {x \over y} + 8 + 11 {y \over x} \biggr) 
                               ( \Xt^2 + \pi^2 )
   	\PlusBreak{\null - {1 \over 18} \BigglBl}
                     2 \biggl( {22 \over y} - 3 (x-y) \biggr) 
                       ( \Xt^2 + \pi^2 )  
    	\MinusBreak{\null - {1 \over 18} \BigglBl }
            \biggl(11 {x^3 \over y} + 8 x^2 + 6 x y \biggr)
                  \Bigl( (X-Y)^2 + \pi^2 \Bigr)
                    + {11 \over 2} \, {\pi^2 \over x y} \Biggr]   
       	\MinusBreak{}
    {2 \over 27} \, {23 + 9 x y \over y} \, \Xt
           + {1 \over 4} 
  +   \Bigl\{ t \leftrightarrow u \Bigr\} 
\,, \label{H7mmpp} \\[1pt plus 4pt]
%
%
H^{[8]}_{--++} & = & 
    2 (x^2+y^2) \Biggl[ 
                  3 ( \li4(-x) + \li4(-y) )
                 - 2 \Xt \li3(-x) - 2 \Yt \li3(-y)  
   	\PlusBreak{2 (x^2+y^2) \BigglBl}
                 {1 \over 2} (\Xt^2 - \Yt^2) \li2(-x) 
                 + {1 \over 6} \, X^3 \Yt 
                - {1 \over 4} \biggl( X^2 + {\pi^2 \over 6} \biggr) Y^2  
   	\MinusBreak{2 (x^2+y^2) \BigglBl}
                   {1 \over 3} \biggl( X-i {\pi \over 2} \biggr)   \Yt^3  
              - {\pi^2 \over 8} (X^2 + 6 X Y - 4 \Yt^2 ) 
                 + {7 \over 60} \, \pi^4 \Biggr]
   	\PlusBreak{}
           4 x {} (x - 3 y) \Biggl[ \li4(-x) + \li4(-y) 
               + \li4 \biggl( -{x \over y} \biggr)  
               - \Yt {} ( \li3(-x) + \li3(-y) ) {}
   	\MinusBreak{ {\null} + 4 x  (x - 3 y) \BigglBl } 
                  {\pi^2 \over 6} \li2(-x) 
                 + {1 \over 12} Y^4 
                - {1 \over 3} \biggl(X-i {\pi \over 2} \biggr) Y^3 
   	\MinusBreak{\null + 4 x  (x - 3 y) \BigglBl}
              i \, {\pi \over 2} \, X Y^2 
                 + i {\pi^3 \over 6} Y - {\pi^4 \over 24} \Biggr] 
   	\PlusBreak{}
            {1 \over 3} \biggl( 2 \, {x^3 \over y} - x^2 + 14 y^2 \biggr) 
                   \Biggl[ \li3(-x) + \zeta_3  
       - \Xt \li2(-x) 
           + i \, {\pi \over 2} \, X^2
   	\MinusBreak{\null + {1 \over 3} \biggl( 2 {x^3 \over y} 
                                            - x^2 + 14y^2 \biggr) \BigglBl }
                {11 \over 8} \, 
                       \pi^2 X - {3 \over 8}\, i \pi^3 \Biggr] 
   	\PlusBreak{}
           {1 \over 3} \biggl(4 {x^3 \over y} + 17 x^2 - 30 x - 2 y^2 \biggr) 
                  \Biggl[ \li3(-y) - \zeta_3 - \Yt \li2(-y)  
                  + i {\pi \over 2} \, Y^2 
   	\MinusBreak{ \null +{1 \over 3} \biggl(4 \, {x^3 \over y} 
                             + 17 x^2 - 30 x - 2 y^2 \biggr) \BigglBl}
                   {1 \over 2} \, \Xt \Yt {} (X-Y) 
                  - {\pi^2 \over 2} (X-i \pi) \Biggr]
   	\PlusBreak{} 
            {2 \over 9} \biggl( {x^2 \over y} + y^2 \biggr) \Xt^3  
           + {1 \over 12} \biggl( 8\, {y^2 \over x} - 15 y^2 
                                                   - 48  x + 27 x^2 \biggr)
                    \Xt^2 \Yt  
   	\MinusBreak{} 
         {1 \over 6}\biggl( 14 \, {y^2 \over x} - 7 y^2 - 36 x + 34 x^2 \biggr)
                    \Xt \Yt^2
      	+
             {1 \over 36} \biggl( 44 {y^2 \over x} - 5 y^2 
                                      - 48 x + 35   x^2 \biggr)  \Yt^3
      	\PlusBreak{} 
             {\pi^2 \over 72} ( 182 y^2 + 96 x - 42 y + 21 x^2 )
                       \Xt 
   	\MinusBreak{} 
           {\pi^2 \over 36} \biggl( 29 \, {y^3 \over x} - 2 y^2 - 63 x 
                     + 155   x^2 - {48 \over y} \biggr) \Yt 
           - 2 \zeta_3 (x-y)  
      	\PlusBreak{} 
        {1 \over 36} \biggl( {9 \over y^2} - {44 \over y} 
                          + 3 (1-3 x) (1-7 x) \biggr)
                    ( \Xt^2 + \pi^2 )   
      	\PlusBreak{} 
         {1 \over 36} \biggl( {36 \over x^2} - {212 \over x} 
                       + 3 (y^2 + 4 x y + 24 x^2) \biggr)
                                ( \Yt^2 + \pi^2 )  
   	\MinusBreak{} 
          {1 \over 18} \biggl( {44 \over y} - {22 \over x} 
                               - 3  (2 x^2 + 52 x y + 29 y^2) \biggr)
                    \Xt \Yt 
       	\PlusBreak{} 
       {\pi^2 \over 36} \biggl( {44 \over y} + {176 \over x} 
                            - 55 x^2 + 8 x y - 144 y \biggr)
           - {1 \over 54} \biggl( {83 \over y} + 45 x - 9 \biggr) \Xt 
   	\MinusBreak{} 
           {1 \over 54} \biggl( {92 \over x} + 45 y - 144 \biggr) \Yt 
\,, \label{H8mmpp}\\[1pt plus 4pt]
%
%
I^{[7]}_{--++} &=& 
   {1\over 9 y} \, \Xt {} (X-Y) 
  + \Bigl\{ t \leftrightarrow u \Bigr\} 
\,, \label{I7mmpp}\\[1pt plus 4pt]
%
%
I^{[8]}_{--++} &=& 
{1\over 18}  \Biggl[ - 3 (x^2 + y^2) \Bigl((X-Y)^2 + \pi^2 \Bigr) \Yt
             - \biggl({y^2 \over x} + 4 x - y \biggr) (\Xt^2 + \pi^2)
	\PlusBreak{{1\over 18}  \BigglBl }
              {4 \over y} \biggl(\Xt^2 + {\pi^2 \over 2} \biggr)
             + \biggl(2 \, {x^2 \over y} + 7 x - y + {7\over x} \biggr) \Yt^2
        \PlusBreak{{1\over 18} \BigglBl }
               \biggl( {y^2 \over x} - 3 y - 2 {x^2 \over y} \biggr) 
                             \Bigl((X-Y)^2 + \pi^2 \Bigr)
                     - 6 \Yt \Biggr] 
\,. \label{I8mmpp}
\end{eqnarray}

For $-$+$-$+, the functions are
\begin{eqnarray}
A^{[1]}_{-+-+} & = & 
      - {1 \over 24} (1+x^2) \biggl( {3 \over x} - {2 \over y^2} \biggr) \Xt^4 
      - {\pi^2 \over 6}  {(1 - x y)^2 \over x y^2} \, \Xt^2
      + {1 \over 2} \, \zeta_3 \, {y^2 \over x} \,\Xt 
                    - {\pi^4 \over 80}\, {y^2 \over x}
	\PlusBreak{}
            {11 \over 6} \,{y^2 \over x}  \Biggl[ \li3(-x) 
                      - \Xt \li2(-x)
                      - {1 \over 2} ( \Xt^2 + \pi^2 )   \Yt
         + {1 \over 3} \, X^3 + {3 \over 2}\,  i \pi  X^2 
 	\MinusBreak{\null + {11 \over 6} {y^2 \over x}  \BigglBl }
                {31 \over 24} \,\pi^2 X 
                   + {3 \over 8} \,i \pi^3 - {19 \over 6} \,\zeta_3  \Biggr]
       \PlusBreak{ }
        \biggl({11 \over 12}\, {x^3 \over y^2} 
                   + {2 \over 9} \,{1-x \over y} \biggr) 
                ( \Xt^2 + \pi^2 )   \Xt  
	+ {\pi^2 \over 18}\biggl( {6 \over y} - 3 x + 8 y \biggr) \Xt
       \MinusBreak{}
          {1 \over 36} \biggl( 90 \, {x^3 \over y^2} - 114 \, {x \over y} 
                     - 31 y \biggr) 
               ( \Xt^2 + \pi^2 )  
   	- {\pi^2 \over 36} \biggl( 66 \,{x^2 \over y} - 17 x + 4 + 
                                 {38 \over x} \biggr)  
        \PlusBreak{}
         \biggl( {101 \over 27} \, y + 5\, {x \over y} + {19 \over 12} 
                  + {337 \over 54}  \, {y \over x} \biggr) (X-Y)
	\PlusBreak{}
         \biggl({337 \over 54}\, {y \over x} - {5 \over y} 
                   - {41 \over 12} + {101 \over 27} \, y \biggr) \Yt
      + {5 \over 2} + {11093 \over 648} \, {y^2 \over x} 
\,, \label{A1mpmp} \\[1pt plus 4pt]
%
%
A^{[2]}_{-+-+} & = & 
      - {1+6 x^2-3 x \over y^2} \Biggl[ \li4 \biggl(-{x \over y} \biggr) 
             + \li4(-x) - \li4(-y) 
             - \Yt \li3(-x)  
	\PlusBreak{}
               {\pi^2 \over 6} \,\li2(-x)  
             - {1 \over 12} \, \Yt^4 
             - {1 \over 6} \, Y {} \Bigl(\Xt  Y^2 - \pi^2   (Y-i \pi) \Bigr)
             + \zeta_3   \Yt + {67 \over 720}  \, \pi^4 \Biggr] 
	\PlusBreak{}
      {x \over y^2} (x^2-3 x+6) \biggl( {1 \over 8} \,\Yt^4 
                      + {\pi^2 \over 6}  \, \Yt^2 + {\pi^4 \over 80} \biggr) 
      + {y \over 2} \, \zeta_3  \Yt  
	\PlusBreak{}
     {1 \over 6} \biggl(11 + 60 {x \over y^2} \biggr) y {} \Biggl[ 
                  \li3(-y) - \Yt   \li2(-y) 
                - {1 \over 3} \, \Yt^3 
	\MinusBreak{\null +{1 \over 6} 
                        \biggl(11 + 60 {x \over y^2} \biggr) y  \BigglBl }
               {1 \over 2} ( \Yt^2 + \pi^2 )   (X - \Yt) 
            - {17 \over 24}\, \pi^2  \Yt - {19 \over 6} \, \zeta_3  \Biggr] 
	\MinusBreak{} 
          {5 \over 12} \, {x \over y} \Bigl( 4 \Yt^3 
                          - 9 \pi^2 \Yt - 52 \zeta_3 \Bigr)  
         + {11 \over 72}\, y {}  ( 8 \Yt^2 + 13 \pi^2 )   \Yt
	\MinusBreak{} 
            {5 \over 4} \biggl( {1 \over x^2} - 4 \, {y \over x} + 1 
         + {121 \over 45} \, y \biggr) ( \Yt^2 + \pi^2 )
   + {\pi^2 \over 36} \biggl( {60 \over y} - 75 x + 8 y \biggr)
	\PlusBreak{}
       \biggl[ {5 \over 2} \biggl( {1 \over x} - 2 x \biggr)
         - {236 \over 27} \, y \biggr]   \Yt 
      - {1 \over 36 y} - {25 \over 18} + {11093 \over 648}\, y 
\,, \label{A2mpmp} \\[1pt plus 4pt]
%
%
A^{[3]}_{-+-+} & = & 
   - {x^2+6-3 x \over y^2} \Biggl[ 
   \li4 \biggl( -{x \over y} \biggr) - \li4(-x) - \li4(-y) 
	\PlusBreak{ - {x^2+6-3 x \over y^2} \BigglBl}
             (X-Y) ( \li3(-x) - \zeta_3 ) 
           - {\pi^2 \over 6} \biggl( \li2(-x) + {1 \over 2} Y^2 \biggr) 
 	\PlusBreak{ - {x^2+6-3 x \over y^2} \BigglBl}
             {11 \over 240} \,  \pi^4  
           - {1 \over 12}   \Bigl( (X-Y)^4
                      - (2 X^2 - 3   X  Y + 4   \pi^2)  X Y \Bigr)
                    \Biggr]   
	\PlusBreak{}
          {1- 3 x + 6 x^2 \over x y^2} \biggl( {1 \over 8} (X-Y)^4 
               +{\pi^2 \over 6} (X-Y)^2 + {\pi^4 \over 80} \biggr) 
      +   {y \over 2 x} \, \zeta_3  (\Xt + \Yt)  
 	\MinusBreak{}
           {1 \over 6} \biggl( 11 {y \over x} + {60 \over y} \biggr)
                \Biggl[ \li3(-x) + \li3(-y) 
                       - (X-Y) \li2(-x) 
 	\MinusBreak{\null - {1 \over 6} \biggl( 11 \, {y \over x} 
                        + {60 \over y} \biggr) \BigglBl  }
            {1 \over 6} (2 X - Y) \Bigl( (X+Y)^2 - 6   Y^2 \Bigr)
   	\MinusBreak{\null - {1 \over 6}\biggl( 11 {y \over x} 
                        + {60 \over y} \biggr) \BigglBl}
             i \pi {} \Bigl( (X-Y)^2 + \pi^2 \Bigr)
   	\MinusBreak{\null - {1 \over 6}\biggl( 11 {y \over x} 
                        + {60 \over y} \biggr) \BigglBl}
      {\pi^2 \over 24} (17 X + 25 Y + 2 i \pi)
           + {13 \over 6} \, \zeta_3 \Biggr]         
	\MinusBreak{}
        {5 \over 12 y} \Biggl[4 (\Xt + 5 \Yt) \Bigl( (X-Y)^2 + \pi^2 \Bigr)
                      + \pi^2 (5 \Xt - 3 \Yt) - 52 \zeta_3 \Biggr]
	\MinusBreak{}
       {5 \over 4} ( x^2 - 4 y + 1 ) \Bigl( (X-Y)^2 + \pi^2 \Bigr)  
	\MinusBreak{}
       {121 \over 36} \, {y \over x} ( \Yt^2 + \pi^2 )
      + {\pi^2 \over 36} \biggl( 83  {y \over x} + 15 - {60 \over y} \biggr)
      - {5 \over 2} \biggl( x - {2 \over x} \biggr) (X-Y)  
 	\MinusBreak{}
        {y \over 54 x} (67 \Xt + 472 \Yt) 
      - {x \over 36 y} - {25 \over 18} + {11093 \over 648} \,{y \over x}
\,, \label{A3mpmp}\\[1pt plus 4pt]
%
%
B^{[1]}_{-+-+} & = & 
    B^{\Susy, [1]}_{-+-+} 
 - H^{[7]}_{-+-+} + 2 H^{[8]}_{-+-+} - H^{[9]}_{-+-+}
 + I^{[7]}_{-+-+} - 2 I^{[8]}_{-+-+} + I^{[9]}_{-+-+}
	\PlusBreak{}
         {1 \over 3} (1-x) \biggl( 2\, {x^2 \over y^2} 
                         - {3 \over y} \biggr) \Xt^3
 - {2 \over 3} \biggl({4 \over y^2} + 3 \, {x^2 \over y} 
              - {1 \over x y} + {y^2 \over x} \biggr) \Xt^2 \Yt
	\MinusBreak{}
           {2 \over 3} (4-y) {y \over x} \, \Xt \Yt^2
         - {4 \over 3} \, {y^2 \over x} \,  \Yt^3
         - {\pi^2 \over 3} \biggl( 4 {x \over y^2} - 5 \biggr) (\Xt - 2 \Yt)
 	\MinusBreak{}
      {2 \over 9}  \biggl( 12 \, {x^2 \over y} - x - 19 \biggr) \Xt^2
         + {52 \over 9} \, {y^2 \over x} \, \Yt^2
         + {2 \over 9}  \biggl( {24 \over y} - 1 + 39 y - {13 \over x}\biggr)
                    \Xt \Yt
	\PlusBreak{}
         {4 \over 3} (\Xt - 2 \Yt) 
\,, \label{B1mpmp} \\[1pt plus 4pt]
%
%
B^{[2]}_{-+-+} & = & 
B^{\Susy,[2]}_{-+-+} - H^{[7]}_{-+-+} - H^{[8]}_{-+-+} + 2 H^{[9]}_{-+-+}
                   + I^{[7]}_{-+-+} + I^{[8]}_{-+-+} - 2 I^{[9]}_{-+-+} 
 	\MinusBreak{}
          {2 \over 3} (1-x) \biggl(2 \, {x^2 \over y^2} 
                                    -  {3 \over y} \biggr)   \Xt^3
         - {1 \over 3}  \biggl(4 \, {x \over y^2} + 5 (x-y) 
                  - 8 \,{y^2 \over x} \biggr)   \Xt^2   \Yt 
   	\PlusBreak{}
           {4 \over 3} (1-y) {y \over x}\, \Xt \Yt^2
         + {2 \over 3} \, {y^2 \over x} \, \Yt^3
         + {\pi^2 \over 3} \biggl(4 \, {x \over y^2} - 5 \biggr) (2 \Xt-\Yt) 
   	\PlusBreak{}
           {4 \over 9} \biggl( 12 {x^2 \over y} - x - 19 \biggr) \Xt^2
         - {26 \over 9} {y^2 \over x} \Yt^2
         - {2 \over 9} \biggl( {12 \over y} - 20 + 39 y 
                      - {26 \over x} \biggr) \Xt \Yt 
 	\MinusBreak{}
           {4 \over 3} (2 \Xt-\Yt) 
\,, \label{B2mpmp} \\[1pt plus 4pt]
%
C^{[1]}_{-+-+} & = & 
          - {1 \over 48} \, {1+x^2 \over y^2} \biggl(3 \Xt^4 + 2 \pi^2 \Xt^2
                                 - \pi^4 \biggr)
 	\MinusBreak{}
            {y^2 \over 3 x} \Biggl[ \li3(-x) - \Xt \li2(-x) 
              + {1 \over 3} \, \Xt^3 
              - {1 \over 2} ( \Xt^2 + \pi^2 ) Y 
              + {17 \over 24}\, \pi^2 \Xt - {1 \over 6} \, \zeta_3 \Biggr]  
 	\MinusBreak{}
         {1 \over 36} \biggl( 39 \, {x^3 \over y^2} 
                         + 11 {1 - x {} (1-3 x) \over y} \biggr) 
                          \Xt {} ( \Xt^2 + \pi^2 )
          - {\pi^2 \over 18} \biggl( {6 \over y} - 2 x + 1 \biggr) \Xt 
	\PlusBreak{}
           {1 \over 72} \biggl(- 198 \, {x^2 \over y^2} 
                          - 42 \, {x \over y} + 33 x - 55 y \biggr)
                              ( \Xt^2 + \pi^2 ) 
 	\MinusBreak{}
           {\pi^2 \over 216} \biggl(468 \, {x \over y} + 17 x - 200 y 
                                                   - {47 \over x} \biggr)
 	- {1 \over 108} \biggl( 594 {x \over y} 
            - 43 x + 13 y - {137 \over x} \biggr)   \Xt 
  	\MinusBreak{}
           {11 \over 4} - {4849 \over 1296} \, {y^2 \over x} 
%
\,, \label{C1mpmp} \\[1pt plus 4pt]
%
%
C^{[2]}_{-+-+} & = & 
  {2-6 x+3 x^2 \over y^2} \Biggl[ \li4 \biggl( -{x \over y} \biggr) + \li4(-x) 
          - \li4(-y) - \Yt {} ( \li3(-x) - \zeta_3 )  
	\PlusBreak{{2-6 x+3 x^2 \over y^2} \BigglBl}
           {\pi^2 \over 6}  \, \li2(-x) 
          - {1 \over 24}  \Bigl( 4 X Y - Y^2 - 2 \pi^2 \Bigr) Y^2 
                         - {7 \over 360}\, \pi^4 \Biggr]  
      	\MinusBreak{}
              {1 \over 6} \biggl( 2 y - 3 \,  {9 - 13 x \over y} \biggr)
                  \Biggl[ \li3(-y) - \Yt  \li2(-y)
                  + {1 \over 3}\, \Yt^3 
                  + {1 \over 2}\, i \pi {} \Bigl( Y^2 + 2 \pi^2 \Bigr)  
         	\MinusBreak{\null -{1 \over 6}  
                    \biggl( 2 y - 3 {(9 - 13 x) \over y} \biggr) \BigglBl}
                    {1 \over 2} ( \Yt^2 + \pi^2 )  (X-Y) 
                  - {5 \over 8}\, \pi^2 \Yt - {\zeta_3 \over 6}  \Biggr]  
      	\MinusBreak{}
            {9 - 13 x \over 48 y} \Bigl(20 \Yt^3 + 17 \pi^2 \Yt 
                                  + 20 \zeta_3 \Bigr)  
      	\PlusBreak{}
       {1 \over 8} \biggl( 11\, {y^2 \over x^2} + {4 \over x} + 26 
              + {88 \over 9} \, y \biggr)  ( \Yt^2 + \pi^2 )
          + {\pi^2 \over 216} \biggl(396 \, {x \over y} + 63 - 217 y \biggr) 
      	\MinusBreak{}
            {1 \over 108} \biggl( 297\,  {x-y \over x} - 56 y \biggr) \Yt 
          - {x \over 18 y} + {115 \over 72} - {4849 \over 1296} \, y 
\,, \label{C2mpmp} \\[1pt plus 4pt]
%
%
C^{[3]}_{-+-+} & = & 
        {2 x^2 - 6 x + 3 \over y^2} \Biggl[ \li4 \biggl( -{x \over y} \biggr) 
            - \li4(-x) - \li4(-y) + (X-Y) \li3(-x)   
      	\MinusBreak{{2 x^2 - 6 x + 3 \over y^2} \BigglBl}
         {\pi^2 \over 6} \biggl( \li2(-x) - X^2 - {1 \over 2} \, Y^2 \biggr)
          + {1 \over 24}\, X^4 - {1 \over 6} \, X Y^3
      	\PlusBreak{{2 x^2-6 x+3 \over y^2} \BigglBl}
                   {1 \over 24}\, Y^4          
                  - \zeta_3 (X-Y) + {7 \over 120} \, \pi^4 \Biggr]
      	\PlusBreak{}
         {1 \over 6} \biggl( 2 \, {y \over x} - 3 \, {9 x - 13 \over y} \biggr)
                  \Biggl[ \li3(-x) + \li3(-y)  
                  - (X-Y) \li2(-x) \Biggr]
     	\MinusBreak{}
           {y \over 18 x} \Biggl[ (2 X - Y) \Bigl( (X+Y)^2 - 6 Y^2 \Bigr)    
                    + 6 i \pi {} \Bigl( (X-Y)^2 + \pi^2 \Bigr) 
      	\PlusBreak{\null - {y \over 18 x} \BigglBl}
                       {\pi^2 \over 4} ( 17 X + 25 Y + 2 i \pi )
                             + 5 \zeta_3 \Biggr]  
      	\PlusBreak{}
         {9 x - 13 \over 12 y} \Bigl( X^3 - 3 X Y^2 + 2 \pi^2 (X+Y) \Bigr) 
          + {11 \over 9} \, {y \over x} ( \Yt^2 + \pi^2 )  
      	\PlusBreak{}
        {1 \over 8} ( 11 y^2 + 4 x + 26 ) \Bigl( (X-Y)^2 + \pi^2 \Bigr)
       - {\pi^2 \over 216} \biggl( 217 \, {y \over x} - 63 
                                              - {396 \over y} \biggr)
       	\PlusBreak{}
          {193 \over 108} \, {y \over x}  \, \Yt
         + {1 \over 108} \biggl( 297 (1-y) + 137 \,{y \over x} \biggr) (X-Y)
         - {1 \over 18 y} 
       	\PlusBreak{}
           {115 \over 72} - {4849 \over 1296} \, {y \over x} 
\,, \label{C3mpmp} \\[1pt plus 4pt]
%
%
D^{[1]}_{-+-+} & = & 
 - A^{\Susy,[1]}_{-+-+} + A^{[1]}_{-+-+} + C^{[1]}_{-+-+} 
                        + E^{[1]}_{-+-+}
      	\PlusBreak{}
            {1 + x^2 y \over 3 y^2} 
                  ( \Xt^2 + \pi^2 ) (\Xt - 2 \, \Yt) 
          -  {y \over 3 x} (4-y)  \Xt \Yt^2
          - {2 \over 3} \, {y^2 \over x} ( \Yt^2 + \pi^2 ) \Yt 
      	\PlusBreak{}
            {2 \over 3} \, {y \over x} ( \Xt^2 - \pi^2 ) \Yt
          - {\pi^2 \over 3} \, y \Xt
          - {1 \over 9}  \biggl( {6 \over y} - 3  x + 8   y \biggr)   \Xt^2
      	\PlusBreak{}
            {y^2 \over 9 x} \biggl( 22 \Yt^2 + {5 \over 16}\, \pi^2 \biggr)   
          +  {1 \over 9} \biggl( {12 \over y} - 6 x 
                         + 16 y - 11 \, {y^2 \over x} \biggr) \Xt \Yt
      	\PlusBreak{}
            {1 \over 3} ( \Xt - 2 \, \Yt ) 
          - {17 \over 54} \, {y^2 \over x} 
\,, \label{D1mpmp} \\[1pt plus 4pt]
%
%
D^{[2]}_{-+-+} & = & 
   - A^{\Susy, [2]}_{-+-+} + A^{[2]}_{-+-+} 
                           + C^{[2]}_{-+-+} + E^{[2]}_{-+-+}
      	\MinusBreak{}
            {1 \over 3}  \, {1 + x^2 y \over y^2} 
               ( \Xt^2 + \pi^2 )  (2 \Xt - \Yt)
          - {y^2 \over 3 x} \,  \Yt^2   (4 X - Y + 3 i \pi)   
      	\MinusBreak{}
            {4 \over 3} \,  {y \over x} \,  \Xt^2 \Yt
          - {y \over 3} \Bigl( 2 \Xt \Yt^2 - \pi^2   (2 \Xt - \Yt) \Bigr)
      	\PlusBreak{}
            {2 \over 9} \biggl( {6 \over y} - 3  x + 8 y \biggr) \Xt^2
          - {11 \over 9} \, {y^2 \over x} \, \Yt^2    
          - {1 \over 9} \biggl( {6 \over y} - 3 x 
                      + 8 y - 22 {y^2 \over x} \biggr) \Xt \Yt
      	\PlusBreak{}
            {5 \over 144}\,  \pi^2   y 
          + {1 \over 3} ( \Yt - 2 \, \Xt ) - {17 \over 54} \, y 
\,, \label{D2mpmp} \\[1pt plus 4pt]
%
%
D^{[3]}_{-+-+} & = & 
     - A^{\Susy, [3]}_{-+-+} 
     + A^{[3]}_{-+-+} + C^{[3]}_{-+-+} + E^{[3]}_{-+-+}
      	\PlusBreak{}
           {1 + x^2 y \over 3 y^2} 
                 ( \Xt^2 + \pi^2 )  (\Xt + \Yt) 
          +  {y^2 \over 3 x}  \Bigl( (\Yt^2 + \pi^2) \Yt
                        + \Xt \Yt^2 \Bigr)   
      	\PlusBreak{}
          {2 \over 3}\, {y \over x} 
               \biggl( \Bigl( \Xt^2 + {\pi^2 \over 2} \Bigr) \Yt
                        + \Xt \Yt^2 \biggr)
          - {\pi^2 \over 3} \,  y \Xt  
      	\MinusBreak{}
            {1 \over 9} \biggl({6 \over y} - 3 x + 8 y \biggr) \Xt^2
          - {11 \over 9} \, {y^2 \over x} \, \Yt^2 
          + {1 \over 9} \biggl( 6 {x \over y} + 3 + 11 {y \over x} \biggr)
                                        \Xt \Yt
          + {5 \over 144} \, \pi^2 \, {y \over x} 
      	\PlusBreak{}
             {1 \over 3} (\Xt + \Yt)
          - {17 \over 54} \, {y \over x}
\,, \label{D3mpmp}\\[1pt plus 4pt]
%
%
E^{[1]}_{-+-+} &=& 
        - {1+x^2 \over 12 y^2} ( \Xt^2 + \pi^2 ) \Xt
         + {1\over 18} \biggl({x^2 + 5 \over y} - x \biggr) \Xt^2
         - \pi^2 \, {y^2 \over 108 \, x}  
         - {1 \over 6} \, \Xt
\,, \label{E1mpmp}\\[1pt plus 4pt]
%
%
E^{[2]}_{-+-+} &=& 
- {y \over 9} \biggl(\Yt^2 + {\pi^2 \over 12} \biggr) 
-  {1-x \over 72 y} 
\,, \label{E2mpmp}\\[1pt plus 4pt]
%
%
E^{[3]}_{-+-+} &=& 
 - {1\over 9} \, {y \over x}  \biggl(\Yt^2 + {\pi^2 \over 12} \biggr) 
     + {1-x \over 72 y} 
\,, \label{E3mpmp}\\[1pt plus 4pt]
%
%
F^{[1]}_{-+-+} &=& 
  {1\over 18}  \Biggl[ {y^2\over x}  (\Xt^2 + 4 \pi^2)
              + {y \over x}  (2-y) (\Yt^2 + \pi^2) 
	\PlusBreak{{1\over 18}  \BigglBl}
               {y \over x} (2 x-y) \Bigl((X-Y)^2 + \pi^2 \Bigr) \Biggr] 
\,, \label{F1mpmp}\\[1pt plus 4pt]
%
%
F^{[2]}_{-+-+} &=& 
  {1\over 18} \Biggl[ {y \over x}  (2-x) (\Xt^2 + \pi^2)
              + y {} ( \Yt^2 + 4 \pi^2 ) 
	\PlusBreak{ {1\over 18} \BigglBl}
               {y\over x} (2 y-x) \Bigl((X-Y)^2 + \pi^2\Bigr) \Biggr] 
\,, \label{F2mpmp}\\[1pt plus 4pt]
%
%
F^{[3]}_{-+-+} &=& 
   {1 \over 18}  \Biggl[ - {y \over x} (1-2 x) (\Xt^2 + \pi^2)
              - {y \over x} (1-2 y) (\Yt^2 + \pi^2) 
	\PlusBreak{{1 \over 18} \BigglBl}
                {y\over x} \Bigl((X-Y)^2 + 4 \pi^2 \Bigr) \Biggr] 
\,, \label{F3mpmp}\\[1pt plus 4pt]
%
%
H^{[7]}_{-+-+} & = & 
           - {1+x^2 \over y^2} \Biggl[6 \li4(-x)
               - 4  \Xt {} ( \li3(-x) + \li3(-y) - \zeta_3 )
               - 2  \Yt {} ( \li3(-x) - \zeta_3 ) 
      	\PlusBreak{- {1+x^2 \over y^2}   \BigglBl}
                \Bigl( \Xt^2 - 2 \Xt \Yt + \pi^2 \Bigr) 
                           \li2(-x) 
               + {1 \over 3} \, X^3 \Yt - 2 \Xt X Y^2 
       	\MinusBreak{- {1+x^2 \over y^2} \BigglBl}
             {\pi^2 \over 12} \Bigl( 9 X^2 - 20 X Y 
                                     - 2 i \pi {} (X + 2 Y) \Bigr)
               - {2 \over 5} \, \pi^4 \Biggr]   
	\PlusBreak{}
            2 \biggl( {14 \over y^2} + {16 \over y} + 5 \biggr) 
                  \Biggl[ \li4 \biggl( -{x \over y} \biggr) + \li4(-x)
                  - \li4(-y) - \Yt {} ( \li3(-x) - \zeta_3 )   
	\PlusBreak{\null + 2 \biggl( {14 \over y^2} + {16 \over y} + 5 \biggr)}
                     {\pi^2 \over 6} \li2(-x) - {1 \over 6} X Y^3 
                  + {1 \over 24} \Bigl( Y^2 + 2 \pi^2 \Bigr) Y^2 
  	\MinusBreak{\null + 2\biggl( {14 \over y^2} + {16 \over y} + 5 \biggr)}
                    {7 \over 360} \, \pi^4 \Biggr]  
	\MinusBreak{}
           {1 \over 3 y^2} \Bigl(14 - x^2 (1-2 x) \Bigl)
                                 \biggl( \li3(-x) - \zeta_3 
                                    - \Xt \li2(-x) 
                + {1 \over 2} \, i \pi {} ( \Xt^2 + \pi^2 )  \biggr) 
  	\MinusBreak{}
            {2 \over 3} \biggl( 30\, {x \over y} - x + 8 \biggr) 
                        \biggl( \li3(-y) - \zeta_3 
                                    - \Yt   \li2(-y)
                      - {1 \over 2}\, X {} ( \Yt^2 + \pi^2 ) \biggr)
  	\PlusBreak{}
          {1+x^2 y \over 18 y^2} \Bigl( 4 \Xt^3 - 7 \pi^2 \Xt
                   + 15 (\Xt^2 + \pi^2) \Yt \Bigr)  
  	\PlusBreak{}
              {y^2 \over 36 x}   \Bigl( 8 \Yt^2 - 12 \Xt \Yt
                        + 24 \Xt^2 + 29 \pi^2 \Bigr) \Yt  
  	\MinusBreak{}
           {1 \over 12} (23 - 10 x ) 
                    \Bigl( \Xt^2 Y + i \pi{} \Bigl( X^2 + \pi^2 \Bigr) \Bigr) 
  	\PlusBreak{}
           {\pi^2 \over 36} \biggl( 50 {x^2 \over y} - 45 x + 117 \biggr) \Xt
           + {\pi^2 \over 36} \biggl( {96 \over y} + 27 x - 7 y \biggr) \Yt
           + 2 \,{1-x \over y} \,  \zeta_3    
  	\PlusBreak{}
        {1 \over 36} \biggl( {63 \over y^2} + {156 \over y} 
                    + 96 - 44 y + 9 y^2 \biggr)
                    ( \Xt^2 + \pi^2 )   
  	\PlusBreak{}
          {1 \over 36} \biggl( {36 \over x^2} + {64 \over x} 
                 + 173 - 26 x + 9 x^2 \biggr)
                    ( \Yt^2 + \pi^2 )  
       	\MinusBreak{}
                {1 \over 18} \biggl( {36 \over y} - {22 \over x} 
                                          + 92 + 9 y^2 \biggr) \Xt \Yt
              - {\pi^2 \over 36} (2 - 3 x) 
                 \biggl( {26 \over y} - {22 \over x} - 5 + 3 y \biggr)  
       	\MinusBreak{}
        {1 \over 54} \biggl( 45 {x \over y} + 9 x + 92 y \biggr) \Xt 
           - {1 \over 54} \biggl( 92 \, {y^2 \over x} - 9 x + 153 \biggr) \Yt
\,, \label{H7mpmp} \\[1pt plus 4pt]
%
%
H^{[8]}_{-+-+} & = & 
             {4 \over 3} \biggl({x \over y^2} - 2 \biggr) 
                     \biggl( \li3(-x) - \zeta_3 
                                   - \Xt \li2(-x)
               + {1 \over 2} \, i \pi {} ( \Xt^2 + \pi^2 ) \biggr) 
  	\PlusBreak{}
         {2 \over 9} \,{1+x^2 y \over y^2} \Bigl(\Xt^3
                                   + 3 ( \Xt^2 + \pi^2 ) \Yt \Bigr)
          -  (1-y) {y \over 3 x} ( \Xt^2 - 2 \pi^2 ) \Yt
  	\PlusBreak{}
               (5 - 2 y) {y \over 3 x} \, \Xt \Yt^2 
           + {11 \over 36} \, {y^2 \over x} ( 4 \Yt^2 + 7 \pi^2 ) \Yt 
           + {29 \over 36}\, y \pi^2 \Xt 
       	\MinusBreak{}
          {1 \over 9} \biggl({9 \over y^2} + {15 \over y} 
                  - (3-y) (7 x - 2 y) \biggr) 
                   ( \Xt^2 + \pi^2 )
  	\PlusBreak{}
             y^2 \biggl( {y^2 \over x^2} - {53 \over 9 x} \biggr)  
                                          ( \Yt^2 + \pi^2 )  
           - {y \over 9} \biggl({11 \over x} + 3 (y - 5 x) \biggr)   
                         \Bigr(\Xt \Yt + 4 \pi^2 \Bigl)
       	\PlusBreak{}
            {\pi^2 \over 3} \biggl( {2 \over y} + x^2 
                                            - 10 y {} (x-y) \biggr)  
           -  {2 \over 27} \biggl(9 {x \over y} + 23   y \biggr) \Xt 
           - {46 \over 27} \, {y^2 \over x} \, \Yt
           + {1 \over 2} 
\,, \label{H8mpmp} \\[1pt plus 4pt]
%
%
H^{[9]}_{-+-+} & = & 
           - {1+x^2 \over y^2}   \Biggl[ 6 \li4(-x)
               - 4 \Xt {} ( \li3(-x) + \li3(-y) - \zeta_3 )
               - 2 \Yt {} ( \li3(-x) - \zeta_3 )  
  	\PlusBreak{ - {1+x^2 \over y^2}   \BigglBl}
                \Bigl( \Xt^2 - 2 \Xt \Yt + \pi^2 \Bigr) 
                           \li2(-x) 
               + {1 \over 3}  \,  X^3 \Yt - 2 \Xt X Y^2 
  	\MinusBreak{ - {1+x^2 \over y^2} \BigglBl}
              {\pi^2 \over 12} \Bigl( 9 X^2 - 20 X Y 
                 - 2 i \pi {} (X + 2 Y) \Bigr)
                  - {2 \over 5} \, \pi^4 \Biggr]   
  	\PlusBreak{}
            2 \biggl(14\, {x^2 \over y^2} + 16\,{x \over y} + 5 \biggr) \Biggl[ 
              \li4 \biggl( -{x \over y} \biggr) 
                   + \li4(-x) - \li4(-y) - \Yt \li3(-x) 
  	\PlusBreak{\null + 2 \biggl( 14 {x^2 \over y^2} + 16 {x \over y} 
                                                        + 5 \biggr) \BigglBl}
                     \Yt \zeta_3  
                   +  {\pi^2 \over 6} \,  \li2(-x) - {1 \over 6} X Y^3 
    	\PlusBreak{\null + 2 \biggl( 14 {x^2 \over y^2} + 16 {x \over y} 
                                                        + 5 \biggr) \BigglBl}
                   {1 \over 24} \Bigl( Y^2 + 2   \pi^2 \Bigr)  
                       Y^2 - {7 \over 360} \, \pi^4 \Biggr] 
  	\MinusBreak{}
             8 {1 - 3 x \over y^2} \Biggl[ \li4(-x)   
                             - {1 \over 2}\, \Xt {} ( \li3(-x) - \zeta_3 )
                               + {\pi^2 \over 6} \, \li2(-x) 
              - {1 \over 48} \, X^4
  	\MinusBreak{\null - 8 {1 - 3 x \over y^2} \BigglBl}
            {\pi^2 \over 12} \, X^2 - {7 \over 180} \, \pi^4 \Biggr]
       	\MinusBreak{}
            {1 \over 3} \biggl( {11 \over y^2} - {30 \over y} 
                  - 2 + {4 \over x} \biggr) \Biggl[ \li3(-x) - \zeta_3 
                                    - \Xt {}  ( \li2(-x) - {\pi^2 \over 6} )  
  	\PlusBreak{\null - {1 \over 3} \biggl( {11 \over y^2} - {30 \over y} 
                  - 2 + {4 \over x} \biggr) \BigglBl}
                    {1 \over 2} \, i \pi {} ( \Xt^2 + \pi^2 ) \Biggr]
       	\MinusBreak{}
            {2 \over 3} \biggl( 30\, {x \over y} - {y \over x} + 21 \biggr)
                \Biggl[ \li3(-y) + \zeta_3 
                       - \Yt {} \biggl( \li2(-y) + {\pi^2 \over 24} \biggr)
       	\MinusBreak{\null - {2 \over 3} \biggl( 30\, {x \over y} 
                           - {y \over x} + 21 \biggr) \BigglBl }
                   {1 \over 2} \, X {} ( \Yt^2 + \pi^2 ) \Biggr]  
  	\PlusBreak{}
          {1 \over 36} \biggl( {30 \over y^2} - {14 \over y} 
                                                   - 44 x + 39 \biggr) 
                    ( \Xt^2 + \pi^2 ) \Xt 
       	\MinusBreak{}
            {1 \over 12} \biggl( 10 {x \over y^2} - 13 x + 3 y \biggr) 
                    ( \Xt^2 + \pi^2 )   \Yt 
   	\PlusBreak{}
           {y^2 \over 9 x} \Bigl( 2 \Yt^3 - 6 \Xt^2 \Yt
                        + 3 \Xt \Yt^2 - \pi^2 \Xt 
                        + 8 \pi^2  \Yt \Bigr)   
   	\PlusBreak{}
            {\pi^2 \over 36} \biggl( {108 \over y} + 19 x + 49 \biggr) \Xt
          + {\pi^2 \over 12} \, {x \over y} ( 22 - 17 y ) \Yt
           - 2 \, {1-x \over y} \, \zeta_3 
   	\PlusBreak{}
         {1 \over 36} \biggl( {63 \over y^2} - {6 \over y} 
              + 3  x^2 - 33 x y - 251 y \biggr)
                    ( \Xt^2 + \pi^2 )  
   	\PlusBreak{}
         {1 \over 36} \biggl( {9 \over x^2} - {26 \over x} 
                          + 173 - 36 x y + 28 x \biggr)
                    ( \Yt^2 + \pi^2 )  
       	\MinusBreak{}
         {1 \over 9} \biggl( {18 \over y} + {22 \over x} 
                  - 73 x - 18 x y - 142 y \biggr)
                       \Xt \Yt 
       	\PlusBreak{}
            {\pi^2 \over 36}\biggl( {106 \over y} + {44 \over x} 
                        + 11 x^2 + 47 x y + 193 y \biggr)
           - {1 \over 54} \biggl( 45 \,  {x \over y} - 144 + 92 y \biggr) \Xt 
         	\MinusBreak{}
             {1 \over 54}\biggl(92 \, {y^2 \over x} 
                  - {9 \over x} + 153 \biggr) \Yt 
\,, \label{H9mpmp} \\[1pt plus 4pt]
%
%
I^{[7]}_{-+-+} &=& 
 {1 \over 18} \Biggl[3 {x^2 + 1 \over y^2} ( \Xt^2 + \pi^2 ) \Yt
                     - \biggl({1 \over x} + 2 x - 6\, {x \over y}  \biggr) 
                                ( \Xt^2 + \pi^2 ) 
	\PlusBreak{{1 \over 18} \BigglBl}
                       \biggl( - {6 \over y} + 2 x + {1\over x} \biggr) 
                               ( \Yt^2 + \pi^2 )
                     + \biggl( {1 \over x} + 2 \, {1-2 x \over y} \biggr) 
                                      \Bigl((X-Y)^2 + \pi^2 \Bigr)
 	\PlusBreak{{1 \over 18} \BigglBl}
                       \biggl( {y \over x} - 3 - 6\, {x\over y} \biggr) \pi^2
                     + 6 \Yt  \Biggr] 
\,, \label{I7mpmp}\\[1pt plus 4pt]
%
%
I^{[8]}_{-+-+} &=& 
 {1 \over 9}  \biggl( y {} (X-Y) (\Xt + 4 \Yt) 
                  + {y\over x} {} (\Xt - 4 \Yt) \Yt \biggr) 
\,, \label{I8mpmp}\\[1pt plus 4pt]
%
%
I^{[9]}_{-+-+} &=& 
 {1 \over 18} \Biggl[ - 3 \, {x^2 + 1 \over y^2} (X-Y) 
                                     ( \Xt^2 + \pi^2 ) 
                     + 2 \biggl( 2 \, {y^2 \over x} + 3 y \biggr) 
                               ( \Xt^2 + \pi^2 )
	   \PlusBreak{{1\over 18} \BigglBl}
                       2 \, {y^2 \over x} ( \Yt^2 + \pi^2 )
                     + 2 \biggl( 2\, {y \over x} - 3 x + {6 \over y} \biggr) 
                                (X-Y) \Xt
	   \PlusBreak{{1\over 18} \BigglBl}
                       6 \pi^2 \, {y \over x}
                     - 6 (X-Y) \Biggr]
\,. \label{I9mpmp}
\end{eqnarray}
%


\section{Finite remainder functions for pure $N=1$ super-Yang-Mills theory}
\label{N=1RemainderAppendix}

In this appendix, we present the independent $N=1$ supersymmetric 
finite remainder functions appearing 
in~\eqns{SUSYTwoloopSingleTrace}{SUSYTwoloopDoubleTrace}.
For $-$$-$++, these functions are
\begin{eqnarray}
\ASusy{[1]}{--++} &=& 
3 y {}\Biggl[ \li4(-x) + \li4(-y) - \li4\biggl(- {x \over y} \biggr)
                   - \Xt {} ( \li3(-y) + \li3(-x) )  
 	\PlusBreak{3 y \BigglBl}
                {\pi^2 \over 6} \, \li2(-x) - {1 \over 6} \, X^3 Y
                   + {1\over 24} (\Xt^4 - \Yt^4 )
                   - {1\over 12} \, \Xt \Yt^2 (3 \Xt - 2 \Yt)
 	\MinusBreak{3 y \BigglBl}
                    {\pi^2\over 12} \Bigl((X+Y)^2 + 2 i \pi Y \Bigr)
        - {\pi^4\over 180} \Biggr]
        \MinusBreak{}
        {1 \over 2 x} \biggl( {1\over 4}\, \Xt^4 
                              + {\pi^2\over 3}\, \Xt^2
              - \zeta_3 \Xt + {\pi^4 \over 40} \biggr)
	\PlusBreak{}        
           {3\over 2}\, {1-2 x \over x} \Biggl[ \li3(-x) - \zeta_3
                        - \Xt {} \biggl(\li2(-x) + {2\over 3}\, \pi^2 \biggr)
                        + {1\over 6}\, \Xt^3 
	\MinusBreak{-\null {3\over 2} \, {1-2 x \over x} \BigglBl}
                          {1\over 2} (\Xt^2 + \pi^2) \Yt 
                        + i {\pi\over 2} (X^2 + 2 \pi^2 ) \Biggr]
	\PlusBreak{}
           {1\over 16 x} \Bigl(4  \Xt^3 + 9 \pi^2 \Xt
                            - 52 \zeta_3 \Bigr)
          + {3\over 2 y} (\Xt^2 + \pi^2) 
	\MinusBreak{}
           {\pi^2 \over 16 x} (13 - 8 x)
          - {85\over 18 x} \, \Xt + {143\over 12 x} 
\,, \label{A1Smmpp}\\[1pt plus 4pt]
%
%
\ASusy{[3]}{--++} &=& 
          - {x \over 8 y} \Bigl((X-Y)^2 + \pi^2\Bigr)^2
          - {\pi^2 \over 12} \biggl({1 \over y} - 3 x \biggr) 
                  \Bigl((X-Y)^2 + \pi^2 \Bigr)
	\PlusBreak{}
             {\zeta_3 \over 2}\, {\Xt \over x y} 
          - {7 \over 240}\,  {\pi^4 \over y}
        - {1 \over 8 y} (1-x) (y-3 x) 
                        \biggl( \Xt^2 - 3 \Xt \Yt
                                 + {\pi^2 \over 3} \biggr) \Xt
	\MinusBreak{}
            {3 \over 2 x y} 
                \Biggl[ \li3(-x) - X {} 
                 \biggl(\li2(-x) - {\pi^2 \over 6} \biggr)
                      -  {1\over 3}\, X^3 - {1\over 2}\, X^2 Y 
	\MinusBreak{\null - {3\over 2}\, {1\over x y} \BigglBl }
                       i {\pi\over 2} (X^2 + X Y + Y^2)
                      + {5 \over 8} \,\pi^2 X + i {\pi^3 \over 24} 
                                 + {13 \over 12} \,  \zeta_3  \Biggr]
	\MinusBreak{}
             {9 \over 4}\, \Xt^2  \Yt 
           - {\pi^2\over 12} (7 - 4 x) \Xt
          + {9\over 4 y} (\Xt^2 + \pi^2) 
          + {3\over 4}  \Bigl((X-Y)^2 + \pi^2\Bigr)
	\PlusBreak{}
           {\pi^2 \over 2} \biggl({23\over 16 x y} - 1 \biggr)
          - {3\over 2 x y} \biggl({58\over27} - y \biggr) \Xt
            - {143\over 12 y} 
                + \Bigl\{t \leftrightarrow u \Bigr\} 
\,, \label{A3Smmpp}\\[1pt plus 4pt]
%
\BSusy{[1]}{--++} &=& 
         2 \, {x {} (1-x) \over y} \Biggl[ \li4 \biggl(- {x\over y} \biggr) 
            - 4 \li4(-x) + \li4(-y) + 3 \Xt \li3(-x)  
	\MinusBreak{2 {x (1-x) \over y} \BigglBl}
               {1 \over 2} \Bigl( \Xt^2 + 2 (X-Y) \Yt + 2 \pi^2 \Bigr)
                        \li2(-x)
             - {1 \over 48}\, \Xt^4
             + {1 \over 6}\, X^3 Y
	\PlusBreak{2 {x (1-x) \over y} \BigglBl}
               {1\over 24} \, Y^4
             - {1 \over 8}\, X^2 Y^2 + {1 \over 12} (X-i \pi) Y^3 
             + i {\pi \over 4} \, X Y {} (3 X - 2 Y)
	\MinusBreak{2 {x (1-x) \over y} \BigglBl}
               {\pi^2 \over 24} (9 X^2 + 18 X Y - 7 Y^2 
                            + 20 i \pi X)
             + {83 \over 360} \, \pi^4 \Biggr]
	\PlusBreak{}
              4 \, {1+x^2 \over x} \Biggl[ \li4\biggl(- {x \over y} \biggr) 
             + 2\li4(-x) - 2\li4(-y) + {3 \over 2}\, \Xt \li3(-y)
	\MinusBreak{\null + 4 {1+x^2 \over x} \BigglBl}
               {3 \over 2}\, \Yt \li3(-x) 
             - {1 \over 2}\, \Bigl ((X-Y)^2 - 2 \Xt \Yt - \pi^2 \Bigr) \li2(-x)
	\MinusBreak{\null + 4 {(1+x^2) \over x} \BigglBl}
               {3\over 8} \, X^3 Y + {17 \over 16} \, X^2 Y^2 
             - {5 \over 24} \, X Y^3
             -  {3 \over 2} \, \zeta_3 (X-Y) - {29 \over 720}\, \pi^4 
	\MinusBreak{\null + 4 {(1+x^2) \over x} \BigglBl}
               i {\pi \over 24} (X^3 + 18 X^2 Y - 42 X Y^2 + Y^3)
	\PlusBreak{\null + 4 {(1+x^2) \over x} \BigglBl}
               {\pi^2 \over 48} \Bigl( 27 X^2 - 36 X Y + 11 Y^2 
                            + 2 i \pi {} (11 X - Y) \Bigr) \Biggr]
	\MinusBreak{}
               6 x \Biggl[ 2 \li4\biggl(-{x \over y} \biggr) + 3 \li4(-x) 
                + 2 \li4(-y) - \Yt {} (3 \li3(-x)  + 2 \li3(-y))  
	\MinusBreak{\null - 6 x \BigglBl}
                    {1 \over 4} \biggl( (X-Y)^2 - 2 \Xt \Yt 
                        + (X-Y) (\Xt + \Yt) 
                + {2\over 3}\, \pi^2 \biggr) \li2(-x)
	\MinusBreak{\null - 6 x \BigglBl}
                  {13 \over 36} \, X^3 Y + {5 \over 12} \, X^2 Y^2
                - {11 \over 18} \,  X Y^3 + {5\over 36} \, Y^4
	\MinusBreak{\null - 6 x \BigglBl}
                  i {\pi \over 36} ( X^3 + 27 X^2 Y - 10 Y^3 )
                + {\pi^2 \over 12} \Bigl(5 X {} (X-Y) + 2 Y^2 \Bigr)
	\PlusBreak{\null - 6 x \BigglBl}
                  i {\pi^3 \over 36} (13 X + 11 Y)
                - \zeta_3 {}(X-Y) - {59 \over 360} \, \pi^4 \Biggr] 
        -    {1 \over 24} (3 x - 2 y) \Xt^4 
	\MinusBreak{}
            2 \Biggl[ 4 \li4\biggl(- {x\over y} \biggr) + 5 \li4(-x)
                - 5 \li4(-y) - 6 \Yt \li3(-x) 
	\MinusBreak{\null -2 \BigglBl}
                 {1 \over 2} \Bigl( (X-Y)^2 - 2 \Xt \Yt - 2 \pi^2 \Bigr) 
                                                                  \li2(-x)
                + X^2 Y^2  - {5 \over 6} \, X^3 Y 
	\MinusBreak{\null -2 \BigglBl}
                  {7 \over 6} \, X Y^3 
                + {5 \over 12} \, Y^4
                - i {\pi \over 6} (3 X^3 + 9 X^2 Y - 6 X Y^2 - 4 Y^3 )
	\PlusBreak{\null -2 \BigglBl}
                  {\pi^2 \over 24} \Bigl(27 X^2 + 18 X Y - 11 Y^2 
                           + 4 i \pi {} ( 4 X - 3 Y) \Bigr)
                - {11 \over 180} \, \pi^4 \Biggr]
	\MinusBreak{}
            3 \biggl( {2 x^2 \over y} - x - {2\over x} \biggr) 
                  \biggl(\li3(-x) - \zeta_3 - \Xt \li2(-x)
               + i {\pi \over 2} \, \Xt^2 + i {\pi^3 \over 2}  \biggr)
	\MinusBreak{}
            3 \biggl({4 \over y} + 2 y - {y^2 \over x} \biggr)  
                 \biggl( \li3(-y) - \zeta_3 - \Yt \li2(-y)
               + i {\pi\over 2}\, \Yt^2 
                 + i {\pi^3 \over 2} \biggr)
	\PlusBreak{}
            {1 \over 4} \biggl(3 x {} \, {1-x \over y} - y \biggr) \Xt^3
          + {3 \over 4} \biggl( 5 (1-y) {y \over x} - 3 x \biggr) 
                    \Xt \Yt^2
          - 3 x {}\, {1-x \over y} \Xt^2 \Yt
	\MinusBreak{}
            {1 \over 2} \, \biggl(3 y {} \, {1-y \over x} - x \biggr) \Yt^3
          + {\pi^2 \over 4} \, \biggl( {4 x^2 \over y} + 2   x - 37 \biggr) \Xt
          + 6 \zeta_3 {} \biggl({y^2 \over x} - 2 {x^2 \over y} \biggr)
	\MinusBreak{}
           {\pi^2 \over 2} \biggl({y^2 \over x} + 2 y - 19 \biggr) \Yt
          + {3 \over 4} \biggl(7 \, {x^2 \over y} + x - 3 \biggr) \Xt^2
          + {3 \over 2} \biggl(4 \, {y \over x} - 3 x \biggr)   \Yt^2
	\PlusBreak{}
            {3 \over 4} (y - 2 x) \biggl({3 \over x y} - 4\biggr) \Xt \Yt 
          - {3 \over 2}\, \pi^2 (y-2 x) \biggl({1 \over x y} - 1 \biggr) 
\,. \label{B1Smmpp} 
\end{eqnarray}

The independent $N=1$ supersymmetric remainder functions for $-$+$-$+ are,
\begin{eqnarray}
\ASusy{[1]}{-+-+} & = & 
          - {1+x^2 \over 8 x} ( \Xt^2 + \pi^2 )^2
         + {\pi^2 \over 12}  \biggl( {y^2 \over x} - 3 \biggr)  
                  ( \Xt^2 + \pi^2 )
         + {y^2 \over 2 x} \biggl(\zeta_3 \Xt
               + {7 \over 120}\, \pi^4 \biggr)
	\PlusBreak{}
           {3 \over 2}  {y^2 \over x}  \Biggl[ \li3(-x) - \Xt \li2(-x)
             + {1\over 3} \, \Xt^3 
             - {1\over 2} (\Xt^2 + \pi^2) Y 
             + {17 \over 24} \, \pi^2 \Xt 
 	\MinusBreak{\null + {3 \over 2}  {y^2 \over x}  \BigglBl}
                {19 \over 6} \, \zeta_3 \Biggr]
          +   {1 \over 8 y} (x-y) (1 - 3 x) 
                          ( \Xt^2 + \pi^2 ) \Xt
           + {\pi^2 \over 2}\, y \Xt
	\MinusBreak{}
            {3 \over 4} (2x - y) ( \Xt^2 + \pi^2 )
          + {\pi^2 \over 16} \biggl(13 {y \over x} - 7 y + 16 x \biggr)
 	\PlusBreak{}
            {3 \over 2} \, {y \over x} \biggl(1 - {58\over27} y \biggr) \Xt
          + {143\over12} {y^2\over x}
\,, \label{A1Smpmp}\\[1pt plus 4pt]
%
%
\ASusy{[2]}{-+-+} & = & 
      3\, {x \over y} \Biggl[ \li4\biggl(- {x\over y} \Bigl) 
                      + \li4(-x) - \li4(-y)
                      - \Yt {} (\li3(-x) - \zeta_3 ) 
 	\PlusBreak{3 {x \over y} \BigglBl}
                        {\pi^2 \over 6}\, \li2(-x) 
                      + {1 \over 24} (Y^2 - 4 X Y + 2 \pi^2) Y^2 
                      - {7 \over 360} \, \pi^4 \Biggr]
 	\MinusBreak{}
            {y \over 8} \biggl( \Yt^4 + {4 \over 3} \pi^2 \Yt^2
                - 4 \zeta_3 \Yt + {\pi^4 \over 10} \biggr) 
         + {3 \over 4} (2-3 x) {y \over x}
              ( \Yt^2 + \pi^2 ) 
 	\MinusBreak{}
            {3 \over 2} (2-y) \Biggl[ \li3(-y) 
                     - \Yt {} \biggl(\li2(-y) - {\pi^2 \over 6} \biggr)
           - {1\over 2} X {} (\Yt^2 + \pi^2) - \zeta_3 \Biggr]
 	\MinusBreak{}
           {\pi^2 \over 16} (8 x - 15 y)
         + {y\over 16} (20 \Yt^3 + 17 \pi^2 \Yt - 52 \zeta_3 )
         - {29\over9} y \Yt + {143\over 12} y
\,, \label{A2Smpmp}\\[1pt plus 4pt]
%
%
\ASusy{[3]}{-+-+} & = & 
   {3 \over y} \Biggl[ \li4\biggl(- {x\over y} \biggr) - \li4(-x) - \li4(-y)
                   + (X-Y) (\li3(-x) - \zeta_3 )
 	\MinusBreak{{3 \over y} \BigglBl}
            {\pi^2 \over 6} \biggl(\li2(-x) - X^2 - {1 \over 2} Y^2 \biggr)
       + {1 \over 6} X^3 Y - {1 \over 4} X^2 Y^2 + {7 \over 120} \pi^4 \Biggr]
 	\MinusBreak{}
         {y \over 2 x} \biggl( {\pi^2 \over 3} (X-Y)^2 
          + {\pi^4 \over 40} - \zeta_3 {} (\Xt + \Yt) \biggr)
          - { 1 - x y \over 8 x y} (X-Y)^4
 	\MinusBreak{}
           {3 \over 2} \biggl( {y\over x} - 2 \biggr)  
                      \Biggl[ \li3(-x) + \li3(-y)
                            - (X-Y) \li2(-x) - {1 \over 6} \, X^3 
	\PlusBreak{\null - {3 \over 2} \biggl( {y\over x} - 2 \biggr)\BigglBl}
            {1 \over 2} \, X Y^2 - {\pi^2 \over 3} (X+Y) \Biggr]
	\PlusBreak{}
             {y \over 16 x} \biggl(4 ( \Xt + 5 \Yt) (X-Y)^2
          + \pi^2 (9 \Xt + 17 \Yt) - 52 \zeta_3 \biggr)
 	\MinusBreak{}
           {9 \over 4} \, {y \over x} ( \Yt^2 + \pi^2 )
         + {3 \over 2} \, y  {} \Bigl((X-Y)^2 + \pi^2 \Bigr) 
          - {\pi^2 \over 16 x} (8-15 y)
 	\MinusBreak{}
            {y \over 2 x}
               \biggl( {85 \over 9} \Xt + {58 \over 9} \Yt \biggr)
          + {143 \over 12} \, {y \over x} 
\,, \label{A3Smpmp}\\[1pt plus 4pt]
%
%
\BSusy{[1]}{-+-+} & = & 
            4 \biggl({y^2 \over x} - 3\biggr) 
                  \Biggl[ \li4\biggl(- {x\over y} \biggr) 
                                          + 2\li4(-x)
                     - 2 \li4(-y)
                     - {3\over 2}\, \Yt  {} (\li3(-x) - \zeta_3 ) 
 	\PlusBreak{4 \biggl({y^2 \over x} - 3\biggr) \BigglBl }
                       {3\over 2}\, \Xt {} (\li3(-y) - \zeta_3 ) 
                     - {1 \over 2} \Bigl( (X-Y)^2 - \pi^2 - 2 \Xt \Yt \Bigr) 
                             \li2(-x)
 	\MinusBreak{4 \biggl({y^2 \over x} - 3\biggr) \BigglBl}
                       {3 \over 8} X^3 Y + X^2 Y^2 
                     -  i {\pi \over 24} \, X^3 + i \pi X Y^2 
 	\PlusBreak{4 \biggl({y^2 \over x} - 3\biggr) \BigglBl}
                       {\pi^2 \over 24} \Bigl(X^2 - 20 X Y + 5 Y^2  
                                 - 2 i \pi {} (X + 2 Y)  \Bigr)
                    - {\pi^4 \over 240} \Biggr]
 	\PlusBreak{}
            12 \biggl({x\over y} + 2 (1-x) \biggr) \Biggl[ \li4(-x)
                     - {1 \over 2} \, \Xt {} (\li3(-x) - \zeta_3)
 	\PlusBreak{\null +12 \biggl({x\over y} + 2 (1-x) \biggr) \BigglBl}
                      {\pi^2 \over 6} \biggl(\li2(-x) - {1 \over 2} X^2 \biggr)
                             - {1\over 48} \, X^4
                     - {7 \over 180}\, \pi^4 \Biggr]
	\MinusBreak{}
            6 \Biggl[ 3\li4\biggl(- {x \over y} \biggr) + 3 \li4(-x) - \li4(-y)
               + \Xt \li3(-y) 
 	\PlusBreak{\null -6 \BigglBl }
               \biggl( (X-Y)^2 - \Xt^2 + {\pi^2 \over 6} \biggr) \li2(-x)
 	\PlusBreak{\null -6 \BigglBl }
               {5\over 24} (2 X^3 - 4 X Y^2 + Y^3) Y 
              -  {1 \over 6} \, X^4 
               - {1\over 4} \, i \pi X^3 + 3 \zeta_3 \Xt \Biggr]
 	\MinusBreak{}
            4 y \Biggl[ \li4\biggl( - {x \over y} \biggr) 
                    + 2 \li4(-x) + \li4(-y)
                    + {1\over 24} (4 X^3 - 8 X Y^2 + Y^3 ) Y 
	\MinusBreak{\null - 4 y \BigglBl}
                      {7\over 48} \, X^4 
                    + {1\over 2} \Bigl((X-Y)^2 - 2 \Xt^2 
                            + \Yt^2 \Bigr) \li2(-x) 
	\MinusBreak{\null - 4 y \BigglBl}
                      i {\pi\over 12} (X-Y) (X^2 + 10 X Y - 2 Y^2 )
                    + 3 \zeta_3 \Xt
	\PlusBreak{\null - 4 y \BigglBl}
                      {\pi^2 \over 24} \Bigl( 7 X {} (X-2 Y) + 8 Y^2 
                            + 2 i \pi {}(4 X + 3 Y) \Bigr)
                  - {23\over 240}\, \pi^4 \Biggr]
 	\MinusBreak{}
            (1-x) {y\over 12 x} \Biggl[ (3 X - 10 Y) X Y^2
                     - 2 i \pi {}\Bigl( 18 (X-Y) X + Y^2 \Bigr) Y
	\PlusBreak{\null - {1\over 12} (1-x) {y\over x} \BigglBl}
                       \pi^2 \Bigl( 25 X^2 + 4 X Y + Y^2 
                              + 2 i \pi {} (13 X + 3 Y) \Bigr)
                     - {26 \over 15} \, \pi^4 \Biggr]
	\PlusBreak{}
            3 {1+y^2 \over x} \Biggl[ \li3(-x) - \Xt \li2(-x) 
                - {3 \over 4}\, \Xt^2 (2 Y + i \pi)
	\PlusBreak{\null +3 {(1+y^2) \over x} \BigglBl}
                  {1 \over 12} \, X^3  
                 + {\pi^2 \over 12} (2 X - 10 Y - 3 i \pi) - 5 \zeta_3 \Biggr]
	\MinusBreak{}
          3 (1-x) {y \over x} \Biggl[ \li3(-y) - \Yt \li2(-y)
                   - {1 \over 8} \, \Xt^3 
                   + {\pi^2 \over 8} \, \Xt
                   + {5\over 4}\, \Xt^2 \Yt 
       	\MinusBreak{\null - 3 (1-x) {y \over x} \BigglBl }
                     {1 \over 2} (2 X + i \pi) \Yt^2
                   + {5\over 12} \, \pi^2  \Yt  
                   + i {\pi^3 \over 2} + 2 \zeta_3 \Biggr]
       	\MinusBreak{}
             {y^2\over 8 x} \Biggl[ \Xt^3
                      - 42 \Xt^2 \Yt 
                     + 36 \Xt \Yt^2 - 24 \Yt^3
                    - \pi^2 (\Xt + 26 \Yt) - 168 \zeta_3 \Biggr]
       	\MinusBreak{}
           {x^2 \over y} ( \Xt^2 + \pi^2 ) \Xt
          + {3 \over 8} \biggl( 3 \, {y^2 \over x} - 4 \biggr)
                        ( \Xt^2 + \pi^2 ) 
          + {3 \over 8} \, {y \over x} (2 - 3 y) ( \Yt^2 + \pi^2 ) 
       	\PlusBreak{}
            {3 \over 8} \, {y \over x}(2 x - 3 y) \Bigl((X-Y)^2 + \pi^2 \Bigr) 
           + {27 \pi^2 \over 8} \, {y^2 \over x}
\,, \label{B1Smpmp}\\[1pt plus 4pt]
%
%
\BSusy{[2]}{-+-+} & = & 
           - {6 \over y} \Biggl[ 3 \li4\biggl(- {x \over y} \biggr) 
                         - \li4(-x) - 3 \li4(-y)
                   + {1 \over 12}\, X^4 - {1 \over 2}\, X Y^3
       	\PlusBreak{- {6 \over y} \BigglBl}
                    {1 \over 8}\, Y^4
                   + \Bigl( 2 \Xt - 3 \Yt \Bigr) \li3(-x)
                   - {\pi^2 \over 6} \biggl(\li2(-x) - 2 X^2 
                                          - {3 \over 2} Y^2 \biggr)
       	\MinusBreak{- {6 \over y} \BigglBl}
                     \zeta_3 (2 \Xt - 3 \Yt) + {7 \over 72} \pi^4 \Biggr]
       	\PlusBreak{}
             4 {y^2 \over x} \Biggl[ \li4\biggl(- {x \over y} \biggr)
                     - \li4(-x) + 4 \li4(-y)
                   - 3 \Yt \li3(-y) 
                    + {1 \over 16} \, Y^4
       	\PlusBreak{\null + 4 {y^2 \over x} \BigglBl}
                     {1 \over 2} \Bigl( 2 \Xt {} (X-Y) 
                   - \Yt^2 - 2 \pi^2 \Bigr) \li2(-x)
                   + {3 \over 4} X^3 Y 
       	\PlusBreak{\null + 4 {y^2 \over x} \BigglBl}
                     i {\pi \over 12} \Bigl(X^3 + 18 X^2 Y 
                                              - 33 X Y^2 + Y^3 \Bigr)
                   - {5 \over 8}\, X^2 Y^2 
                   - {5 \over 6}\, X Y^3
       	\MinusBreak{\null + 4 {y^2 \over x} \BigglBl}
                     {\pi^2 \over 24} \Bigl(9 X^2 - 6 X Y - 10 Y^2 
                               + i \pi {} (4 X - 26 Y) \Bigr)
                   - {77 \over 720} \, \pi^4 \Biggr]
       	\PlusBreak{}
               6 \Biggl[ \li4\biggl(-{x\over y} \biggr) + 2 \li4(-x) 
                 - \li4(-y) + 2 \Xt \li3(-y)
       	\MinusBreak{\null + 6 \BigglBl}
                  {1 \over 2} \biggl(X^2 - 2 \Xt Y 
                     + {\pi^2 \over 3} \biggr) \li2(-x)
                 - {1 \over 12}\, X^4 - {1 \over 6}\, X^3 Y 
       	\PlusBreak{\null + 6 \BigglBl}
                   X^2 Y^2 - {1 \over 6} \, X Y^3
                 + {1 \over 24}\, Y^4
                 - i {\pi \over 6} (X^2 - 6 Y^2) X
       	\PlusBreak{\null + 6 \BigglBl}
                   {\pi^2 \over 24} \Bigl( X^2 - 20 X Y + 2 Y^2 
                             - 2 i \pi {} (X + 2 Y) \Bigr)
                 + {\pi^4 \over 40} \Biggr]
       	\MinusBreak{}
             4 y  \Biggl[ \li4\biggl(- {x\over y} \biggr) 
                    + 2 \li4(-x) - 2 \li4(-y)
                    - {3 \over 2} \, \Yt {} ( \li3(-x) - \zeta_3 ) 
      	\PlusBreak{\null - 4 y  \BigglBl }
                             {3 \over 2}\, \Xt {} (\li3(-y) - \zeta_3 ) 
                    + {1 \over 24}\, X^4 - {11 \over 24}\, X^3 Y 
                    + {19 \over 16} X^2 Y^2
       	\MinusBreak{\null - 4 y  \BigglBl }
                      {1\over 2} \Bigl((X-Y)^2 - \pi^2 
                                  - 2 \Xt \Yt \Bigr) \li2(-x)
       	\MinusBreak{\null - 4 y  \BigglBl }
                      {7 \over 24} \, X Y^3 + {1 \over 24} \, Y^4
                    + i {\pi \over 24} \Bigl(X^3 - 18 X^2 Y 
                                                 + 42 X Y^2 + Y^3 \Bigr)
       	\MinusBreak{\null - 4 y  \BigglBl }
                      {\pi^2 \over 48}  \Bigl( 3 X^2 + 12 X Y + 19 Y^2 
                               + 6 i \pi {} (X + 5 Y) \Bigr)
                    + {103 \over 720} \, \pi^4 \Biggr]
       	\MinusBreak{}
            3 \biggl( 2 {y^2 \over x} - x + 5 \biggr) 
                          \Biggl[ \li3(-y) + \zeta_3 
                  - \Yt \li2(-y) 
                  + i {\pi \over 4} (Y^2 + \pi^2 )
       	\PlusBreak{\null -3 \biggl( 2 {y^2 \over x} - x + 5 \biggr) \BigglBl}
                   {1 \over 12} (3 Y^2 +\pi^2) Y 
                  - X {} ( \Yt^2 + 3 \pi^2 )  \Biggr]
       	\MinusBreak{}
            3 \biggl(4 \, {y^2 \over x} - 2 (1-y) 
                                 + 3 \, {x^2 \over y} \biggr)  
                  \Biggl[ \li3(-x) - \zeta_3 - \Xt \li2(-x)
      	\PlusBreak{\null -3\biggl( 4 {y^2 \over x} - 2 (1-y) 
                                      + 3 {x^2 \over y} \biggr) \BigglBl }
                   i {\pi \over 2} \, X {}(X - i \pi) 
                        + i {\pi^3 \over 6} \Biggr]
  	\MinusBreak{}
             {1 \over 4} \Biggl[ \Bigl( 27 \Xt^2 
                                - 18 (X-Y)^2 + 17 \pi^2 \Bigr) \Yt
                   + 232 \pi^2 X 
  	\MinusBreak{\null - {1 \over 4} \BigglBl}
                  4 i \pi^3  - 48 \zeta_3 \Biggr]
           + {1 \over 2} \biggl( 4 \, {x^2\over y} - x + y \biggr)
                             \Bigl(\Xt^3 + \pi^2 (7 X+i \pi) \Bigr)
  	\MinusBreak{}
            {y \over 4} \Biggl[ 3 (2 X^2 - Y^2) Y 
                   + 3 i \pi {} (2 X^2 + 4 X Y - 3 Y^2 ) + 24 \zeta_3
      	\PlusBreak{\null - {y \over 4} \BigglBl} 
                    \pi^2 (20 X - Y + i \pi) \Biggr]
           + {3 \over 8} (2-x) \biggl(3 \, {y \over x} - {4 \over y} \biggr)
                            (\Xt^2 + \pi^2)
  	\MinusBreak{}
             {3 \over 8} \biggl( {8\over x} + y + 4 x \biggr) 
                          ( \Yt^2 + \pi^2) 
           + {27 \over 8} \, \pi^2 y 
  	\MinusBreak{}
           {3 \over 8} \biggl(6 \, {y \over x} + y - 12 x \biggr) 
                               \Bigl((X-Y)^2 + \pi^2 \Bigr)
\,. \label{B2Smpmp}
\end{eqnarray}
%


\end{document}